\documentclass[usenatbib]{mn2e}
\newcommand{\bib}{\bibitem[\protect\citeauthoryear}
\newcommand{\pa}{\,\rlap{\raise 0.5ex\hbox{$\propto$}}{\lower 1.0ex\hbox{$\sim$}}\,}
\usepackage{epsf}
\begin{document}
\title[Magnetoionic media around FR\,I radio galaxies]
{Structures of the magnetoionic media around
the FR\,I radio galaxies 3C\,31 and Hydra\,A}
\author[R.A. Laing et al.]
   {R.A. Laing \thanks{E-mail: rlaing@eso.org}$^{1}$, A.H. Bridle $^2$,
    P. Parma $^3$,  M. Murgia $^{3,4}$\\
    $^1$ European Southern Observatory, Karl-Schwarzschild-Stra\ss e 2, D-85748 
    Garching-bei-M\"unchen, Germany \\ 
    $^2$ National Radio Astronomy Observatory, Edgemont Road, Charlottesville,
    VA 22903-2475, U.S.A. \\
    $^3$ INAF -- Istituto di Radioastronomia, via Gobetti 101, I-40129 Bologna,
    Italy\\
    $^4$ INAF - Osservatorio Astronomico di Cagliari, Loc. Poggio dei Pini, Strada 54,
    I-09012 Capoterra (CA), Italy}

\date{Received }
\maketitle

\begin{abstract}
We use high-quality Very Large Array (VLA) images of the Fanaroff \& Riley Class
I radio galaxy 3C\,31 at six frequencies in the range 1365 to 8440\,MHz to
explore the spatial scale and origin of the rotation measure (RM) fluctuations
on the line of sight to the radio source.  We analyse the distribution of the
degree of polarization to show that the large depolarization asymmetry between
the North and South sides of the source seen in earlier work largely disappears
as the resolution is increased.  We show that the depolarization seen at low
resolution results primarily from unresolved gradients in a Faraday screen in
front of the synchrotron-emitting plasma. We establish that the residual degree
of polarization in the short-wavelength limit should follow a Burn law and we
fit such a law to our data to estimate the residual depolarization at high
resolution. We discuss how to interpret the structure function of RM
fluctuations in the presence of a finite observing beam and how to address the
effects of incomplete sampling of RM distribution using a Monte Carlo
approach. We infer that the observed RM variations over selected areas of
3C\,31, and the small residual depolarization found at high resolution, are
consistent with a power spectrum of magnetic fluctuations in front of 3C\,31
whose power-law slope changes significantly on the scales sampled by our data.
The power spectrum $P(f)$ can only have the form expected for Kolmogorov
turbulence [$P(f) \propto f^{-11/3}$] on scales \la 5\,kpc. On larger scales we
find $P(f)$ \pa $f^{-2.3}$. We briefly discuss the physical interpretation of
these results.  We also compare the global variations of RM across 3C\,31 with
the results of three-dimensional simulations of the magnetic-field fluctuations
in the surrounding magnetoionic medium.  We infer that the RM variation across
3C\,31 is qualitatively as expected from relativistic-jet models of the
brightness asymmetry wherein the apparently brighter jet is on the near side of
the nucleus and is seen through less magnetoionic material than the fainter
jet. We show that our data are inconsistent with observing 3C\,31 through a
spherically-symmetric magnetoionic medium, but that they are consistent with a
field distribution that favors the plane perpendicular to the jet axis --
probably because the radio source has evacuated a large cavity in the
surrounding medium.  We also apply our analysis techniques to the case of
Hydra\,A, where the shape and the size of the cavities produced by the source in
the surrounding medium are known from X-ray data.  We emphasise that it is
essential to account for the potential exclusion of magnetoionic material from a
large volume containing the radio source when using the RM variations to derive
statistical properties of the fluctuations in the foreground magnetic field.
\end{abstract}


\section{Introduction}
\label{Introduction}

The hot intracluster medium in rich clusters of galaxies is magnetised, with
field strengths which are important for the suppression of heat conduction, even
if not dynamically significant \citep[and references therein]{CT02}. The fields
are apparent via synchrotron radiation from cluster halo sources and Faraday
rotation of polarized emission from embedded and background sources (the subject
of the present paper). At high spatial resolution, the rotation of the {\bf
E}-vector position angle is accurately proportional to the square of the
wavelength and depolarization with increasing wavelength is small, indicating
that a foreground medium is responsible.  Imaging of these foreground Faraday
rotation measure (RM) variations across radio sources in clusters has recently
been used to estimate the power spectrum of RM and, by implication, of 
magnetic-field fluctuations, in turn leading to improved estimates of field
strength \citep{EV03,VE03,VE05,Murgia,Govoni06,Guidetti08}. RM fluctuations are
also seen against radio sources in sparser environments such as galaxy groups
(e.g.\ \citealt*{PBW84,LB87}), but the structure and strength of the magnetic
fields in these systems have not been studied in detail.

\begin{figure}
\begin{center}
\epsfxsize=6cm
\epsffile{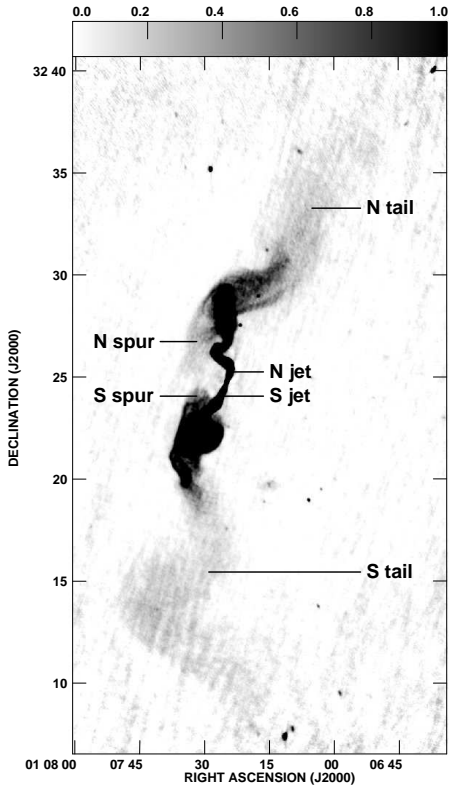}
\caption{Grey scale at 5.5~arcsec FWHM resolution showing the main regions of
3C\,31 in total intensity at 1.4 GHz as defined by \citet[fig.~1b]{lb08a}.  This
image shows the radio source in equatorial co-ordinates, while the displays of
depolarization and rotation measure below have been rotated anti-clockwise by
19.7 deg to make the inner jet axis vertical.
\label{layout}
}
\end{center}
\end{figure}

RM variations in clusters of galaxies have normally been modelled assuming that
the magnetic field is tangled either on a single scale \citep{Burn66,Felten96}
or over a range of scales (e.g.\ \citealt{EV03,Murgia}) and embedded in a
smoothly varying, spherically symmetric density distribution with parameters
derived from X-ray observations. This interpretation is complicated by three 
effects:
\begin{enumerate}
\item It is usually impossible to locate a radio galaxy along the line of sight
  within a cluster, and the path length over which Faraday rotation occurs is
  therefore not well determined.  It often has to be assumed that the sources
  are in the cluster mid-plane.
\item The sources are inclined to the line of sight and may be extended on
  scales comparable with changes in external density, so the Faraday depth will
  vary systematically over the radio structure. The observed correlation between
  depolarization and jet sidedness (in the sense that the side of the source
  containing the brighter jet depolarizes less rapidly with increasing
  wavelength) is most easily interpreted in this way: if the nearer jet is
  brighter as a result of Doppler boosting, then its emission is seen through
  less magnetoionic material and therefore suffers less Faraday rotation
  \citep{L88}.  Although the general association between depolarization
  asymmetry and jet sidedness is well documented in both weak and powerful
  sources \citep{Morg97,GC91}, there are as yet few sources that exhibit
  asymmetric depolarization for which it has been clearly established that the
  Faraday rotation is indeed due to a foreground medium (although there are no
  known counter-examples). The best-observed examples are all in central cluster
  galaxies with cooling cores: Cygnus\,A \citep*{DCP87}; M\,87 \citep*{OEK};
  Hydra\,A \citep{HydraA} and Hercules\,A \citep{HerA}.
\item Radio sources must interact with their environments, leading to local
  deviations from spherical symmetry in the density distribution. There is
  direct evidence for the presence of evacuated cavities in the hot plasma
  coincident with the radio lobes of sources in clusters, as predicted by models
  of radio-source evolution \citep{Scheuer74,McNN2007}. Magnetic fields in the
  surrounding medium may be important in stabilizing the cavities against
  Kelvin-Helmholtz and Raleigh-Taylor instabilities \citep{DP08}.
\end{enumerate}

In the present paper, we image and model the Faraday rotation distribution
across the bright FR\,I \citep{FR74} class radio source 3C\,31. This source is
identified with the nearby elliptical galaxy NGC\,383, which is the brightest
member of a rich group of galaxies \citep{Arp66,Zwi68}. We therefore probe a
different environment from the Abell clusters studied by previous authors. Hot
gas associated with the galaxy has been detected on both the group and galactic
scales by X-ray imaging \citep{KB99,Hard02}. The nucleus of the source is
located at the centre of the hot gas distribution, simplifying the geometry, and
the orientation is well constrained, at least for the inner jets \citep{LB02a}.
3C\,31 shows a pronounced depolarization asymmetry in the expected sense
\citep{Burch79,Strom83,And92}. Because it is bright, highly polarized, and
well resolved in both dimensions \citep[][and references therein]{lb08a}, the
variation of the Faraday rotation measure across it can be studied in unusually
rich detail.  We can make a stringent test of  the hypothesis of foreground
rotation and evaluate the spatial statistics of RM over different parts of the
source.  As the jet orientation is well constrained, we have a unique
opportunity to compare the global changes in RM fluctuation amplitude across the
structure with the predictions of three-dimensional models, including the
effects of departures from spherical symmetry.  3C\,31 therefore provides an
opportunity to examine the kiloparsec-scale structure of the magnetoionic medium
in an environment that may be more characteristic of the FR\,I population as a
whole than the clusters which have been the subject of many earlier studies.

We will show that our three-dimensional analysis of 3C\,31 is most consistent
with the hypothesis that the radio source has evacuated large cavities which are
essentially devoid of thermal plasma, although existing X-ray observations of
the NGC\,383 group are not deep enough to image such cavities if they are
present.  This motivated us to apply our techniques to the RM distribution of a
second source, Hydra\,A, in which the cavities are well characterised
\citep{Wise2007}.

\begin{figure*}
\epsfxsize=17cm
\epsffile{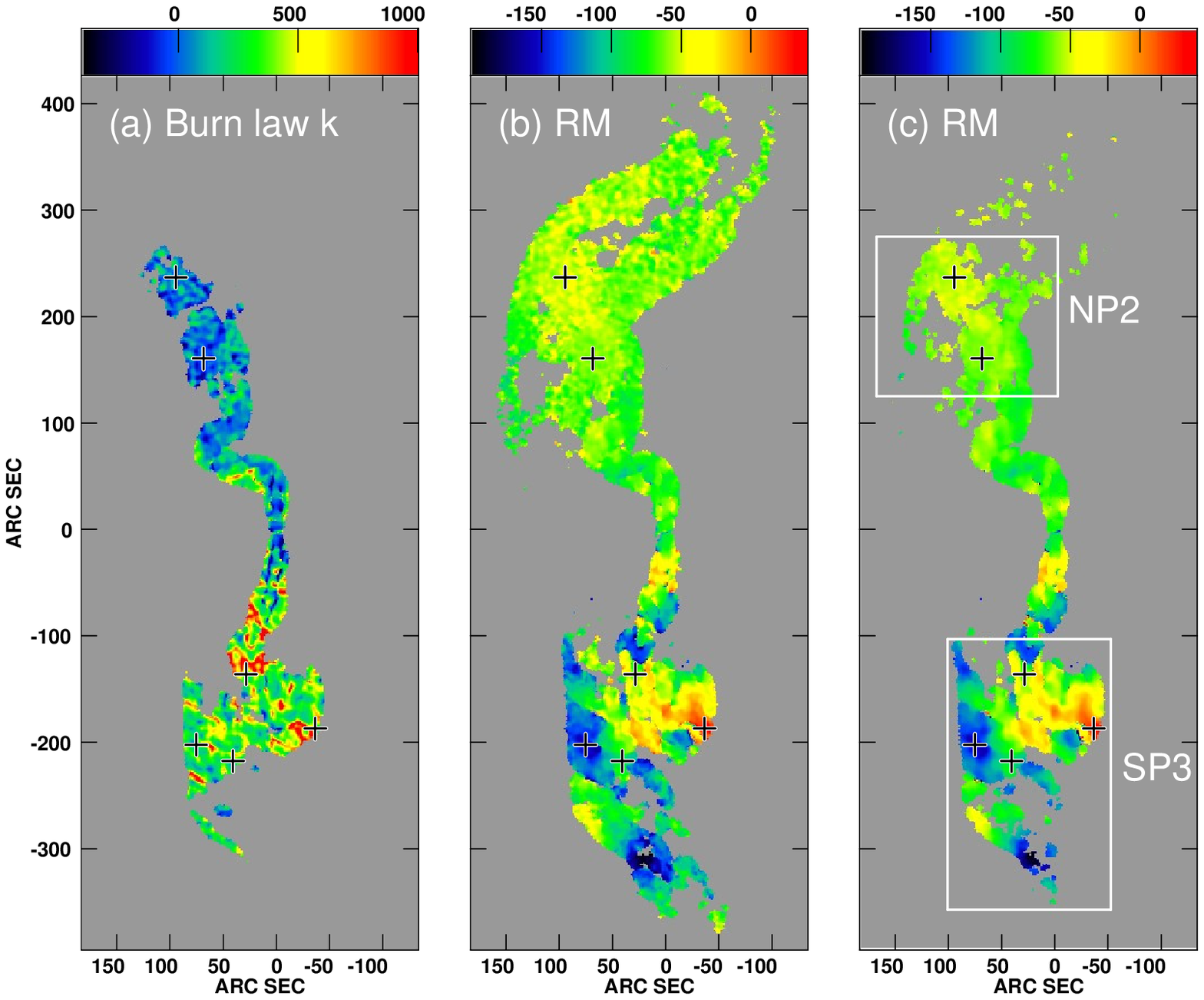}
\caption{Depolarization and rotation measure at a resolution
of 5.5\,arcsec FWHM. (a) Burn law $k$ in
rad$^2$\,m$^{-4}$, derived from a weighted least-squares fit to the relation
$\ln [p(\lambda)] = \ln [p(0)] - k\lambda^4$ between 4985 and 1365\,MHz.  The
colour range is from $-400$ to 1000\,rad$^2$\,m$^{-4}$. (b) RM from a fit to the four
frequencies between 1636 and 1365\,MHz, primarily to show the RM in the North
tail and spur. The range is $-$200 to $-$33\,rad\,m$^{-2}$. (c) as (b), but
using 5 frequencies between 4985 and 1365\,MHz. The errors are smaller than in
panel (b), but fewer points in the North of the source have sufficient signal to
determine the RM.  The rectangles marked `NP2' and `SP3' denote the regions for
the RM histograms in Fig.~\ref{RMhist}(a) and (e) and the structure-function
analysis of Section~\ref{Sfunc_observe}.  The crosses mark the positions for the 
plots in Figs~\ref{p-lambdasq-5.5} and \ref{PA5.5}. 
\label{fig:RMDP5.5}}
\end{figure*}

\begin{figure*}
\epsfxsize=17cm
\epsffile{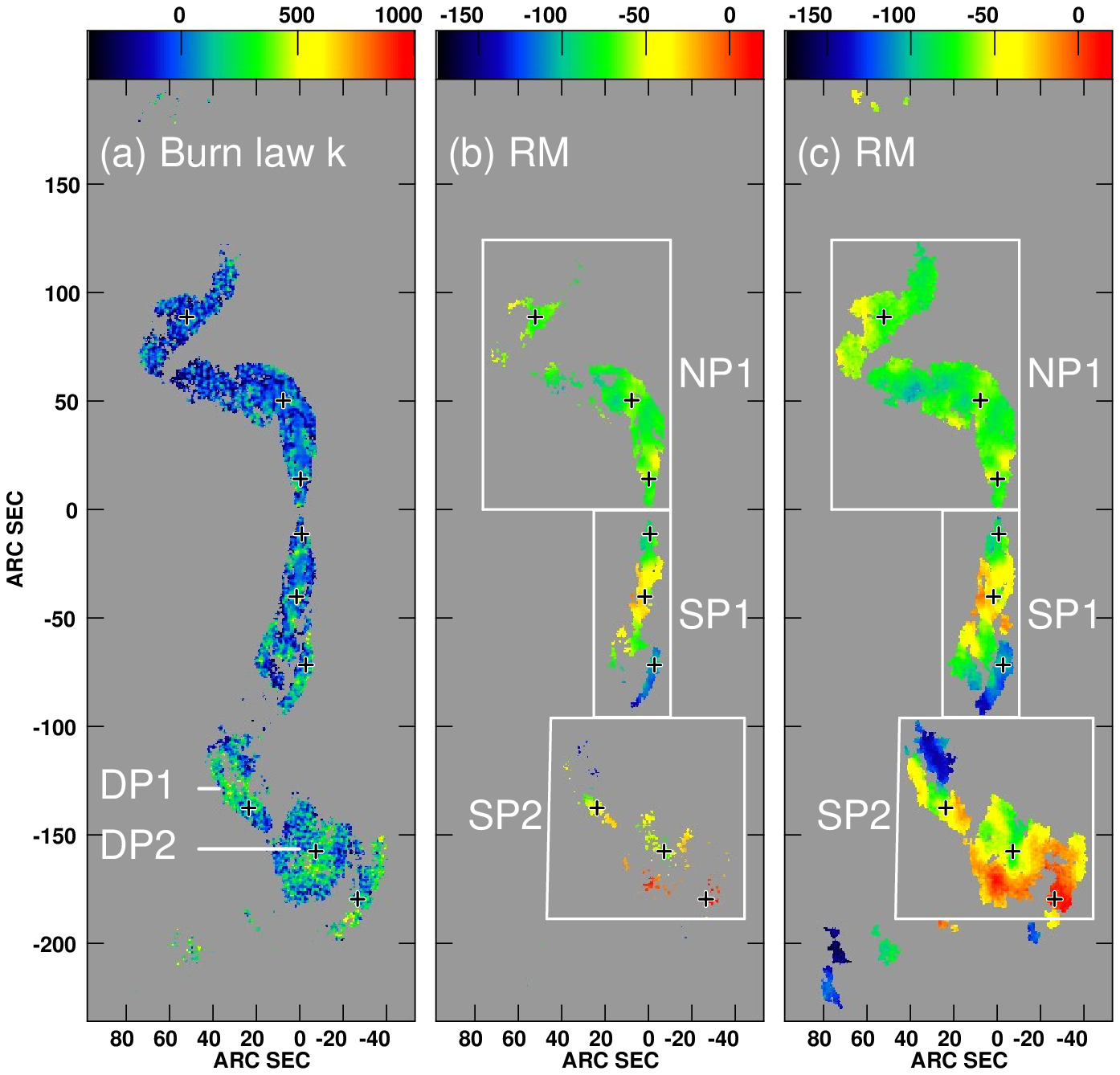}
\caption{Depolarization and rotation measure at a resolution of 1.5\,arcsec
FWHM.  (a) Burn law $k$ in rad$^2$\,m$^{-4}$, derived from a weighted
least-squares fit to the relation $\ln [p(\lambda)] = \ln [p(0)] - k\lambda^4$
for six frequencies between 8440 and 1365\,MHz.  The colour range is from $-400$
to 1000\,rad$^2$\,m$^{-4}$, as in Fig.~\ref{fig:RMDP5.5}(a).  (b) RM determined
by a fit to position-angle images at 6 frequencies between 1365 and 8440\,MHz
using a version of the {\sc aips} task {\sc rm} modified by G.B. Taylor.  The
colour range is $-$180 to $+$20\,rad\,m$^{-2}$. The boxes NP1, SP1 and SP2 show
the areas used for the RM histograms in Fig.~\ref{RMhist} and for the
structure-function analysis in Section~\ref{Sfunc_observe}. (c) As (b), but
using the {\sc pacerman} algorithm of \citet*{Pacerman}.  The crosses mark the
positions for the plots in Figs~\ref{p-lambdasq} and \ref{PA-lambdasq}.
\label{fig:RMDP1.5}}
\end{figure*}

Section~\ref{Pol} of the present paper presents the main features of our data on
the distributions of the Faraday rotation and depolarization of the polarized
emission from 3C\,31, and Section~\ref{over} gives an overview of our analysis
and general assumptions.  Sections~\ref{2d} and \ref{3d} discuss the structure
of the magnetic field in the foreground screen that is responsible for the
Faraday rotation measure fluctuations.  Section~\ref{2d} describes our analysis
of the two-dimensional spatial statistics of the RM
fluctuations. Section~\ref{3d} explores three-dimensional models of the
magnetoionic medium consistent with this two-dimensional analysis.
Section~\ref{Hydra} presents an analysis of the RM distribution in
Hydra\,A. Section~\ref{Conclusions} summarises our conclusions.  Some
mathematical details are given in Appendices~\ref{App-theory} and
\ref{psformulae} and an analysis of the effects of partially-resolved foreground
rotation is described in Appendix~\ref{SP3depol}.

Throughout this paper, we take NGC\,383 to have a redshift of 0.0169 (the mean
value from \citealt{Smith2000}, \citealt*{HVG} and \citealt{RC3}).  We assume a
concordance cosmology with a Hubble Constant $H_0$ = 70\,km
s$^{-1}$\,Mpc$^{-1}$, $\Omega_\Lambda = 0.7$ and $\Omega_M = 0.3$. At the
redshift of NGC\,383, this gives a linear scale of
0.344\,kpc\,arcsec$^{-1}$. For Hydra\,A, we take z = 0.0549 \citep{Smith2004} and
a linear scale of 1.067\,kpc\,arcsec$^{-1}$.

\section{The Observed Depolarization and Faraday Rotation in 3C\,31}
\label{Pol}

The observations and their reduction have been presented by \citet{lb08a}, along
with discussions of spatial variations in the radio spectrum and salient
features of the total intensity and the apparent magnetic-field structure -- all
intrinsic to the source.  The principal large-scale features of the source
discussed in that paper are shown schematically in Fig.~\ref{layout}.  The data
were obtained under NRAO observing proposals AF236 and AL405 at six frequencies
- 1365, 1435, 1485, 1636, 4985 and 8440\,MHz - with angular resolutions of 1.5
and 5.5 arcsec.  The high quality and good sampling of the $\lambda^2$ and
angular domains in these data make them well suited for analysis of the
statistics of the Faraday rotation and depolarization across 3C\,31.

Magnetised thermal plasma along the line of sight causes Faraday rotation of the
plane of polarization of linearly polarized radiation, quantified by the
rotation measure, RM [equations~(\ref{eq-RMdef}) -- (\ref{eq-RMnumbers})]. If
the thermal plasma is mixed with the emitting material or if there are
variations of foreground Faraday rotation across the beam, the emission tends to
depolarize with increasing wavelength.  We will show that all of our
observations are consistent with the expected variation of polarization with
wavelength for an almost completely resolved foreground screen. In
Section~\ref{partres}, we establish some essential theoretical results which
underpin this analysis. We discuss the observed depolarization, the evidence
that the Faraday rotating medium associated with 3C\,31 is entirely in front of
the radio source, and the variations of RM across the structure in
Sections~\ref{Depol}, \ref{Faraday-foreground} and \ref{RM-variations},
respectively.

\subsection{The effects of partial resolution for a foreground screen}
\label{partres}

The observing beam modifies the derived spatial statistics of RM. In practice,
we derive {\bf E}-vector position angle from $Q$ and $U$ images convolved with
the beam and use the position angles at multiple frequencies to calculate RM.
The beam has the effect of suppressing high spatial frequencies but for
wavelengths close to 0 (and consequently low depolarization), this effect can be
evaluated simply. The argument, which is fundamental to the analysis in 
Section~\ref{2d}, is as follows \citep{Tribble91a}. For a complex polarization
\begin{eqnarray*}
{\bf p}(\lambda) & = & |{\bf p}(\lambda)| \exp[2i\chi(\lambda^2)] \\
\end{eqnarray*}  
defined to have zero position angle for $\lambda = 0$, the RM at any wavelength is given by
\begin{eqnarray*}
{\rm RM} & = & \frac{\partial \chi}{\partial\lambda^2} \\
         & = & \frac{1}{2} \Im \left (\frac{\partial \ln{\bf p}}{\partial\lambda^2}
         \right ) \\ 
\end{eqnarray*}
In the case of pure foreground rotation, $|{\bf p}|$ is independent of
$\lambda$. If, in addition, it does not vary over a beam $W({\bf r_\perp})$ of
unit area, i.e.\ ${\bf p}({\bf r_\perp},\lambda) = {\bf p}_0$, where ${\bf p}_0$
is a constant and ${\bf r_\perp}$ is a vector in the plane of the sky, then the
measured polarization is
\begin{eqnarray*}
\overline{{\bf p}({\bf r_\perp},\lambda)} & = &|{\bf p}_0|  \int W({\bf r_\perp}-{\bf r^\prime_\perp})\exp[2i{\rm
    RM}({\bf r^\prime_\perp})\lambda^2] d^2{\bf r^\prime_\perp} \\
\end{eqnarray*}
The observed RM is derived by differentiating under the integral sign and taking
the zero-wavelength limit:
\begin{eqnarray}
\overline{{\rm RM}({\bf r_\perp})} & = & \int W({\bf r_\perp}-{\bf r^\prime_\perp}){\rm
    RM}({\bf r^\prime_\perp})d^2{\bf r^\prime_\perp} \label{eq:rm-conv}
\end{eqnarray}
i.e.\ {\em the measured RM distribution is closely approximated by the convolution of
the true RM distribution with the observing beam}.
We use this result to include the observing beam explicitly in
our analysis, verifying post hoc that the approximation is valid.

This approach also makes it easy to see why $\lambda^2$ rotation persists even
in the presence of small but significant depolarization.
Consider a
circular Gaussian beam $W({\bf r_\perp}) = \exp(-r_\perp^2/2\sigma^2)/2\pi\sigma^2$.  To
first order
\[
{\rm RM}({\bf r_\perp}) \approx {\rm RM}({\bf 0}) + {\bf r_\perp \cdot} \nabla {\rm RM}
\]
where $\nabla {\rm RM}$ is also evaluated at ${\bf r_\perp} = {\bf 0}$.
The integral for ${\bf p}(\lambda)$ can then be evaluated analytically to give:
\begin{eqnarray}
\overline{{\bf p}(\lambda)} & \approx &
|{\bf p}|\exp[2i({\rm RM}({\bf 0}) + {\bf r_\perp \cdot}\nabla {\rm RM})) \lambda^2]\nonumber  \\
&\times& \exp[-2|\nabla {\rm RM}|^2\sigma^2\lambda^4] \label{eq:partial-depol}
\end{eqnarray}
The first exponential term corresponds to $\lambda^2$ rotation and the second to
depolarization. Thus if the RM varies by a small amount across the beam, then we
expect $\lambda^2$ rotation even though there is significant (albeit small)
depolarization.  The wavelength dependence of depolarization has the functional
form derived by \citet{Burn66}:
\begin{equation}
p(\lambda) = p(0)\exp(-k\lambda^4) \label{eq-burn}
\end{equation}
with $k = 2|\nabla {\rm RM}|^2\sigma^2$, but note that this will not in general
persist at longer wavelengths \citep[see Section~\ref{innerscale} and 
Appendix~\ref{SP3depol}]{Tribble91a}.

\subsection{Depolarization}
\label{Depol}

The effect of depolarization between two wavelengths $\lambda_1$ and $\lambda_2$
is usually expressed in terms of the ratio ${\rm DP}_{\lambda_1}^{\lambda_2} =
p(\lambda_2)/p(\lambda_1)$, where $p(\lambda)$ is the degree of polarization.
Our six-frequency data set is capable of providing more sophisticated
information about the depolarization, but to make effective use of it we need a
suitable fitting function. As we have shown in Section~\ref{partres}, a Burn law
variation [\citealt{Burn66}; equations~(\ref{eq:partial-depol}) and
(\ref{eq-burn})], is expected for partially resolved foreground Faraday rotation
in the short-wavelength limit. The data at the vast majority of individual
pixels are fit to within the errors by this relation and we show images of $k$
at 5.5 and 1.5-arcsec resolution in Figs.~\ref{fig:RMDP5.5}(a) and
\ref{fig:RMDP1.5}(a).\footnote{The wavelength dependences at a small minority of
points in the 5.5-arcsec images of the Southern part of the source show
significant deviations from a Burn law.  We return to this and related issues in
Section~\ref{innerscale} and Appendix~\ref{SP3depol}.}  Errors in $k$ due to
rotation across the observing bands are negligible for the bandwidths used for
these observations \citep[][table~1]{lb08a} and the observed Faraday rotations
(see Section~\ref{Faraday-foreground}).

\begin{figure}
\epsfxsize=6.25cm
\epsffile{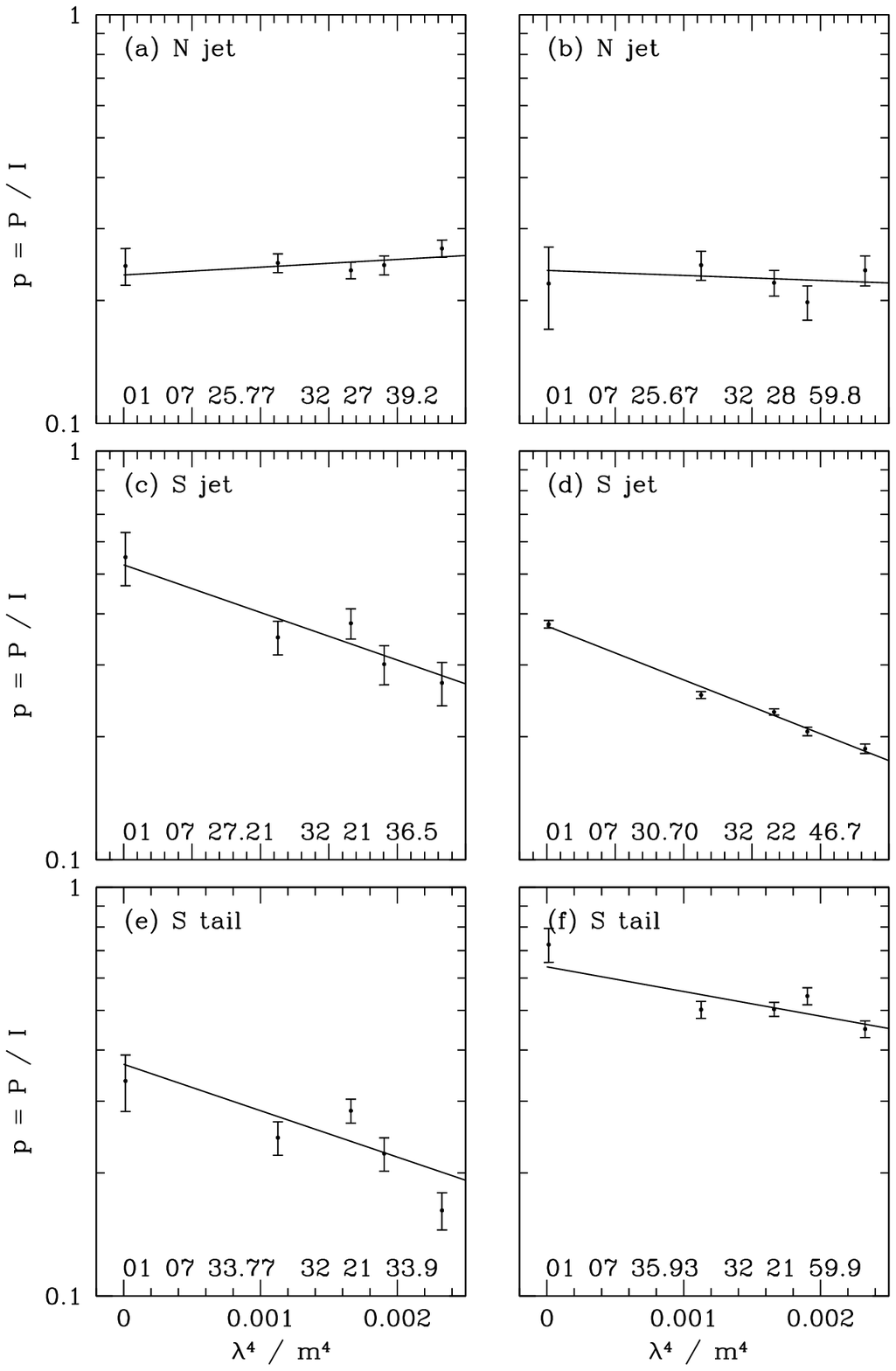}
\caption{Plots of degree of polarization, $p$ (log scale), against $\lambda^4$
at the positions indicated by crosses in Fig.~\ref{fig:RMDP5.5} and quoted in
the individual panels. The resolution is 5.5\,arcsec FWHM. Burn law fits
[equation~(\ref{eq-burn})] are also plotted.
\label{p-lambdasq-5.5}}
\end{figure}

\begin{figure}
\epsfxsize=8.5cm
\epsffile{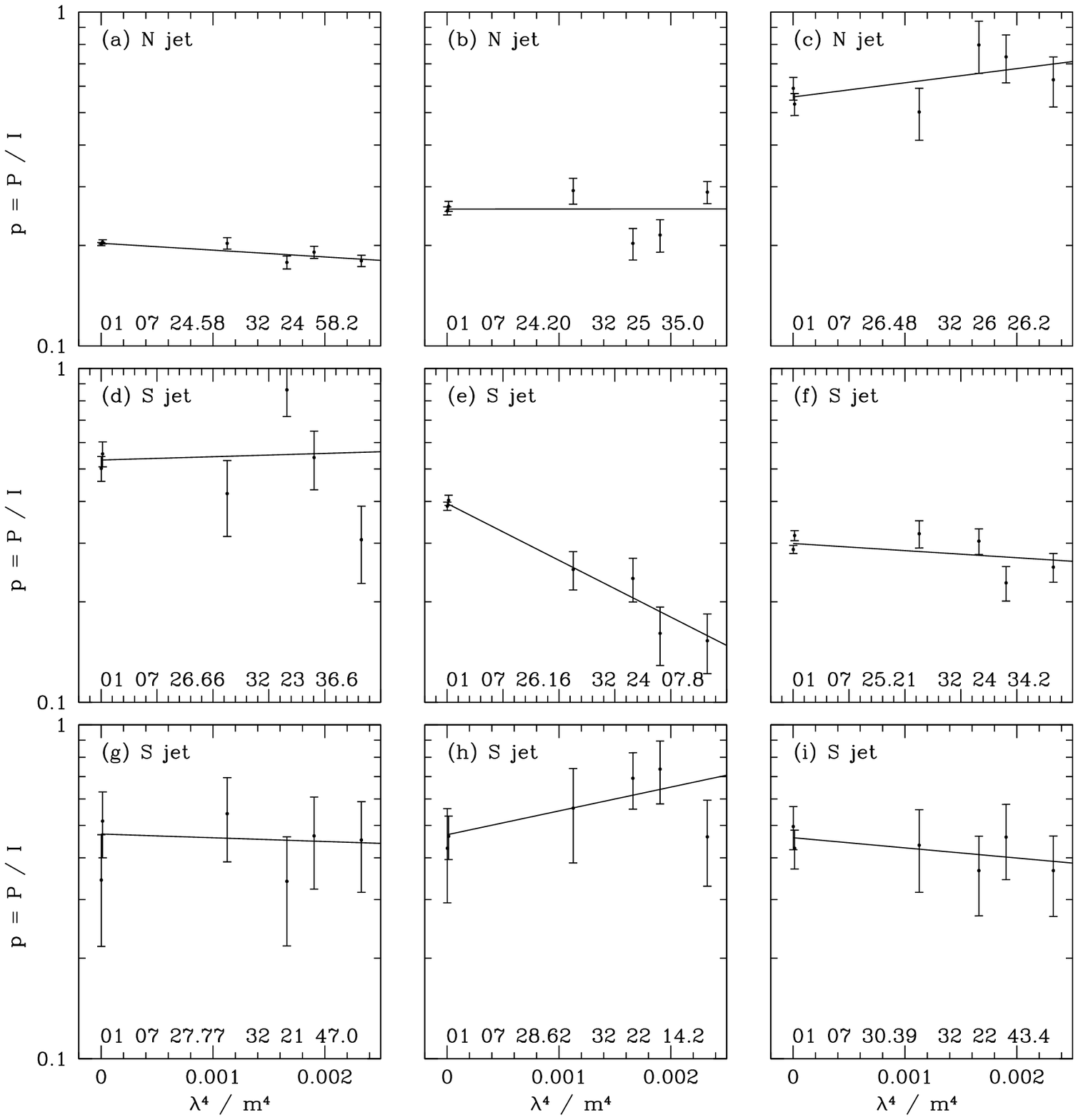}
\caption{Plots of degree of polarization, $p$ (log scale), against
$\lambda^4$
at the positions indicated
by crosses
in Fig.~\ref{fig:RMDP1.5} and quoted in the
individual panels.  The resolution is 1.5\,arcsec. Burn law fits
[equation~(\ref{eq-burn})] are also plotted. Of the fits in this figure,
only
(a) and (e) show significant depolarization.
\label{p-lambdasq}}
\end{figure}

\begin{figure}
\epsfxsize=6.25cm
\epsffile{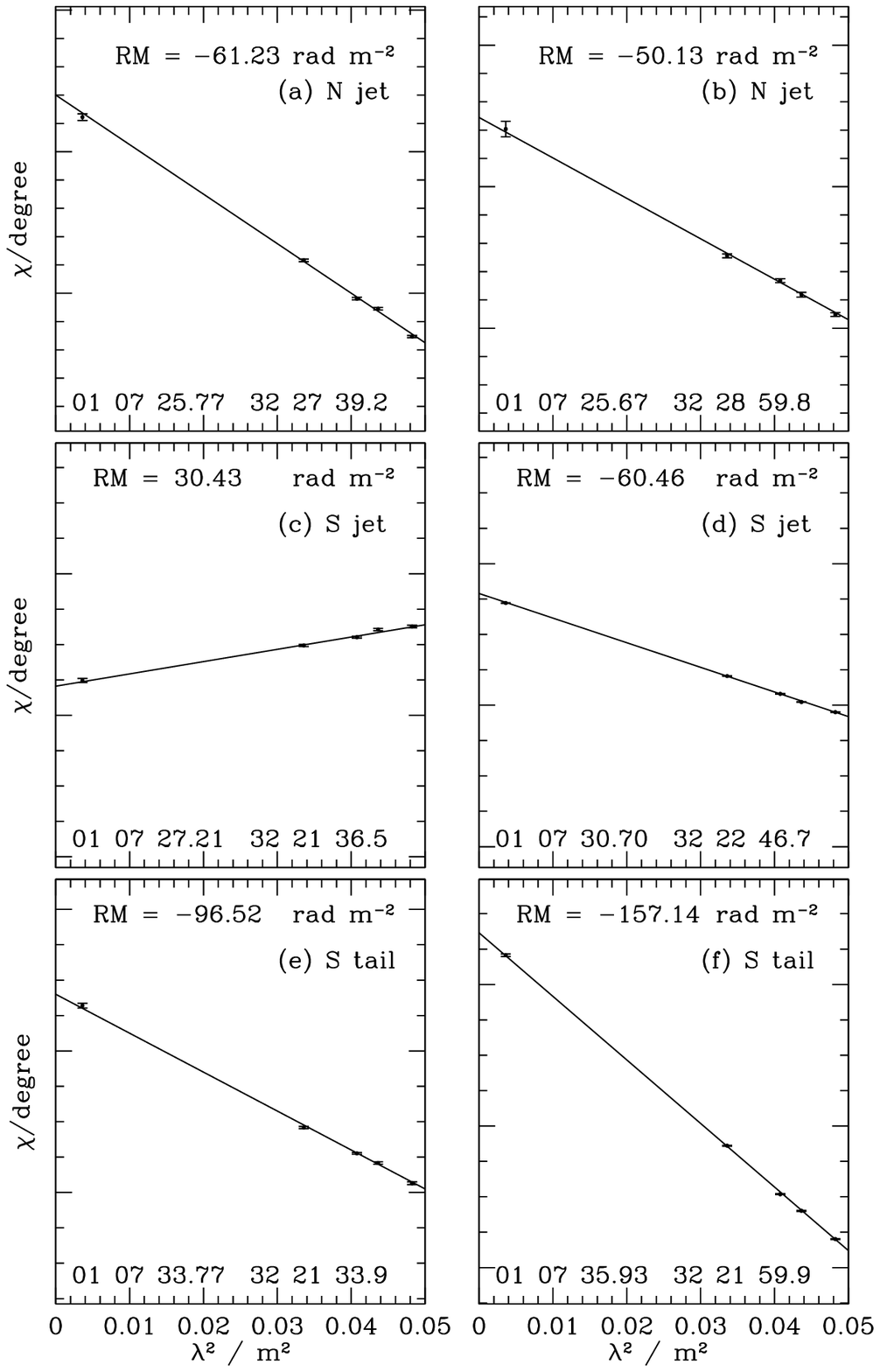}
\caption{Plots of {\bf E}-vector position angle $\chi$ against $\lambda^2$ at
the positions indicated in Fig.~\ref{fig:RMDP5.5}. The resolution is 5.5\,arcsec
FWHM. Fits to $\chi(\lambda^2) = \chi(0) + {\rm RM}\lambda^2$ are shown, and the
values of RM are given in the individual panels.  The ranges in angle are
300$^\circ$ in panels (a) -- (b) and 600$^\circ$ elsewhere.\label{PA5.5}}
\end{figure}

\begin{figure}
\epsfxsize=8cm
\epsffile{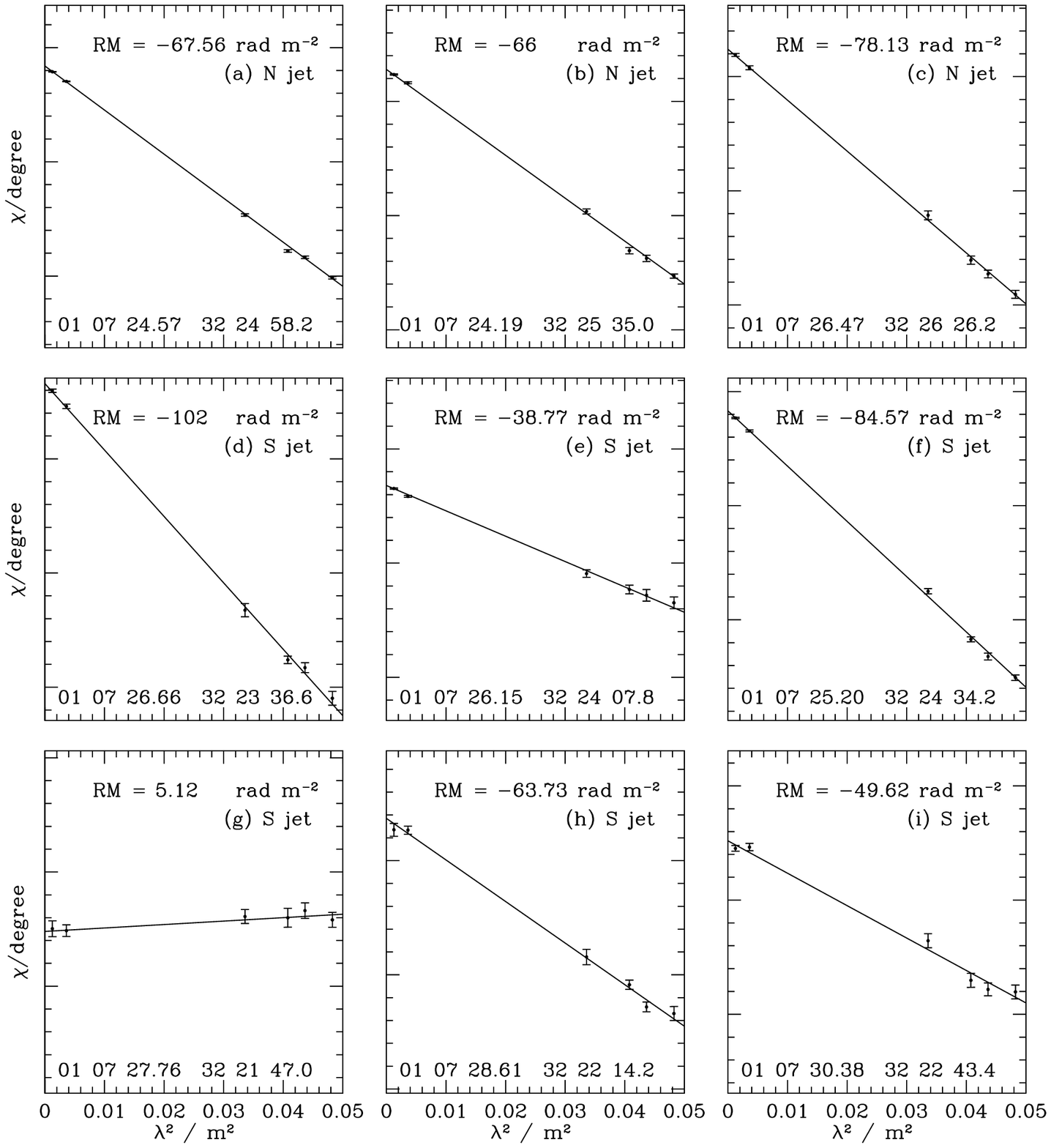}
\caption{Plots of {\bf E}-vector position angle $\chi$ against $\lambda^2$ at
the positions indicated by crosses in Fig.~\ref{fig:RMDP1.5}.  Fits to
$\chi(\lambda^2) = \chi(0) + {\rm RM}\lambda^2$ are shown, and the values of RM
are given in the individual panels. The resolution is 1.5\,arcsec FWHM and the
position-angle range is 300\,deg.
\label{PA-lambdasq}}
\end{figure}
The well-established depolarization asymmetry between the North and South sides
of 3C\,31 \citep{Burch79,Strom83} is still present at a resolution of
5.5\,arcsec, although reduced in amplitude from that in the lower resolution
data.  Fig.~\ref{fig:RMDP5.5}(a) shows $k$ derived from a fit to 5-frequency
data between 1365 and 4985\,MHz at this resolution. Example plots of $p$ against
$\lambda^4$ are shown in Fig.~\ref{p-lambdasq-5.5} in order to demonstrate that
the Burn law gives adequate fits for this frequency range and resolution.
Significant depolarization ($k > 0$) is found at most individual points, but a
minority have $k < 0$ due to noise or the effects of Faraday rotation on a
non-uniform distribution of intrinsic polarization. The mean values for the
North and South of the source are $\langle k \rangle = 83$\,rad$^2$\,m$^{-4}$
and $\langle k \rangle = 364$\,rad$^2$\,m$^{-4}$, respectively, corresponding to
DP$^{\rm 22cm}_{\rm 6cm} \approx$ 0.8 and 0.4, compared with $\sim$1 and 0.1 --
0.3 at resolution of 24\,arcsec $\times$ 47\,arcsec FWHM \citep{Strom83}.

Comparison between Figs~\ref{fig:RMDP5.5}(a) and \ref{fig:RMDP1.5}(a) shows
further that there is much less depolarization at 1.5-arcsec resolution.  The
data at this resolution are noisy and the fitted values of $k$ are consistent
with zero at the majority of individual points (many points therefore have $k <
0$). Of the sky positions shown in Fig.~\ref{p-lambdasq}, only (a) and (e) show
significant depolarization, for example. Positions with $k > 0$ tend to occur in
coherent clumps (the two most prominent are labelled DP1 and DP2 in
Fig.~\ref{fig:RMDP1.5}a), indicating that the residual depolarization in the
South is still significant, however.  Simulation of our image analysis and
fitting procedures shows that the distribution of errors in $k$ is
position-dependent and significantly asymmetrical over much of the source at
1.5-arcsec resolution, leading to biases in spatial averages. For quantitative
comparisons, we therefore use only data with $p > 4\sigma_p$ at all frequencies,
in which case we estimate systematic offsets in the mean values to be negligible
compared with the quoted random errors.  Averages over these positions confirm
not only the decrease in depolarization with increasing resolution, but also the
persistence of a residual difference between the two jets. In a high
signal-to-noise region within 60\,arcsec of the nucleus in the North jet, we
find $\langle k \rangle = 3 \pm 7$\,rad$^2$\,m$^{-4}$ compared with
32\,rad$^2$\,m$^{-4}$ at 5.5\,arcsec resolution. Similarly, for the inner
60\,arcsec of the South jet we find $\langle k \rangle = 55 \pm
7$\,rad$^2$\,m$^{-4}$ (1.5\,arcsec) and 166\,rad$^2$\,m$^{-4}$ (5.5\,arcsec).

Apart from the large-scale North-South asymmetry, there is no compelling
evidence for any detailed correlation between depolarization and source
structure (as might be expected if the effect occurred in a thin surface layer
just in front of the emitting material).

\begin{figure}
\epsfxsize=8.5cm
\epsffile{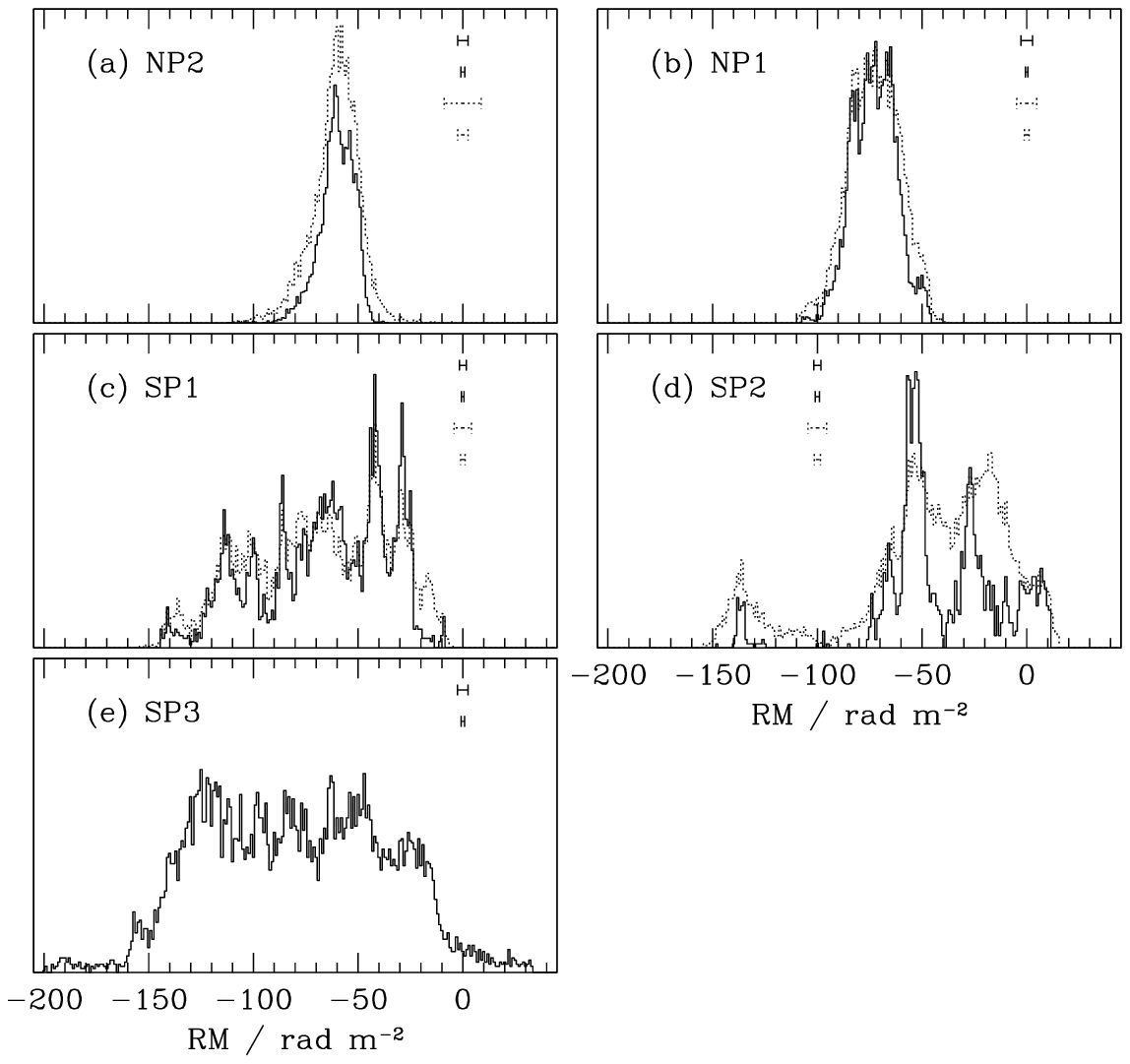}
\caption{Histograms of the RM distributions within the regions marked NP2, SP3
(Fig.~\ref{fig:RMDP5.5}), NP1, SP1 and SP2 (Fig.~\ref{fig:RMDP1.5}).  (a) NP2. The
RM distributions from 5- and 4-frequency fits at 5.5\,arcsec FWHM are shown by
full and dotted lines, respectively. (b) -- (d) NP1, SP1 and SP2. The
distributions at 1.5\,arcsec FWHM from the {\sc aips} and {\sc pacerman} fitting
algorithms are shown by the full and dotted lines, respectively.  (e) SP3, showing
the RM distribution from a 5-frequency fit at 5.5\,arcsec FWHM. The histograms
have been scaled independently to approximately the same maximum value and the
horizontal error bars show the mean and maximum rms errors derived from the RM
fits.\label{RMhist}}
\end{figure}

\subsection{Faraday rotation: images and evidence for foreground rotation}
\label{Faraday-foreground}

RM images at 5.5-arcsec resolution made using a version of the {\sc aips} task
{\sc rm} modified by G.B. Taylor are shown in Fig.~\ref{fig:RMDP5.5}(b) for the
1365$-$1636 MHz data alone and in Fig.~\ref{fig:RMDP5.5}(c) for all five
frequencies between 1365 and 4985\,MHz.  The former is noisier in the region of
overlap, but covers a wider area: the two images are consistent with each other
where they overlap.  Example fits of ${\bf E}$-vector position angle $\chi$
against $\lambda^2$ are shown in Fig.~\ref{PA5.5}.

There is a clear asymmetry in RM fluctuation amplitude, which is much greater in
the South.  In the outer North jet, spur and tail, the low depolarization
(Fig.~\ref{fig:RMDP5.5}a) and $\lambda^2$ rotation together indicate mostly
resolved foreground Faraday rotation.  For the parts of the North tail and spur
which show significant polarized signal at 1365 to 1636 MHz but which are not
detected at 4985\,MHz, the depolarization found at lower resolution between
these frequencies is also negligible \citep{Strom83}.  Large-scale coherent RM
variations ($\approx$10 \,rad\,m$^{-2}$ on scales from $\approx$10\,arcsec up to
$\approx$1\,arcmin) are observed in this part of the source
[Figs~\ref{fig:RMDP5.5}(b) and (c)].  These are large enough to account for the
depolarization of 0.3--0.4 observed between 1.4 and 0.6\,GHz with a beam of 29
$\times$ 55\,arcsec$^2$ \citep{Strom83}.  In the South spur and tail, despite
the large RM gradients and significant depolarization, the rotation remains
accurately proportional to $\lambda^2$ except at a few positions. These are
usually associated with abrupt changes in RM on scales comparable with the
observing beam and high depolarization.  We show in Appendix~\ref{SP3depol}
that these effects are quantitatively consistent with Faraday rotation by a
partially resolved foreground screen.

Images of rotation measure in the jets at 1.5-arcsec resolution using all 6
frequencies are shown in Fig.~\ref{fig:RMDP1.5}. We used two different fitting
programs: the modified {\sc rm} task [Fig.~\ref{fig:RMDP1.5}(b)] and the {\sc
pacerman} algorithm of \citet[Fig.~\ref{fig:RMDP1.5}c]{Pacerman}. In the former
case, we required that all {\bf E}-vector position angles at a given point had
rms error $<10^\circ$, but still found that a few points had wildly discrepant
RM values due to erroneous resolution of $n\pi$ ambiguities. Unlike the effect
noted at 5.5-arcsec resolution (where the RM anomalies are extended over several
beamwidths), the errant values occurred at isolated pixels and cannot represent
real RM structure. They were therefore edited out.  The {\sc pacerman} algorithm
solves for the $n\pi$ ambiguities in regions of high signal-to-noise ratio and
uses this information to resolve them in adjoining, fainter areas.  Where RM
solutions were found by both algorithms, they were in excellent agreement, but
{\sc pacerman} gave credible results for many more points with low signal-to-noise
ratio, especially in the South.

\begin{figure*}
\epsfxsize=12cm
\epsffile{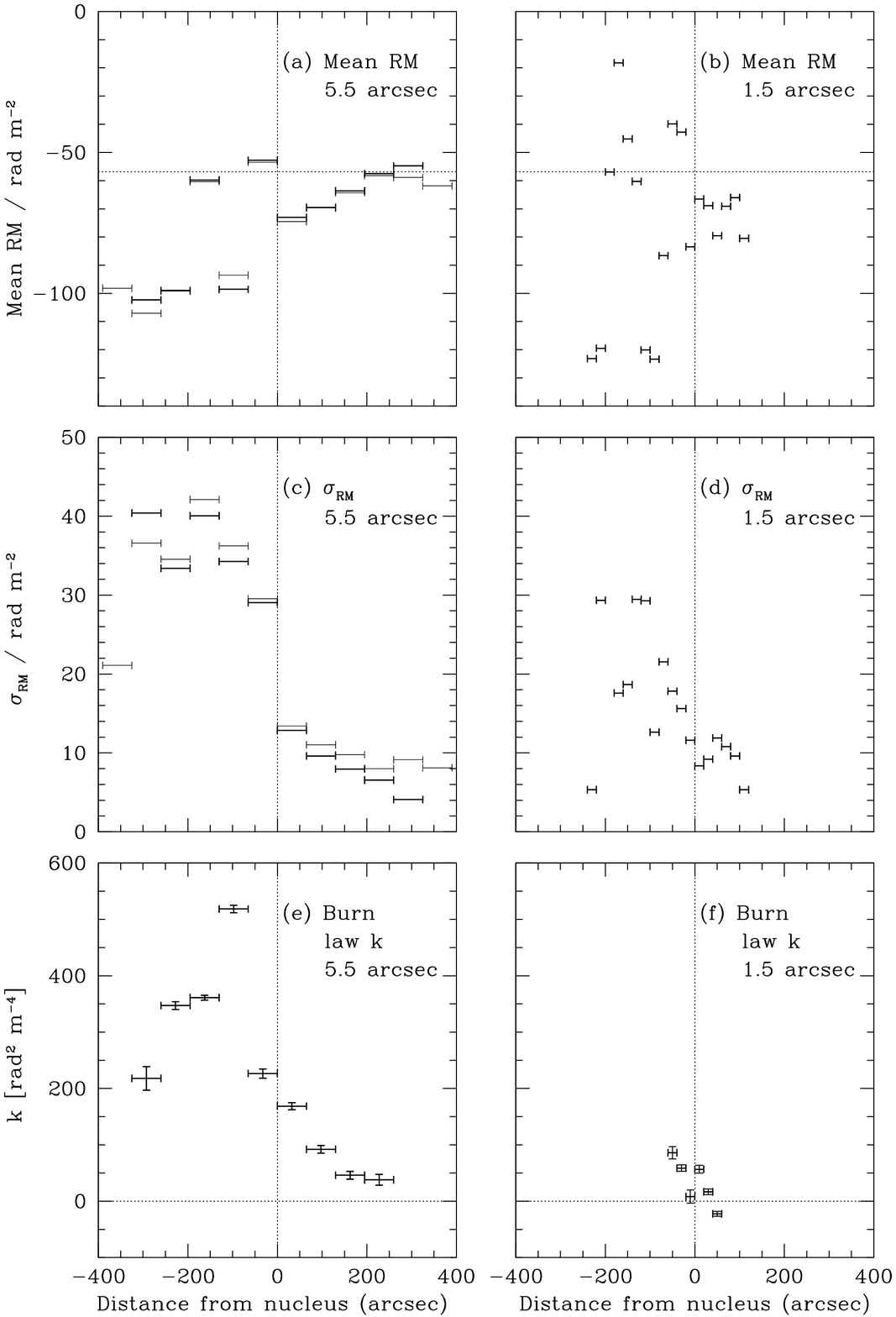}
\caption{Profiles along an axis defined by the inner jets in 3C\,31 (PA
  19.7$^\circ$). Positive distances are in the direction of the North jet and
  the vertical dotted lines show the position of the nucleus.  Quantities are
  evaluated for boxes of length 65 arcsec along the axis for the 5.5\,arcsec
  FWHM data (panels a, c and e) and 20 arcsec for the 1.5\,arcsec\ FWHM data
  (panels b, d, f), using the {\sc pacerman} fitting algorithm for the RM in the
  latter case.  The boxes are wide enough to include all unblanked points. (a)
  and (b) Mean RM, $\langle{\rm RM}\rangle$.  The horizontal dotted line shows
  our adopted value of the foreground rotation measure, RM$_0$. (c) and (d)
  $\sigma_{\rm RM}$, determined with respect to the local mean in the box.  The
  values of $\sigma_{\rm RM}$ have been corrected to first order for the offset
  due to noise on the fit.  (e) and (f) Mean value of the Burn law parameter
  $k$, with statistical uncertainties estimated from the least-squares
  fits. Positive values imply depolarization with increasing wavelength and $k =
  100$\,rad$^2$\,m$^{-4}$ corresponds to a depolarization ratio of 0.79 at
  1365\,MHz.\label{fig:RMprofiles}}
\end{figure*}

At 1.5-arcsec resolution, there are no significant deviations from the relation
$\Delta\chi \propto \lambda^2$ over a range of rotation $\Delta\chi$ up to
$\sim$300$^\circ$ (fits at representative points are shown in
Fig.~\ref{PA-lambdasq}). Taken with the observation that the depolarization is
very slight over our frequency range even in the South (Section~\ref{Depol}),
this implies that most, if not all of the Faraday rotation of the jet emission
must be produced by foreground material and that most of the depolarization
observed at lower resolutions results from RM fluctuations on scales larger than
1.5\,arcsec. The residual depolarization at 1.5\,arcsec resolution is probably
caused by RM fluctuations on even smaller scales. We cannot establish this
conclusively without higher-resolution data, but we show consistency of the
observed wavelength dependence and magnitude of the depolarization with models
of the spatial variation of foreground RM in Sections~\ref{partres} and
\ref{innerscale}, respectively. The most extreme depolarization (e.g.\ regions
DP1 and DP2) tend to be associated with large RM gradients, as expected for
foreground models [Figs~\ref{fig:RMDP1.5}(a) and (c)].

\subsection{Variation of Faraday rotation and depolarization along the jets}
\label{RM-variations}

The RM distributions for regions NP1, NP2, SP1, SP2 and SP3 are plotted as
histograms in Fig.~\ref{RMhist}.  The results for 4- and 5-frequency fits are in
good agreement for region NP2 [Fig.~\ref{RMhist}(a)], although the former are
inevitably noisier. The {\sc aips} and {\sc pacerman} algorithms give very
similar results for NP1 and SP1 but they disagree significantly in SP2, where
the former fits many fewer points [Fig.~\ref{RMhist}(d)]. We use the 1.5-arcsec
{\sc pacerman} images and the 5.5-arcsec, 5-frequency fits for quantitative
analysis. Note that the fitting errors (indicated by the error bars in
Fig.~\ref{RMhist}) are generally much smaller than the widths of the histograms.

In the North, the RM distributions are sharply peaked with a total range from
$-100$ to $-40$\,rad\,m$^{-2}$.  The inner N jet shows a similar range. In the
South, the RM range is much larger ($-200$ to $+30$\,rad\,m$^{-2}$), and the
distributions appear more uniform.

Fig.~\ref{fig:RMprofiles} shows profiles of the mean rotation measure, $\langle
RM\rangle$, the rms with respect to this local mean, $\sigma_{\rm RM}$ and the
parameter $k$ in the Burn depolarization law along the inner jet axis (position
angle 19.7 deg). The 5.5- and 1.5-arcsec data were averaged over boxes wide
enough to contain all unblanked pixels and of lengths 65 and 20\,arcsec along the
axis, respectively.  In order to avoid systematic errors, only points with $p >
4\sigma_p$ were used to generate the profiles of $k$.  We
made a first-order correction to the rms RM, $\sigma_{\rm RM raw}$, by
subtracting the fitting error $\sigma_{\rm fit}$ in quadrature to give
$\sigma_{\rm RM} = (\sigma_{\rm RM raw}^2 - \sigma_{\rm fit}^2)^{1/2}$, but the
effect of this correction is in any case very small.

There is a systematic gradient in $\langle RM\rangle$ along the North jet, but
the values vary erratically in the South. $\sigma_{\rm RM}$ increases from North
to South from a minimum of $\approx$5\,rad\,m$^{-2}$ at 5.5 arcsec resolution
(300\,arcsec North of the nucleus) to a maximum of 30--40\,rad\,m$^{-2}$
(depending on resolution) at 200\,arcsec to the South.  There is only evidence
for a decrease in the last box 400\,arcsec from the nucleus in the South, where
our data are limited in signal-to-noise and restricted to four observing
frequencies. The profile of $\langle k \rangle$ at 5.5-arcsec resolution
[Fig.~\ref{fig:RMprofiles}(e)] shows that depolarization also increases
monotonically from North to South across the nucleus, but there is evidence for
a decrease in the South at larger distances. At 1.5\,arcsec resolution,
depolarization is significant only in the South jet.  The differences between
the two resolutions are as expected for partially resolved RM fluctuations: at
higher resolution there is less depolarization and lower rms but variations
between box means are larger.  The increase in RM fluctuation amplitude going
from North to South across the nucleus is striking.

The very small variation in $\langle RM\rangle$ in the North of the source
suggests a Galactic origin.  Estimates of the Galactic foreground RM derived
from observations of nearby sources range from $-$64 to $-$39\,rad\,m$^{-2}$
\citep{And92} and the value predicted by the smoothed models of \citet{DC05} is
$-$49.3\,rad\,m$^{-2}$, both consistent with this idea. The best estimate of the
foreground RM from our own data comes from the region more than 200\,arcsec\
North of the nucleus of 3C\,31, where fluctuations are smallest
[Fig.~\ref{fig:RMprofiles}(a)].  In what follows, we adopt the unweighted mean RM
from the 5-frequency fits over this region, RM$_0 = -$56.9\,rad\,m$^{-2}$, as
the Galactic contribution. Note, however, that we cannot exclude a significant
large-scale component of RM local to this part of the source
(Section~\ref{Sfunc_observe}).

\subsection{Limits on internal Faraday rotation}

Our limits on depolarization and deviations from $\lambda^2$ rotation, within
60\,arcsec of the nucleus in the North jet can potentially constrain the
internal density of thermal electrons.  For a line of sight intersecting the jet
axis 28\,arcsec from the nucleus in projection (the maximum extent of the
conservation-law model of \citealt{LB02b}), our kinematic model gives a path
length through the jet of 7.3\,kpc and an on-axis equipartition magnetic field
of 1.5\,nT \citep{LB02a}.  We adopt these as fiducial numbers, noting that there
will be variations in path length and field along and transverse to the jet axis
and that the component of the magnetic field along the line of sight will be
reduced due to projection.

We use the formulae given by \citet{Burn66}, taking a conservative upper limit
of $\langle k \rangle <$ 20\,rad$^2$\,m$^{-4}$ (equivalent to a depolarization
of $>$0.95 at our lowest observing frequency of 1365\,MHz).  For a uniform
magnetic field $B$ along the line of sight and a path length $L$, this
corresponds to a limit on the internal density of
\begin{equation}
n/{\rm m}^{-3} \la 120 (B/1.5{\rm nT})^{-1}(L/7.3{\rm kpc})^{-1}
\label{eq-vec}
\end{equation}
The lack of deviation from a $\lambda^2$ law requires that no more than
$\approx$45$^\circ$ of the total rotation is due to thermal matter distributed
uniformly within the jet, giving $n \la$ 370\,m$^{-3}$ for the fiducial
parameters and a uniform field.  If, on the other hand, we assume that the field
component along the line of sight has a Gaussian distribution with rms $B$ and
$N$ reversals through the jet, we find
\begin{equation}
n/{\rm m}^{-3} \la 40 (B/1.5{\rm nT})^{-1}(L/7.3{\rm kpc})^{-1}N^{1/2}
\label{eq-dis}
\end{equation}
Our modelling of the magnetic-field structure in the jets \citep{LB02a} implies
that the magnetic field is dominated by the toroidal component, with a smaller
longitudinal contribution.  There are two extreme possibilities \citep*{LBC}:
\begin{enumerate}
\item The dominant toroidal component is vector-ordered, but the longitudinal component
  has many reversals. We would then expect systematic variations in Faraday
  rotation transverse to the jet axis, which we do not observe.
\item The field is entirely disordered, but anisotropic.
\end{enumerate}
In either model, there would be systematic (and so far unobserved) variations of
depolarization across the jets, together with a gradient in RM if the toroidal
component is ordered.  A pure toroidal field has no component along the line of
sight on the projected jet axis, so we would expect to see little Faraday
rotation and depolarization there.  The density limits for there more realistic
field configurations would be a few times larger than those given by
equations~(\ref{eq-vec}) and (\ref{eq-dis}) for cases (i) and (ii),
respectively. The number of reversals, $N$, in the fully disordered case could
be very large.

Even in the case of a fully ordered field, these limits are fully consistent
with the densities inferred from our conservation-law analysis on the assumption
that the thermal electrons originate entirely from entrained plasma with normal
cosmic abundances: $n \approx$ 0.7\,m$^{-3}$, also at a projected distance of
28\,arcsec from the nucleus \citep{LB02b}.
 
\section{Overview of the analysis}
\label{over}

\subsection{Description}

We seek to interpret the depolarization and Faraday rotation 
results in terms of the magnetic field structure in the Faraday-rotating medium,
first by deriving the two-dimensional statistics of the magnetic-field fluctuations 
that produce the observed Faraday rotation (Section~\ref{2d}) and then by 
constructing models of three-dimensional distributions of gas and magnetic field 
which are consistent with these statistics (Section~\ref{3d}).

Our analysis proceeds in three stages:
\begin{enumerate}
\item The two-dimensional fluctuations of Faraday rotation measure on scales
larger than the observing beamwidth can be quantified in several ways.  We will
justify use of the RM structure function (Section~\ref{Sfunc_theory}) as the
principal statistic. This restricts us to regions small enough that lateral
variations of the foreground density and rms field strength can be neglected.
Our approach is to guess a form for the magnetic power spectrum, derive the
corresponding RM structure function (modified by the observing beam) and compare
it with the data, estimating errors for the poorly-sampled large separations by
making a set of artificial RM images with the theoretical power spectrum sampled
on the same grid as our data.
\item Residual depolarization can be used to estimate the variations of Faraday
rotation across the beam \citep{Burn66,Tribble91a}, thus extending the range of
spatial scales we can study. We evaluate the depolarization expected from
different power spectra numerically, by making finely-sampled realisations of
RM images, which we then use to generate artificial polarization maps at the
observing resolution.
\item For a large-scale foreground medium, the variations of RM fluctuation
amplitude and depolarization with position depend on the orientation of the
source \citep{GC91,Tribble92}.  Knowing the density distribution from the X-ray
observations and the geometry, the strength of the magnetic field can be
estimated by fitting to the observed distributions using analytical single-scale
models (e.g.\ \citealt{Felten96}), appropriately weighted estimates of the
autocorrelation function or power spectrum \citep{EV03,VE05} or 
three-dimensional simulations \citep{Murgia,Govoni06,Guidetti08}.  We use
simulations, incorporating constraints on the magnetic-field power spectrum from
(i) and (ii).  Unique features of our analysis are that the orientation of
3C\,31 (at least close to the nucleus) is well determined and we have enough
data to test whether the Faraday-rotating medium is spherically symmetric with a
radial profile related to the density of the surrounding hot plasma.
\end{enumerate}
The advantage of the approach using two-dimensional RM and depolarization
analyses [steps (i) and (ii)] is that when making artificial RM images, we can
use large computational grids to sample a full range of spatial scales
accurately and efficiently, reserving three-dimensional simulations for studies
which include global variations of density and magnetic-field strength.  We can
also use the Hankel transform relation (Section~\ref{Hankel}) to predict the
structure function expected for a given power spectrum, allowing explicitly for
the effects of the observing beam.

\subsection{Simplifying assumptions}
\label{RMstats}

We make a number of simplifying assumptions, as follows.
\begin{enumerate}
\item Faraday rotation is produced entirely by a foreground screen. We have
  argued in Section~\ref{Faraday-foreground} that this is a very good
  approximation for 3C\,31. 
\item The magnetic field is a Gaussian random variable, and its spatial
  distribution can therefore be described by the power spectrum or its Fourier
  transform, the autocorrelation function. This may well be incorrect if the
  field is intermittent, e.g.\ in the form of flux tubes or filaments \citep[and
  references therein]{EO02}, but our data are too sparse to allow us to
  determine higher-order correlations.
\item The field is isotropic, in the sense that it has no preferred direction
  when averaged over a sufficiently large volume. Our data show no evidence for
  a preferred direction in the RM distribution on scales $\la$100\,arcsec, but
  we note that this may not always be the case: for example the RM distributions
  in M\,84 \citep{LB87} and 3C\,465 \citep{EO02} appear to have banded
  structures. We do see RM gradients on very large scales in 3C\,31, but these
  are consistent with sampling of an isotropic magnetic power spectrum with
  significant amplitude on these scales together with spatial variations in
  density.
\item The amplitude of the magnetic-field power spectrum varies with the thermal
  gas density in the foreground screen, which is in turn a smooth function of
  position. The form of the power spectrum, in contrast, is independent of
  position. We test this assumption by evaluating the spatial statistics in
  different regions of the source.
\end{enumerate}
The importance of these assumptions is that the power spectra of the magnetic
field and RM are then proportional \citep{EV03}. We work in terms of spatial
frequency ${\bf f} = (f_x, f_y, f_z)$ with ${\bf f}_\perp = (f_x, f_y)$ in
the plane of the sky. In the isotropic case, we can define the three-dimensional
magnetic power spectrum $\hat{w}(f)$ such that $\hat{w}(f)df_x df_y df_z$ is the
power in a volume $df_x df_y df_z$ of frequency space.\footnote{Note that some
authors use the one-dimensional form $4\pi f^2 \hat{w}(f)$. For the
three-dimensional form, as used in this paper, a Kolmogorov power spectrum is
$\hat{w}(f) \propto f^{-11/3}$ whereas the one-dimensional form is $4\pi
f^2 \hat{w}(f) \propto f^{-5/3}$.} 

Similarly, the two-dimensional RM power spectrum $\hat{C}(f_\perp)$ is defined
such that $\hat{C}(f_\perp)df_x df_y$ is the RM power for an element $df_x
df_y$. For a foreground screen of uniform depth and constant electron density,
$\hat{w}(f)$ and $\hat{C}(f_\perp)$ are simply related
[equation~(\ref{eq-what-chat})].  If the electron density and the normalization of
$\hat{w}(f)$ vary significantly along the line of sight but not perpendicular to
it, the functional forms of $\hat{w}(f)$ and $\hat{C}(f_\perp)$ are still
identical \citep{EV03}.  Thus we may use the RM structure function evaluated
over limited areas to determine the {\em form} of $\hat{w}(f)$, but estimating its
{\em normalization} as a function of position requires a three-dimensional analysis.

\section{Two-Dimensional Analysis}
\label{2d}

\subsection{Choice of analysis technique}
\label{Methods-2D}

Our goal is to characterise the RM over areas of
the radio source where the parameters which define its spatial statistics do not vary
significantly.  Descriptions in terms of spatial or frequency variables are
equivalent and several different techniques have been used in the
literature.  One approach is to work directly in frequency space
\citep{EV03,VE03} to determine the power spectrum.  Real-space alternatives
are: 
\begin{enumerate}
\item the {\it autocorrelation function} 
\[
C(r_\perp) = \langle{\rm RM}({\bf r_\perp} +{\bf r_\perp^\prime)}{\rm RM}({\bf
r_\perp^\prime})\rangle
\]
\citep{EV03,VE03} and
\item
the related {\it structure function} defined by:
\[
S(r_\perp) = \langle [{\rm RM}({\bf r_\perp}+{\bf r^\prime_\perp})-{\rm RM}({\bf r^\prime_\perp})]^2\rangle 
\]
(e.g.\
\citealt*{SCS,Leahy,MS96}).
\end{enumerate}
Here, the vectors ${\bf r_\perp}$ and ${\bf r^\prime_\perp}$ are again in the
plane of the sky and $\langle\rangle$ denotes an average over ${\bf
r^\prime_\perp}$, defined by the area of the radio source.  For a stationary
random process, the structure and autocorrelation functions are related by
$S(r_\perp) = 2[C(0)-C(r_\perp)]$, although in practice, this relation will be
approached only if the largest scale of fluctuations is significantly smaller
than the size of the measurement area.

We work in real space and use the structure function rather
than the autocorrelation function for the following reasons:
\begin{enumerate}
\item Most of the useful properties of the autocorrelation function only appear
  when $C(r_\perp)$ drops to 0 at some outer scale small compared with the
  sampled area. If this is not the case (as in 3C\,31), the form of the function can vary
  wildly, typically oscillating with large amplitude at large separations.
\item A related point is that sensible use of the autocorrelation function
  requires an accurate knowledge of the zero level. This is uncertain for
  3C\,31: we have argued that the mean RM in the North tail is primarily Galactic,
  but there could still be significant contributions from local material. The
  zero-level for a particular region cannot be determined merely by averaging in
  the presence of large-scale fluctuations.
\item This problem is made worse by the irregular shapes of the sampling
  regions in 3C\,31.  This leads to particular problems with the frequency-space
  approach, as the corresponding window functions \citep{EV03} are very complicated.
\item By comparison, the structure function $S(r_\perp)$ is independent of the mean
  level and (to first order) of structure on scales larger than the measurement
  area. 
\end{enumerate}
$S(r_\perp)$ is therefore relatively robust and can use all of the available
evidence without giving systematic biases at the largest separations.

\subsection{Theoretical structure functions}
\label{Sfunc_theory}

\subsubsection{Hankel transform relations}
\label{Hankel}

For isotropic fluctuations, the RM power spectrum $\hat{C}(f_\perp)$ is the
Hankel transform of $C(r_\perp)$. Using the Fourier transform convention of
\citet{Bracewell},
\begin{eqnarray}
C(r_\perp) & = & 2\pi \int^\infty _0 \hat{C}(f_\perp) f_\perp J_0(2\pi f_\perp
r_\perp) df_\perp \label{eq-Hankel}
\end{eqnarray}
where $\hat{C}(f_\perp)$ depends only on the scalar spatial frequency
$f_\perp$. \footnote{Note that \citet{EV03} use wavenumber rather than spatial
frequency and a different Fourier transform convention, but their power spectra
are the same as ours with the substitution $f_\perp = k_\perp/2\pi$.}  In the
short-wavelength limit for a beam of FWHM = $2\sigma (\ln 2/\pi)^{1/2}$, the
relation becomes
\begin{eqnarray}
C(r_\perp) & = & 2\pi \int^\infty _0 \hat{C}(f_\perp) f_\perp J_0(2\pi f_\perp
r_\perp) \nonumber \\
& \times  & \exp(-2\pi \sigma^2 f_\perp^2) df_\perp \nonumber \\ \label{eq-Hankel-convolve}
\end{eqnarray}

\subsubsection{Power-law power spectra and the effects of the observing beam}

\begin{figure}
\epsfxsize=8.5cm
\epsffile{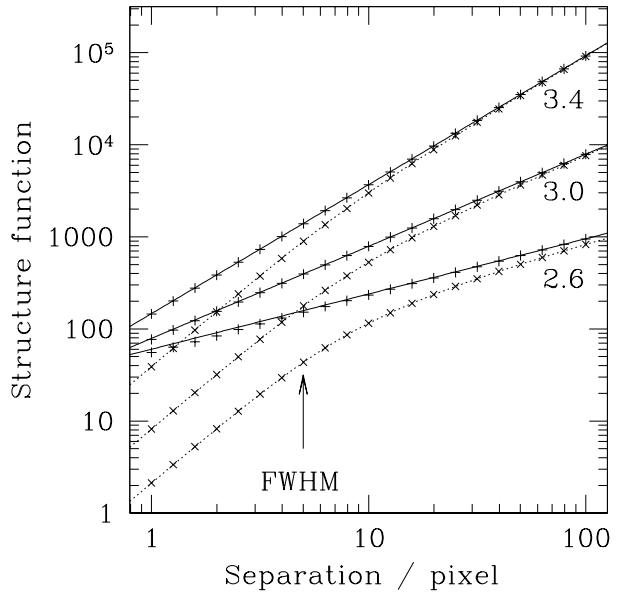}
\caption{Full lines: plots of the structure functions corresponding to power
  spectra $\hat{C}(f_\perp) \propto f_\perp^{-q}$ (for $p$ = 2.6, 3.0 and 3.4)
  over an infinite frequency range, from equation~(\ref{eq-asymp}). Dotted:
  structure functions for the same power spectra, but for images convolved with
  a Gaussian of FWHM 5\,pixels [equation~(\ref{eq-hyper})]. The crosses are the
  results of numerical integration. The agreement between analytical and
  numerical results is very good except for the case of a flat power spectrum
  without convolution, where there are small but significant errors at small
  separations (this case is not relevant to the observations, however).
\label{fig:convolve}}
\end{figure}

In what follows, we use finite frequency intervals and therefore need to
evaluate the Hankel transform integrals [equations~(\ref{eq-Hankel}) and
(\ref{eq-Hankel-convolve})] numerically. As a test of our procedure,
Fig.~\ref{fig:convolve} shows a comparison of analytical and numerical estimates
for structure functions corresponding to power-law $\hat{C}(f_\perp)$ for an
infinite frequency range, both before and after Gaussian convolution of the
image.  The analytical expressions are given in equations~(\ref{eq-asymp}) and
(\ref{eq-hyper}).

The effects of convolution on the observed structure function are very
noticeable, especially for flat power spectra. For example, the structure
function for $q = 2.6$ deviates significantly from its asymptotic form even for
separations $r_\perp \approx 5 \times $FWHM (Fig.~\ref{fig:convolve}).  After
convolution, the structure function approaches $S(r_\perp) \propto r_\perp^2$
for small $r_\perp$ (Section~\ref{smallsep}).

\subsection{Realisations of the RM distribution}
\label{realise}

\subsubsection{Method}
\label{realise-general}

Errors for the structure function of poorly sampled data are impossible to
evaluate analytically: they are correlated between adjacent position bins and
depend on the details of the window function, particularly for large
separations.  For these reasons, we choose to generate multiple realisations of
Gaussian random RM fields with given power spectra on the observed grids. The
dispersion in the structure functions calculated for these realisations gives an
estimate of the errors and allows us to determine the range of scales over which
the observed structure functions are reliable.  Our method for generating a
realisation of a Gaussian random variable with a particular power spectrum is as
follows.
\begin{enumerate}
\item Make an array in the frequency plane whose elements are complex numbers
  whose real and imaginary parts are random numbers selected from a Gaussian
  distribution with zero mean and unit variance.
\item Multiply the elements of this array by the square root of the power
  spectrum, $[\hat{C}(f_\perp)]^{1/2}$.
\item Optionally, multiply by the Fourier transform of the convolving beam.
\item Perform a two-dimensional inverse Fourier transform to the image plane.
\item Evaluate the structure function.
\end{enumerate}
This was implemented using a FFT routine on a grid of size $n = 2^m$, setting
implicit limits $1/n \leq f_\perp \leq 1/2$ to the maximum and minimum
frequencies which can be sampled in either dimension (the upper limit is the
Nyquist frequency).  In practice, we used $n = 8192$ to ensure that the
difference between the structure functions for minimum frequencies of 0 and
$1/n$ is negligible over the range of spatial separations sampled by our data.
  
\subsubsection{Modified power-law $\hat{C}(f_\perp)$}
\label{modPL}

\begin{figure}
\epsfxsize=8.5cm
\epsffile{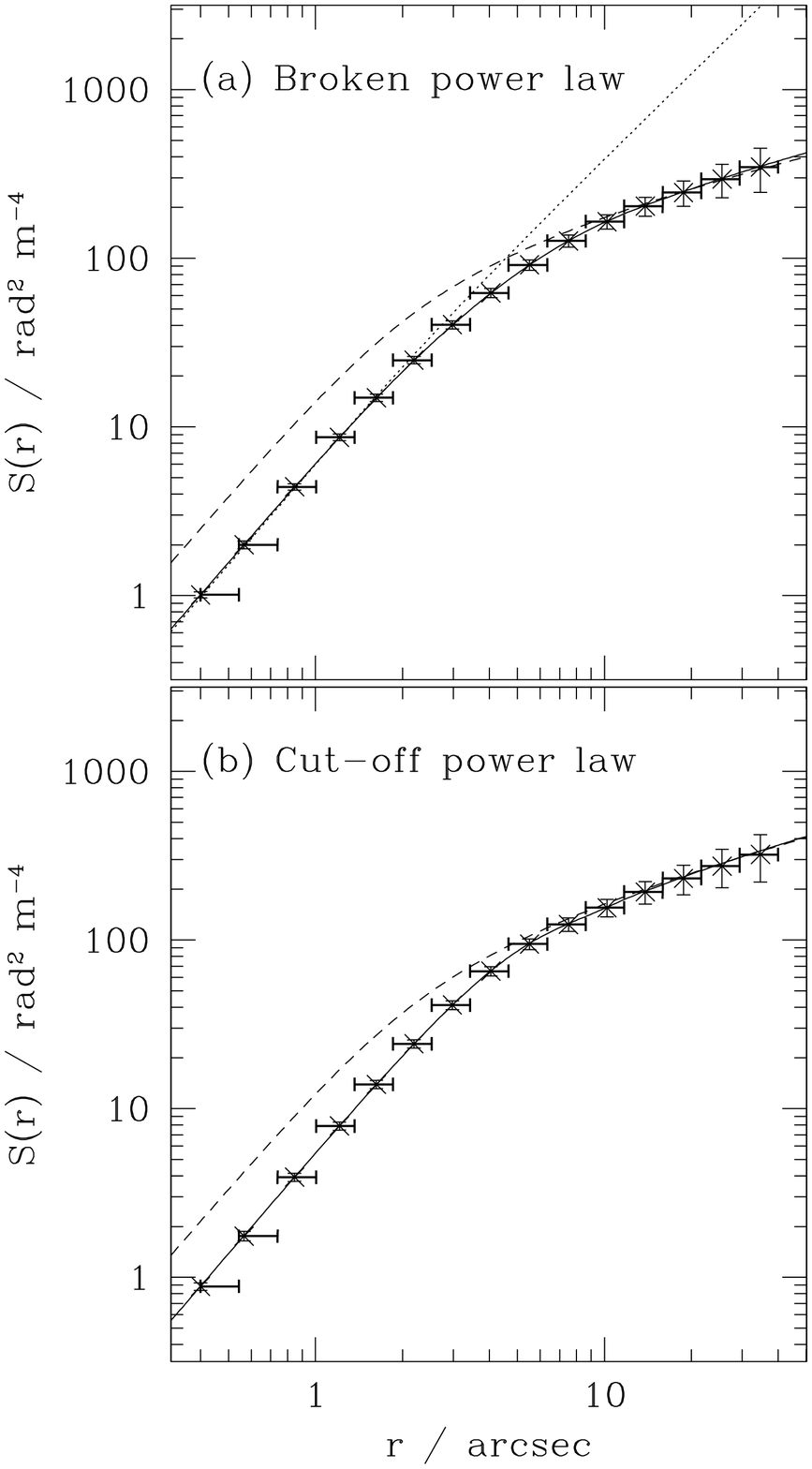}
\caption{Comparison of structure functions derived by simulation and Hankel
    transform for an observing beam of 1.5\,arcsec FWHM.  (a) Broken power-law
    with indices $q_{\rm high} = 11/3$, $q_{\rm low}$ = 2.32 and a break
    frequency $f_{\rm b} = 0.062$\,arcsec$^{-1}$
    [equation~(\ref{eq-broken-pl})]. (b) Power law with index $q = 2.39$ and a
    high-frequency cut-off at $f_{\rm max} = 0.144$\,arcsec$^{-1}$
    [equation~(\ref{eq-cutoff-pl})].  The full lines are the Hankel transform
    relations [equation~(\ref{eq-Hankel-convolve})], assuming no low-frequency
    cut-off in the power spectrum. The points and error bars are derived from
    100 realisations of the power spectrum on a 8192$^2$ grid with the sampling
    appropriate for region NP1. The horizontal bars represent the bin widths and
    the crosses the mean separation for data included in each of the bins.  The
    error bars represent the rms variations for the model structure functions.
    The dashed curves are the structure functions for pure power-law power
    spectra with the appropriate low-frequency indices [$q = 2.39$ and 2.32
    respectively for panels (a) and (b)]. In panel (a), we also show the
    structure function for a power-law power spectrum with $q = 11/3$ (dotted
    line).
    \label{fig:modpl_sf}}
\end{figure}

In order to fit the observed structure functions, we require power spectra of
the cut-off and broken power-law forms given in equations~(\ref{eq-broken-pl}) and
(\ref{eq-cutoff-pl}), below.  For both cases, we have verified that the mean of
the structure functions from realisations of the power spectrum converges to the
Hankel transform result and that the finite size of the grid for the
realisations has a negligible effect on the structure functions for the
separations we sample.  Example calculations for both types of structure
function are shown in Fig.~\ref{fig:modpl_sf}.

\subsection{Observed structure function}
\label{Sfunc_observe}

The areas of 3C\,31 where we have enough contiguous data points to evaluate the
RM structure function are disconnected from each other and have very different
variances. We have therefore analysed the areas NP1, SP1 and SP2
(Fig.~\ref{fig:RMDP1.5}c) separately at 1.5-arcsec resolution. Use of the
5-frequency, 5.5-arcsec RM image (Fig.~\ref{fig:RMDP5.5}c) allows us to examine
the larger areas NP2 and SP3, but adds little extra information for NP1 and SP1,
where essentially all of the emission is detected at the higher resolution (SP2
and SP3 overlap). Systematic errors are likely to be important for SP2: the
signal-to-noise ratio is relatively low and independent RM fitting algorithms
give significantly different structure functions. We plot the data for this
region, but do not use them in our determination of the form of the structure
function.

\begin{figure*}
\epsfxsize=14.5cm
\epsffile{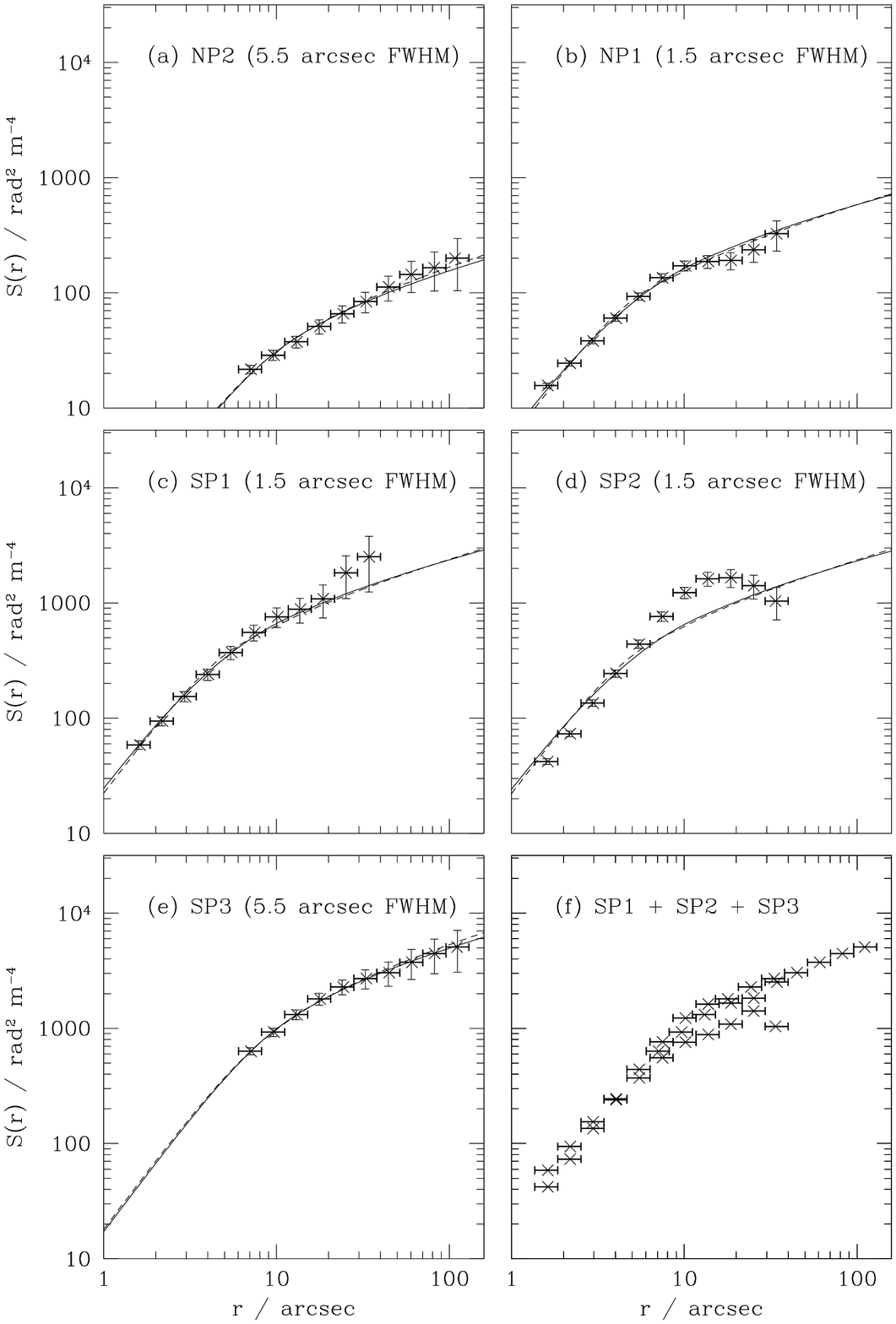}
\caption{(a) -- (e): plots of the RM structure function for the regions
  indicated in Figs~\ref{fig:RMDP5.5} and \ref{fig:RMDP1.5}. The horizontal bars
  represent the bin widths and the crosses the mean separation for data included
  in the bins. The curves are the predictions for the model power spectra
  described in the text, including the effects of the convolving beam. The full
  and dashed curves denote the broken and cut-off power-law models,
  respectively. The error bars represent the rms variations for structure
  functions derived using the broken power law power spectrum on the observed grid of
  points for each region. Panel (f) shows the superposition of structure
  functions for all three regions in the South of the source.\label{sfplots}}.
\end{figure*}

\begin{table*}
\caption{Structure function parameters for individual regions and for a combined
  fit to all four regions. The last line of the table refers to the combined fit
  in which the values of $q_{\rm low}$ and $f_{\rm b}$ (or $q$ and $f_{\rm
  max}$) are the same for all regions, but the individual normalizations are
  varied to minimise the overall chi-squared (these normalizations are listed in
  Table~\ref{Depol-table}). \label{tab:fitparams}}
\begin{tabular}{llrlrrrrrlrrrr}
\hline
&&&&&&&&&&&&&\\
Region & FWHM &\multicolumn{6}{c}{BPL}&\multicolumn{6}{c}{CPL}  \\
&arcsec&\multicolumn{2}{c}{Best fit} & \multicolumn{2}{c}{Min slope}&\multicolumn{2}{c}{Max slope}&
\multicolumn{2}{c}{Best fit} &\multicolumn{2}{c}{Min slope}&\multicolumn{2}{c}{Max slope} \\  
& & $q_{\rm low}$&$f_{\rm b}$&$q_{\rm low}$&$f_{\rm b}$&$q_{\rm
  low}$&$f_{\rm b}$
& $q$&$f_{\rm max}$&$q$&$f_{\rm max}$&$q$
  &$f_{\rm max}$\\
&&&&&&&&&&&&&\\
NP2 & 5.5 & 2.35 & 0.178 & 2.14    & 0.062 & 2.54 & $\infty$
& 2.34 & 0.231 & 2.10    & 0.086 & 2.55 & $\infty$ \\
NP1 & 1.5 & 1.84 & 0.051 &$-0.2$ & 0.034 & 2.45 & 0.076
& 2.35 & 0.148 & 2.00    & 0.122 & 2.62 & 0.190  \\
SP1 & 1.5 & 2.76 & 0.073 &0.50    & 0.024 & 3.10 & 0.155 
& 2.89 & 0.210 & 2.46    & 0.126 & 3.12 & $\infty$  \\
SP3 & 5.5 &2.27 & 0.049 &1.70    & 0.026 & 2.51 & $\infty$ 
& 2.32 & 0.083& 2.03    & 0.058 & 2.63 & $\infty$  \\
&&&&&&&&&&&&&\\
Comb&&2.32 & 0.062 & 2.15 & 0.053 & 2.47 & 0.072 
     & 2.39 & 0.144& 2.28    & 0.134 & 2.49 & 0.156  \\
&&&&&&&&&&&&&\\
\hline
\end{tabular}
\end{table*}

In the presence of noise, the structure function has a positive bias, and for
uncorrelated random noise with rms $\sigma_{\rm fit}$, its expectation value is
$2\sigma_{\rm fit}^2$.  We have made a first-order correction for this bias by
evaluating $S(r_\perp)=S_{\rm raw}(r_\perp)-2\sigma_{\rm fit}^2$, where $S_{\rm
raw}(r_\perp)$ is the structure function determined directly from the images and
$\sigma_{\rm fit}$ is set to be the rms error in the RM for the area in
question. In practice, $\sigma_{\rm fit}$ lies between 1.6 and 3\,rad\,m$^{-2}$
so the correction is small in the North and insignificant in the South.  The
noise is essentially uncorrelated on scales larger than the beam. The measured
structure functions for the 5 regions are shown in Fig.~\ref{sfplots}(a) -- (e);
in Fig.~\ref{sfplots}(f), we show a superposition of the structure functions for
all three southern regions.  

We have investigated the hypothesis that the power spectrum (and hence the
structure function) has the same form for all of the regions, but with varying
normalization.  It is clear from Fig.~\ref{sfplots} and from
Figs~\ref{fig:RMprofiles}(a) and (b) that there is significant power on a range
of spatial scales and that a Gaussian power spectrum, (for which we expect a
flat structure function on large scales) is inadequate. Motivated by the
approximately power-law form for the structure function and
equation~(\ref{eq-asymp}), we initially tried an RM power spectrum of the form
$\hat{C}(f_\perp) \propto f_\perp^{-q}$ over an infinite frequency range. Even
after including the effects of the beam [equation~(\ref{eq-hyper}),
Fig.~\ref{fig:convolve}], the model structure functions showed insufficient
curvature to fit our 1.5-arcsec resolution data: all three regions imaged at
1.5-arcsec resolution require a change in structure-function slope at $r_\perp
\approx 10$\,arcsec; this is also apparent from the superposition of data for
the South of the source in Fig.~\ref{sfplots}(f).  Two functional forms for
$\hat{C}(f_\perp)$ which have been discussed in the literature and which fit the
data reasonably well are a {\sl broken power law} (hereafter BPL)
\begin{eqnarray}
\hat{C}(f_\perp) &= & D_0 f_{\rm b}^{-11/3} (f_\perp/f_{\rm
     b})^{-q_{\rm low}} \makebox{~~~$f_\perp \leq f_{\rm b}$} \nonumber \\
&  = & D_0 f_\perp^{-11/3} \makebox{~~~~~~~~~~~~~~~~~~~~$f_\perp > f_{\rm
     b}$} \label{eq-broken-pl}\\  \nonumber 
\end{eqnarray}
and a {\em cut-off power law} (hereafter CPL)
\begin{eqnarray}
\hat{C}(f_\perp) & = & C_0 f_\perp^{-q} \makebox{~~~$f_\perp \leq f_{\rm max}$}
     \nonumber \\
     & = & 0 \makebox{~~~~~~~~~~$f_\perp > f_{\rm max}$} \label{eq-cutoff-pl} \\ \nonumber 
\end{eqnarray}
The former is motivated by the expectation that the power spectrum will have a
Kolmogorov form on small scales. Fig.~\ref{fig:modpl_sf} compares the structure
functions expected for the high- and low-frequency asymptotic forms of these
power spectra with those corresponding to equations~(\ref{eq-broken-pl}) and
(\ref{eq-cutoff-pl}), emphasising that there must be a change in slope over the
observable frequency range.
 
For each region, we optimised the normalization over a grid of models specified
by $q_{\rm low}$ and $f_{\rm b}$ for BPL and $q$ and $f_{\max}$ for CPL.  In
order to estimate the errors from undersampling, we made multiple realisations
of convolved RM images corresponding to the BPL model power spectrum on the
observed grid and evaluated the rms deviations of their structure functions.
These represent the errors expected purely from statistical fluctuations in the
absence of noise or systematic errors if the model power spectrum is valid
everywhere.  They are much larger than the errors expected from the RM fits, and
we therefore used them to evaluate chi-squared, which we then summed over all
bins with separation $>$FWHM as a measure of the goodness of fit. Note, however,
that the errors for adjacent bins are not independent.

In Table~\ref{tab:fitparams}, we quote the optimised parameters for NP2, NP1,
SP1 and SP3 separately. Uncertainties are approximate $1\sigma$ (68\%
confidence) limits for two interesting parameters. Acceptable fits have tightly
correlated values of $q_{\rm low}$ and $f_{\rm b}$ (or $q$ and $f_{\rm max}$),
so we quote pairs of values defining the extrema of contours of constant
chi-squared.

We then determined a best overall fit by minimising chi-squared summed over all
four regions, giving equal weight to each and allowing the normalizations to
vary independently.  We find best-fitting parameters of $q_{\rm low}$ = 2.32 and
$f_{\rm b}$ = 0.062\,arcsec$^{-1}$ for the BPL model and $q$ = 2.39 and $f_{\rm
max}$ = 0.144\,arcsec$^{-1}$ for the CPL model. The corresponding structure
functions, including the effects of the convolving beam, are plotted in
Fig.~\ref{sfplots}, where the amplitudes have been scaled as in
Table~\ref{Depol-table} to fit the observations (we also include the fit for
SP2).  The rms deviations from multiple realisations are plotted as error bars
attached to the observed points.

\begin{figure}
\epsfxsize=8cm 
\epsffile{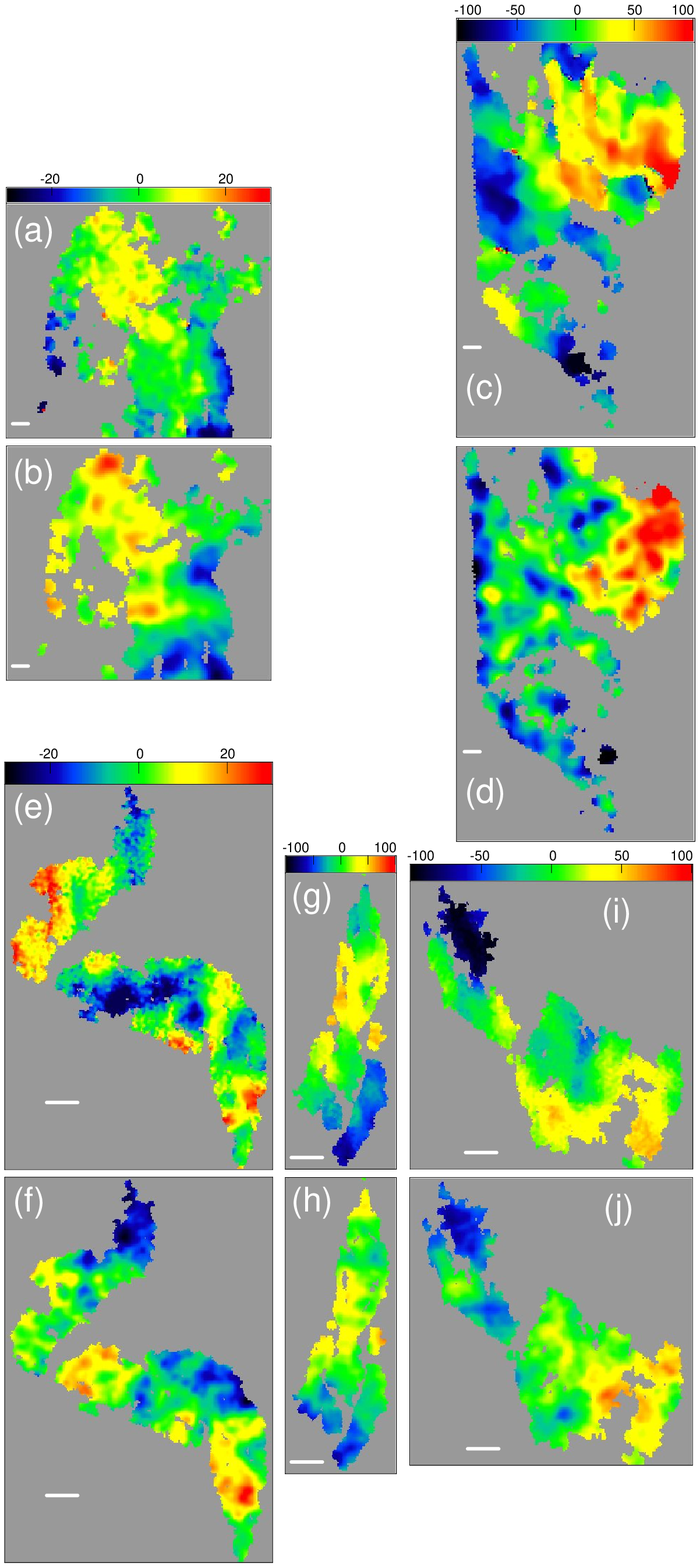}
\caption{A comparison between the RM data in the regions used for
  structure-function analysis and example realisations generated using the 
  BPL power spectrum described in the text. The mean values for each panel were
  subtracted before plotting to facilitate comparison of smaller-scale
  structure. The normalizations are given in Table~\ref{Depol-table}.  (a) and
  (b): data and model for region NP2. (c) and (d): SP3; (e) and (f): NP1; (g)
  and (h): SP1; (i) and (j): SP2.  Note that noise in the RM fits is generally
  uncorrelated on small scales, and the observed RM images therefore appear to
  have slightly more small-scale structure than the models. This is more obvious
  in panels (a), (b), (e) and (f), where the amplitude of the noise relative to
  the plotting range is larger.  The colour scales ($\pm$30\,rad\,m$^{-2}$ for
  panels a, b, e and f); $\pm$100\,rad\,m$^{-2}$ for the remaining plots) are
  shown by labelled wedges. The horizontal lines give a scale of 10\,arcsec.
\label{fig:rm_mod_data}}
\end{figure}

Fig.~\ref{sfplots} shows that the observed structure functions for NP2, NP1, SP1
and SP3 are in very good agreement with the models. Together with the absence of
any compelling evidence for non-Gaussianity, anisotropy or correlation with the
small-scale structure of the source (confirmed by visual comparison between
example realisations, such as those in Fig.~\ref{fig:rm_mod_data}, and the
data), this indicates that our model power spectra are acceptable
representations of the observed RM fluctuations.  The hypothesis that only the
amplitude of the power spectrum varies with position is therefore consistent
with our results, especially given that the approximation of constant path
length through a region inevitably introduces some inaccuracies (these are
corrected in the 3-d simulations of Section~\ref{3d}).  A more rigorous error
analysis (best done using a Bayesian maximum likelihood technique, e.g.\
\citealt{VE05}) is outside the scope of this paper.

\begin{figure}
\epsfxsize=8.5cm 
\epsffile{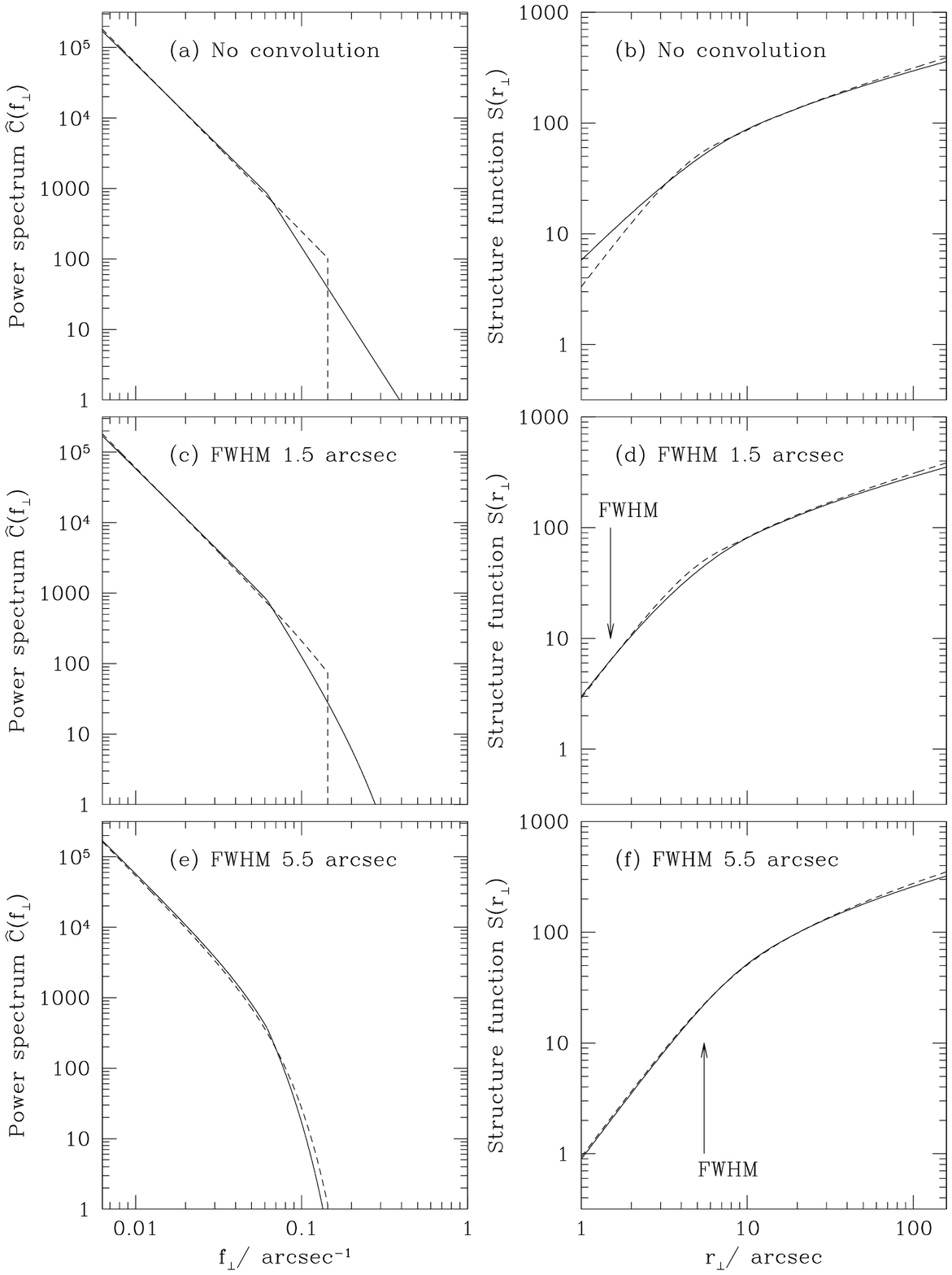}
\caption{(a), (c), (e): the two model RM power spectra discussed in the
  text. Solid line, BPL with indices $q_{\rm high}$ = 11/3, $q_{\rm
  low}$ = 2.32 and a break frequency of 0.062\,arcsec$^{-1}$
  [equation~(\ref{eq-broken-pl})]. Dashed line: CPL with $q$ = 2.39 and a
  high-frequency cut-off at $f_{\rm max}$ = 0.144\,arcsec$^{-1}$
  [equation~(\ref{eq-cutoff-pl})]. (b), (d), (f): structure functions computed for
  the power spectra in panels (a), (c) and (d), with the same line codes. (a)
  and (b) no convolution; (c) and (d) 1.5\,arcsec FWHM convolving beam; (e) and
  (f) 5.5\,arcsec FWHM convolving beam.
\label{ps_sf_comp}}
\end{figure}

The BPL model fits the combined observations very slightly better, and we adopt
it as our standard, but we cannot differentiate between the two functional forms
(or similar ones) using our RM data alone. The two model power spectra have
almost the same form at low spatial frequencies.  The CPL form predicts
$S(r_\perp) \propto r^2$ at small separations (Section~\ref{smallsep}), quite
close to the asymptotic $r^{5/3}$ expected for the Kolmogorov spectrum
[equation~(\ref{eq-asymp})], which is itself steepened by the effects of
convolution. In Fig.~\ref{ps_sf_comp}, we show the two power spectra and their
associated structure functions.  The latter are almost identical, particularly
after convolution (Figs~\ref{ps_sf_comp}d and f), explaining our inability to
distinguish between them. The power in fluctuations on scales $\la$5\,arcsec
differs significantly for the two models, however (Fig.~\ref{ps_sf_comp}b) so in
Section~\ref{innerscale} we attempt to use the measured depolarizations at
1.5-arcsec resolution to select between them.

\begin{table}
\caption{Normalizations $D_0$ and $C_0$ for the combined fit parameters from
  Table~\ref{tab:fitparams}.  Observed and calculated depolarization are also
  given for regions NP2, NP1, SP1 and SP3. 1: region name. 2: normalization
  constant $D_0$ for the model power spectrum of
  equation~(\ref{eq-broken-pl}). 3: normalization constant $C_0$ for the model
  power spectrum of equation~(\ref{eq-cutoff-pl}). 4. Observed Burn law $\langle
  k \rangle$, with errors.  Only data with $p > 4\sigma_p$ are included, and for
  NP1 and SP1 we only include points within 60\,arcsec of the nucleus.  5:
  $\langle k \rangle$ predicted for the BPL power spectrum. 6: As 5, but for the
  CPL power spectrum. Rms sampling errors for the models are also given in
  columns 5 and 6. \label{Depol-table}}
\begin{tabular}{lrrrrr}
\hline
&&&&&\\
Region & $D_0$ & $C_0$ & \multicolumn{3}{c}{$\langle k\rangle$ / rad$^2$\,m$^{-4}$} \\
       &       &       & Observed & BPL & CPL \\ 
&&&&&\\
NP2 &  0.019 &   0.61& $46\pm  8$  &  $21.8 \pm 0.7$  &  $23.2 \pm 0.8$  \\
NP1 &  0.064 &   1.9 & $3 \pm  7$  &  $14.7 \pm 0.1$  &  $9.7 \pm 0.3$  \\
SP1 &  0.26  &   7.8 & $55\pm 13$  &  $59.7 \pm 0.9$  &  $40.0 \pm 1.1$  \\
SP2 &  0.25  &   7.7 &             &        &        \\
SP3 &  0.61  &   19  & $372\pm 18$ & $569 \pm 13$  & $542 \pm 7$  \\
&&&&&\\
\hline
\end{tabular}
\end{table}

\subsection{Constraints from observed depolarization and the validity of the
  short-wavelength approximation}
\label{innerscale}

Measurement of residual depolarization constrains the amplitude of RM
fluctuations on scales smaller than the beam. In order to test consistency with
our adopted model for the power spectrum and to check whether the fitted RM is
close to the convolution of the true RM distribution with the beam [the
short-wavelength approximation of equation~(\ref{eq:rm-conv})], we proceed as
follows.
\begin{enumerate}
\item We make realisations of the RM distributions for our adopted power-spectrum
  parameters on a fine grid (0.05\,arcsec pixels) and scale the amplitude as
  appropriate.
\item We then calculate $Q$ and $U$ at each of our observing frequencies for a
  uniformly polarized source.
\item We convolve the resulting images  to the observing resolution and
  derive the polarized power $P = (Q^2+U^2)^{1/2}$ and {\bf E}-vector
  position angle, $\chi$. 
\item We then fit $\chi$ against $\lambda^2$ and $\ln p$ against $\lambda^4$ to
make images of RM and $k$ as for the observations.
\item We average over the images to derive the mean values of the Burn
  parameter $\langle k \rangle$ and the degree of
  polarization $\langle p(\lambda) \rangle$ at each of our observing wavelengths.
\item Finally, we compare the derived RM image with the convolution of 
  the full-resolution version with the observing beam.
\end{enumerate}

We have verified that the short-wavelength approximation holds to high accuracy
for the model power spectra in regions NP1, NP2, SP1 and SP2 by taking the
differences between simulated RM images and convolutions of the true RM
distributions with the appropriate observing beams. The rms difference is
$<$0.1\,rad\,m$^{-2}$ (negligible compared with the noise) and the means differ
by $<$0.01\,rad\,m$^{-2}$ in all four regions.  Even in SP3, where the
short-wavelength approximation breaks down, the mean structure function for a
large number of simulated RM images is in precise agreement with that derived
from the Hankel transform relation. The reason is that areas of high
depolarization and deviation from $\lambda^2$ rotation occupy a small fraction
of the RM image and therefore have very little effect on the spatial statistics
(see Appendix~\ref{SP3depol} for a more detailed analysis of these effects).

\begin{figure}
\epsfxsize=8.5cm 
\epsffile{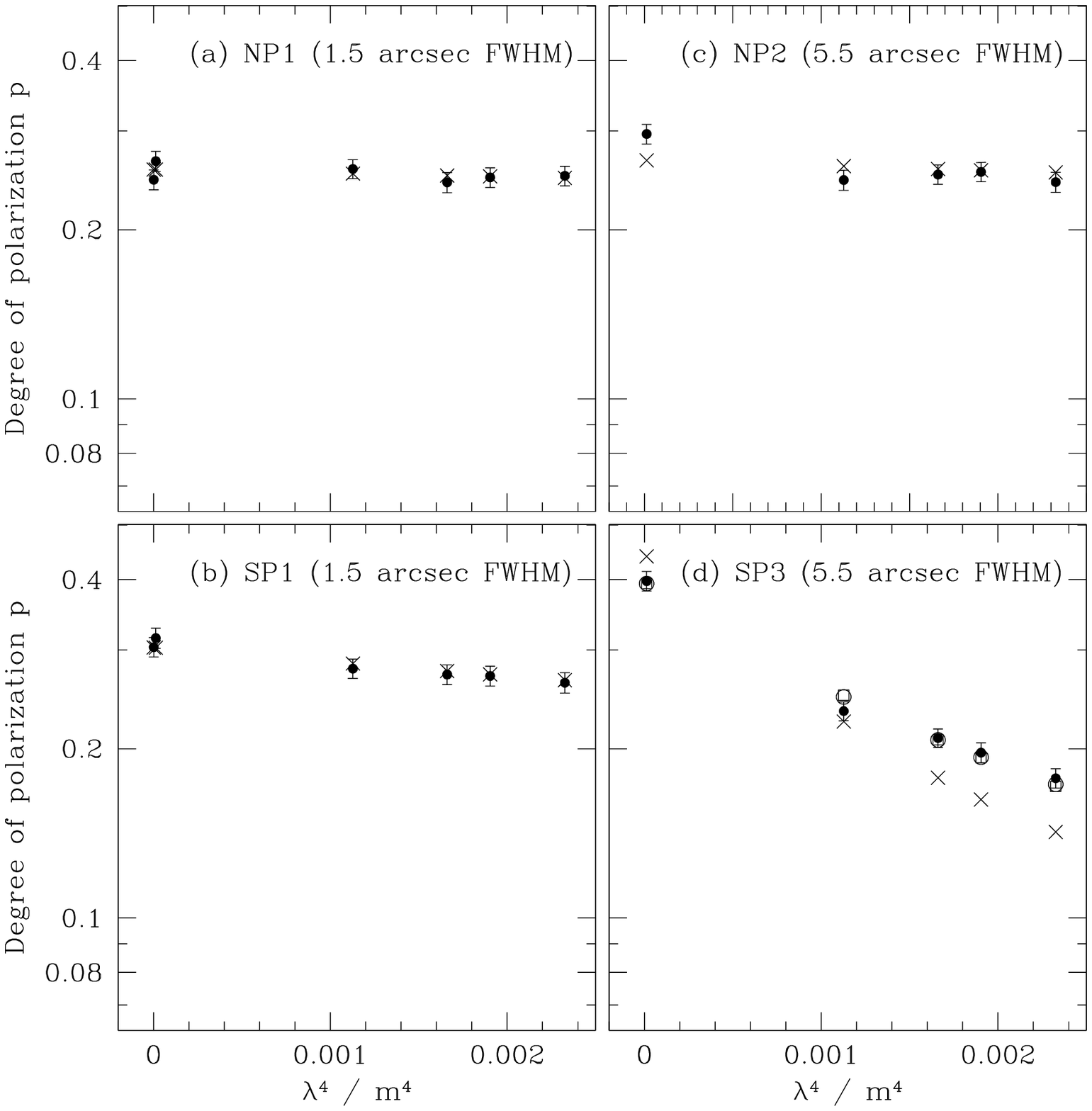}
\caption{Plots of the degree of polarization (on a log scale) against
  $\lambda^4$ for the regions: (a) NP1, (b) SP1, (c) NP2 and (d) SP3.  The
  filled circles with error bars represent the observed mean degrees of
  polarization. The crosses show the predicted variations of polarization with
  wavelength for the BPL model with the amplitudes from
  Table~\ref{Depol-table}.  Panel (d) also shows predicted values for the BPL
  models with $D_0 = 0.38$ and $f_{\rm b} = 0.062$\,arcsec$^{-1}$ (open squares)
  and $D_0 = 0.61$ and $f_{\rm b} = 0.037$\,arcsec$^{-1}$ (open circles). The
  predicted values for these two cases are almost identical, so the symbols
  appear superposed. In all cases, the intrinsic degree of polarization, $p(0)$,
  has been set to give the best agreement (minimum chi-squared) between
  predicted and observed values.
\label{p-lambdasq-theory}}
\end{figure}

The mean values $\langle k \rangle$ predicted for regions NP1 and SP1 by the BPL
[equation~(\ref{eq-broken-pl})] and CPL [equation~(\ref{eq-cutoff-pl})] models
for $\hat{C}(f_\perp)$ are given in Table~\ref{Depol-table}. Both models predict
values consistent with the observations, so we cannot use the measured
depolarization to discriminate between them. In Figs~\ref{p-lambdasq-theory}(a) and
(b), we compare the observed mean degrees of polarization at our six observing
wavelengths with the predictions of the BPL model derived directly from multiple
realisations and find very good agreement (the CPL model again gives very
similar results).

In regions NP2 and SP3, where we have only low-resolution data, the predicted
variations of $\langle p \rangle$ with $\lambda$ are not in such good agreement
with the observations (Table~\ref{Depol-table}). For NP2, we predict slightly
too much depolarization [Fig.~\ref{p-lambdasq-theory}(c)]. The observed degree
of polarization does not fall monotonically with increasing wavelength, however,
indicating either that the approximation of constant intrinsic polarization is
invalid or that we have underestimated the systematic errors. In contrast, we
predict significantly more depolarization in SP3 than is observed
[Fig.~\ref{p-lambdasq-theory}(d)]\footnote{ The observed and predicted
depolarizations at longer wavelengths in SP3 are less pronounced than predicted
by the Burn law, as first pointed out by \citet{Tribble91a}.}. We show in that
figure that we can reproduce the run of polarization with wavelength by simple
modifications to the parameters of the BPL power spectrum, for example, by
reducing the amplitude to $D_0 = 0.38$ but keeping the break frequency and
low-frequency slope at the values in Table~\ref{tab:fitparams} or by maintaining
the amplitude and slope at low spatial frequencies but decreasing the break
frequency to $f_b = 0.037$\,arcsec$^{-1}$.  These discrepancies suggest that the
form as well as the amplitude of the RM power spectrum may vary slightly across
the source, but we cannot quantify the variation further without
higher-resolution observations of the RM fluctuations in the
low-surface-brightness regions NP2 and SP3.

\subsection{The outer scale and autocorrelation lengths}

We have so far implicitly assumed that the outer scale of the magnetic-field and
RM fluctuations is large compared with the size of the regions over which we
evaluate the structure function.  This is justified by the fact that the
structure functions continue to rise with increasing separation up to $r_\perp
\approx 100$\,arcsec. The outer scale cannot be arbitrarily large, however, and
its value has important implications for the estimation of magnetic-field
strength and autocorrelation length. 

Our constraints on the outer scale of the magnetic-field fluctuations from the
structure-function analysis are weak, because of the poor sampling of large
separations. For both the broken and cut-off power law models, we infer $f_{\rm
min} \la 0.005$\,arcsec$^{-1}$ in regions NP2 and SP3.  If the apparent RM
gradient over $\sim$800\,arcsec (Fig.~\ref{fig:RMprofiles}a) is due to
large-scale structure in the local field (rather than a gradient in Galactic RM)
then $f_{\rm min} \la 0.001$\,arcsec$^{-1}$.  The autocorrelation lengths for
the magnetic field [$\lambda_B$; equation~(\ref{eq-lambdaB})] and RM
[$\lambda_{\rm RM}$; equation~(\ref{eq-lambdaRM})] depend on the outer
scale. Equations~(\ref{eq-lambdaB-fourier}) and (\ref{eq-lambdaRM-fourier}) allow us
to determine the autocorrelation lengths for both model power spectra
analytically; values of $\lambda_B$ for three representative minimum
frequencies are given in Table~\ref{tab:autocorr}.

\begin{table}
\caption{Magnetic autocorrelation lengths for the broken and cut-off power-law
models [equations~(\ref{eq-broken-pl}) and (\ref{eq-cutoff-pl}), respectively].
The model used for three-dimensional simulations is the broken power law (BPL)
with $f_{\rm min}$ = 0.00075\,arcsec$^{-1}$. \label{tab:autocorr}}
\begin{tabular}{lrr}
\hline
&&\\
$f_{\rm min}$ & \multicolumn{2}{c}{$\lambda_B$ / kpc} \\
arcsec$^{-1}$ & BPL  & CPL \\
&&\\
\hline
&&\\
0.005   & 4.9 & 2.8  \\
0.001   & 9.0 & 5.7  \\
0.00075 &10.0 & 6.5  \\
&&\\
\hline
\end{tabular}
\end{table}

\subsection{Origins of the power spectrum}
\label{PS-physics}

There is no generally accepted theory of magnetic-field fluctuations in
intra-cluster plasma, although a turbulent cascade from large to small scales is
expected. Dissipation of turbulence is predicted to occur on the resistive
scale, which is far below our resolution limit (e.g.\ \citealt{SC06}). Energy
input might come from sub-cluster mergers or galaxy motions within the group and
from interactions between the radio source and the surrounding plasma. Possible
input scales $L$ range from $L \approx 25$\,kpc (the projected separation between
NGC\,383 and its nearest neighbour and the bending radius of the inner jets) to
$L \approx 500$\,kpc (the total size of the radio source).  The magnetic power
spectrum is expected to have a power-law form with slope $q = 11/3$ over some
inertial range if the turbulence is primarily hydrodynamic \citep{Kol41} or for
the type of MHD cascade analysed by \citet{GS97}. Our results show that the
power spectrum could have this form on scales $\la$5\,kpc, but must flatten on
larger scales. Such a break in slope might occur at a characteristic field
reversal scale $l_\perp \sim 0.01$ -- $0.1L$ if the field is generated by some
form of fluctuation dynamo \citep{SC06,EV06}.

\section{Three-dimensional Models}
\label{3d}

\subsection{General considerations}

We have established that essentially all of the Faraday rotation
associated with
3C\,31 is due to foreground material
(Section~\ref{Faraday-foreground}).  The
only known ionized gas component with the requisite scale size is the
hot plasma
associated with the surrounding group of galaxies, which is centred close to
3C\,31 \citep{KB99}.  Although a component of Faraday rotation
associated with a
thin shell of material around the lobes of extragalactic radio sources
has been
postulated by \citet*[see also \citealt{RB03}]{BCG90}, this component cannot
dominate, as there would then be no reason for the path lengths to differ
between the approaching and receding lobes. There
would then be
no explanation for the
systematic asymmetry in RM variance and/or depolarization with jet prominence seen in
3C\,31 and many other FR\,I sources \citep{Morg97}. Cooler ($\sim$10$^4$\,K)
ionized gas has
been detected only
in the inner few arcsec of 3C\,31
\citep{OOK,Ionized} and may be expected to clump on scales much smaller than
those of the observed large-scale RM gradients. We therefore model the
Faraday
rotation as an effect of magnetic field distributed throughout the group
gas.

\subsection{Simulations with a realistic power spectrum}
\label{3Dsim}

Realistic predictions of the RM distribution can be made using Monte Carlo
simulations.  Our method is essentially that of \citet{Murgia}: as in our
two-dimensional analysis, we assume that the magnetic field is an isotropic,
Gaussian random variable. To construct a numerical realisation of such a
magnetic field in a cubical box, we start from a magnetic vector potential ${\bf
A}({\bf r})$ with a given power spectrum \citep{Tribble91b}.
\begin{enumerate}
\item At each point of a $2^m \times 2^m \times 2^m$ grid in frequency space, we
  select the real and imaginary parts of the vector components of the Fourier
  transform of the vector potential $\hat{{\bf A}}({\bf f})$ from a Gaussian
  random distribution with unit variance.
\item We then multiply by the square root of the power spectrum of the vector
  potential. For a magnetic-field power spectrum $\hat{w}(f) \propto f^{-q}$,
  this is $\propto f^{-q-2}$.
\item Our 2-d analysis was consistent with a magnetic power spectrum which has
the same form everywhere, but varying normalization, so all of the
3-d simulations assume the BPL power spectrum of equation~(\ref{eq-broken-pl})
with $q_{\rm low} = 2.32$ and $f_{\rm b} = 0.062$\,arcsec$^{-1}$. The minimum
frequency is set to $f_{\rm min} = 0.00075$\,arcsec$^{-1}$, matching the size of
the sampling grids. The magnetic autocorrelation length is  $\lambda_B =
29.6$\,arcsec (10.0\,kpc).
\item The corresponding magnetic field in Fourier space, $\hat{{\bf B}}({\bf f})
  = 2\pi i{\bf f} \times \hat{{\bf A}}({\bf f})$, is then divergence-free.
\item We transform the line-of-sight field component $B_z$ to real space
and multiply it by the assumed density function $n({\bf r})$ to generate a
Faraday depth model.\footnote{As noted by \citet{Murgia}, this does not
  precisely preserve the divergence-free character of the field, but is an
  adequate approximation for our purpose.}
\item  $nB_z$ can then be integrated along the line of sight from
the emitting surface to the front surface of the cube to generate a simulated RM
image.  
\item Finally, we convolve the RM image with the observing beam, assuming that
  the short-wavelength approximation holds, and sample it as for the
  observations.
\end{enumerate}
In practice, we used 512$^3$ and 1024$^3$ grids for the low- and high-resolution
images, respectively, with cell sizes in real space set to 2.6 and 1.3\,arcsec.
For each of the models described below, we made 25 realisations at low
resolution, allowing us to explore the variation in predicted profiles, together
with one realisation at high resolution.  We have verified that we recover the
input power spectrum from two-dimensional structure-function analyses for
regions over which the fluctuation amplitudes are reasonably uniform.

\subsection{Comparison between predicted and observed RM distributions}
\label{3d-comparison}

Our criterion for an acceptable 3-d model is that it fits both the form and the
amplitude of the RM structure function to within errors set by a combination of
sampling variance and noise in the observations.  In order to assess the
goodness of fit, we need to average over areas which are small enough to resolve
spatial variations but large enough to give robust estimates of the RM
fluctuation amplitude. Our approach is to make binned profiles of
$\sigma_{\rm RM}$ at 5.5 arcsec resolution from the multiple realisations
(sampled in the same way as the observations) and then to optimise the
overall normalization. Our measure of goodness of fit is chi-squared, summed
over all bins with a significant number of points. As errors, we use the rms of
the fitting error for the observations (from the least-squares analysis of
Section~\ref{Faraday-foreground}) and the sampling error for the simulations
(from the multiple realisations), combined in quadrature.  We have fit both the
4 and 5-frequency RM profiles. The simulations are more tightly constrained by
the latter, whose fitting errors are significantly smaller.

Our assumptions are as follows:
\begin{enumerate}
\item The two sides of the source are intrinsically identical.
\item The emitting and Faraday-rotating media are axisymmetric with an axis
orientated at 52.4$^\circ$ to the line of sight, as in the model of
\citet{LB02a} for the inner jets.
\item The surrounding gas density distribution has a radial density profile as
  characterised by \citet{KB99}, with a central density of 1900\,m$^{-3}$, a core
  radius of $r_c = 52$\,kpc and a form factor $\beta_{\rm atm} = 0.38$. We 
  ignore the effects of the gas component associated with the galaxy \citep[see
  below]{Hard02}.  We distinguish two types of model:
    \begin{description}
    \item [{\em Spherically symmetric}] The source can be represented as a plane
    containing the inner jet axis, i.e.\ it has negligible extent along the line
    of sight. This is physically unrealistic, as we expect thermal matter to be
    excluded from the volume occupied by the relativistic plasma, but should be a
    good approximation where the source is narrow, and is a useful comparison with
    earlier work \citep{GC91,Tribble92}.
    \item [{\sl Cavity}] Thermal plasma is excluded from the volume occupied by
relativistic plasma, as expected from theoretical models (e.g.\
\citealt{Scheuer74}). Cavities apparently devoid of X-ray emitting material are
indeed observed around some FR\,I sources \citep[e.g.][and references
therein]{McNN2007} We have experimented with various geometries, and show the
results for ellipsoidal cavities.
    \end{description}
\item The normalization of the RM power spectrum then depends only on the
density $n(r)$ and the rms field $B(r)$. We assume that the magnetic
field is
given by $B \propto n^\mu$.  We consider a range from $\mu$ = 0.1 --
$\mu$ =
1.5. $\mu=0.5$ corresponds to equipartition of magnetic and thermal
energy and 
was assumed by \citet{GC91}; $\mu = 2/3$ is expected for flux-freezing and
$\mu \approx 1$ is consistent with the correlation of X-ray surface
brightness
and RM
variance for clusters of galaxies \citep{Dolag01,Dolag06}.  The
best-fitting values of $B_0$ and $\mu$ are correlated for any model, since a
higher value of $\mu$ gives a relatively lower Faraday rotation at large
distances from the nucleus, requiring an increase in central field strength.
\end{enumerate}
We show the results for the spherically-symmetric and cavity models in
Sections~\ref{spherical} and \ref{cavities}, respectively.

\subsection{Spherically-symmetric distributions of density and field}
\label{spherical}

\begin{figure*}
\epsfxsize=15.5cm
\epsffile{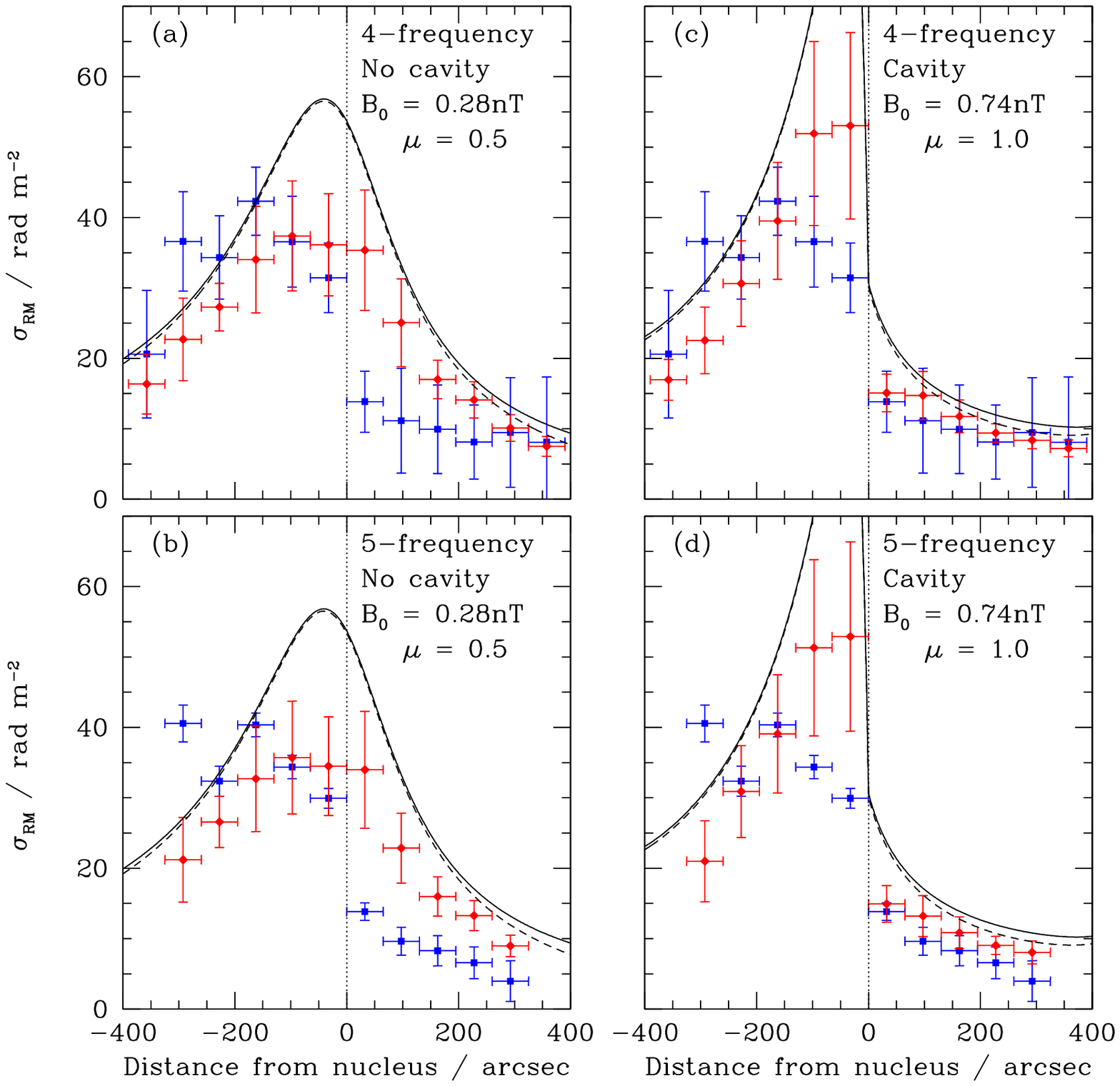}
\caption{Comparison of observed and simulated profiles of rms Faraday rotation
measure $\sigma_{\rm RM}$ at 5.5~arcsec resolution. The blue squares represent
observed values, with vertical bars corresponding to the rms errors on the RM
fit. The red diamonds show the mean values from 25 simulated profiles and the
vertical bars represent the rms scatter in these profiles. The normalizations of
the model profiles have been determined by chi-squared minimisation. The curves
are the predictions of a single-scale model with $D = \lambda_B$, as described
in Section~\ref{1-scale}.  The full curve is for an infinite upper integration
limit and the dashed line corresponds to truncation of the integral at the near
surface of the simulation volume.  (a) and (b): 3-d model with spherical
symmetry (Sec.~\ref{spherical}).  The central rms field strength is $B_0$ =
0.28\,nT and $\mu = 0.5$. (c) and (d): 3-d model with cavity geometry
(Sec.~\ref{cavities}), $B_0$ = 0.74\,nT and $\mu = 1.0$.
\label{fig:3Drms-profiles}
}
\end{figure*}

\begin{figure*}
\epsfxsize=12.5cm 
\epsffile{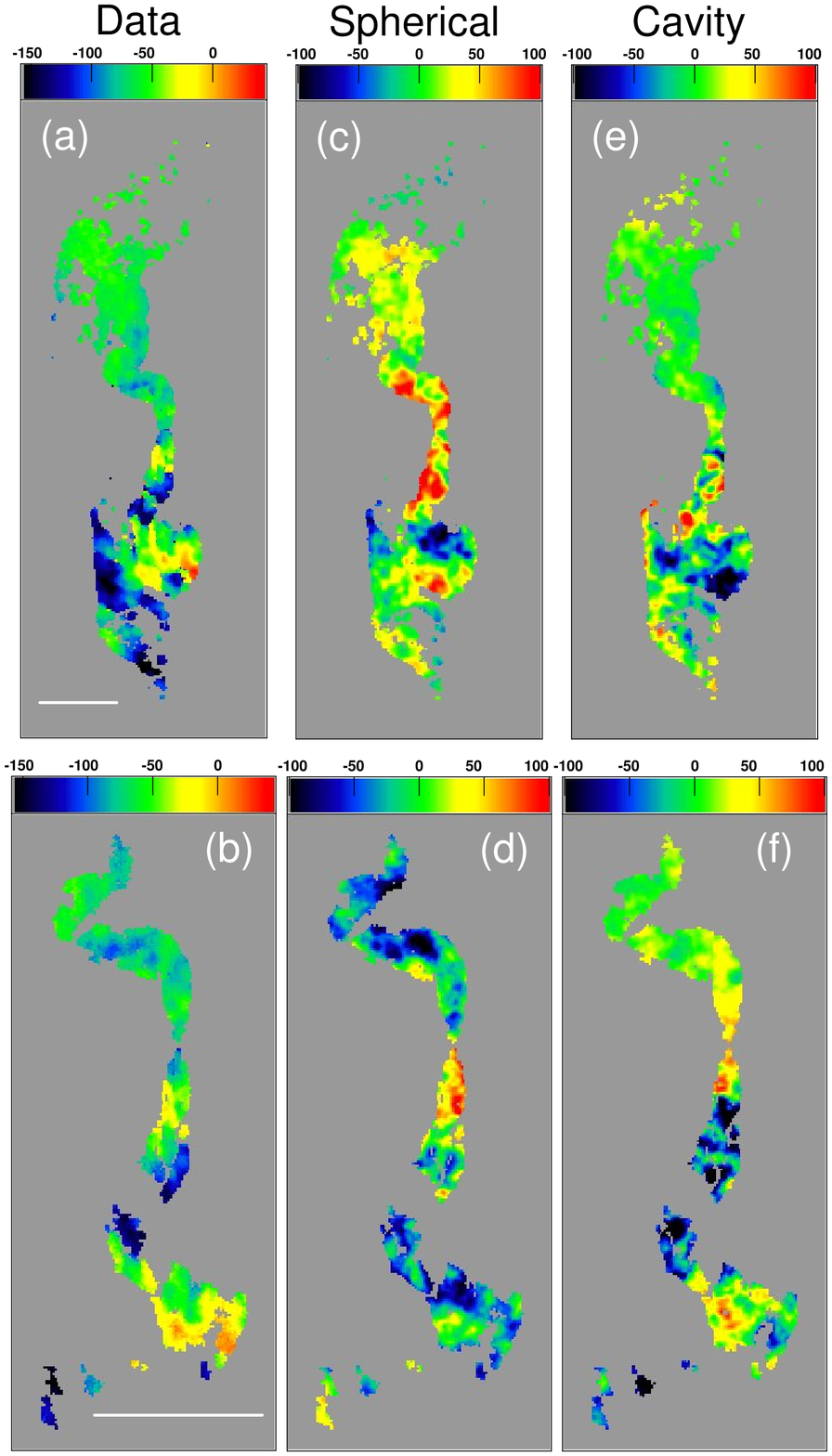}
\caption{Comparison of observed and simulated distributions of Faraday rotation
measure RM at 5.5 and 1.5~arcsec resolutions.  (a) and (b): observed. (c) and
(d): 3-d model with spherical symmetry (Sec.~\ref{spherical}).  (e) and (f): 3-d
model with cavity geometry (Sec.~\ref{cavities}).  The colour scale is the same
for all displays but the RM scales for the data are offset by the Galactic RM
contribution (Section~\ref{RM-variations}).  The bar at the base of panels (a)
and (b) shows a 100~arcsec scale.
\label{fig:3Dsim-images}
}
\end{figure*}

\begin{figure*}
\epsfxsize=14cm 
\epsffile{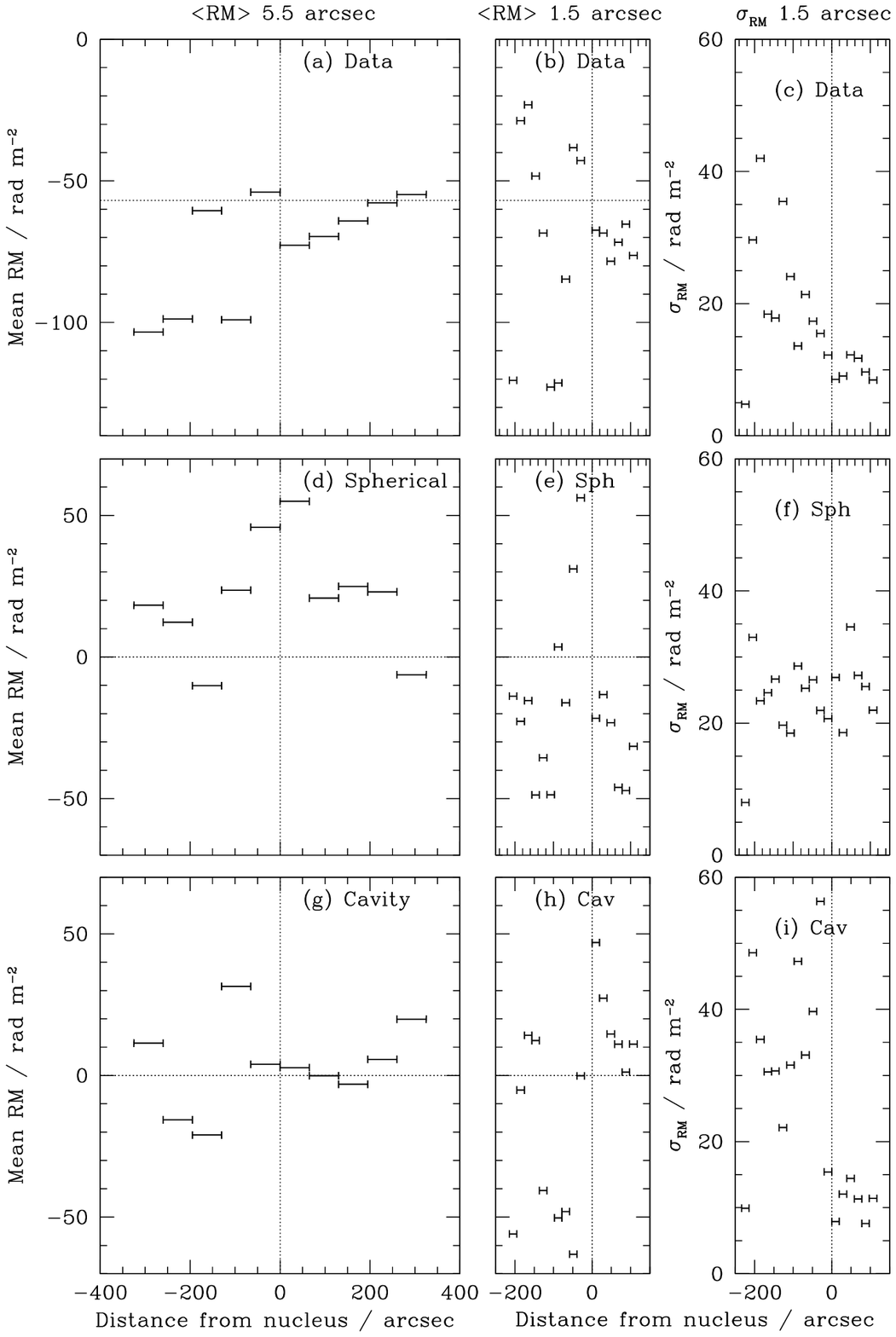}
\caption{Observed and simulated profiles of $\langle{\rm RM}\rangle$ and
$\sigma_{\rm RM}$, with the same binning as in Fig.~\ref{fig:RMprofiles}. The
observed data at 5.5 arcsec resolution are from the 5-frequency fits
[Fig.~\ref{fig:RMDP5.5}(c)] and the simulated profiles correspond to the example
realisations shown in Fig.~\ref{fig:3Dsim-images}. (a) -- (c): observed; (d) --
(f): 3-d model with spherical symmetry (Sec.~\ref{spherical}); (g) -- (i): 3-d
model with cavity geometry (Sec.~\ref{cavities}). Panels (a), (d) and (g) show
the mean RM values at 5.5 arcsec resolution; (b), (e) and (h) show mean values
at 1.5-arcsec resolution and (c), (f) and (i) give the rms RM at the higher
resolution.
\label{fig:3Dsim-profiles}
}
\end{figure*}

Figs~\ref{fig:3Drms-profiles}(a) and (b) show the comparison between the
observed profiles of $\sigma_{\rm RM}$ and those for the spherically-symmetric
model.  The best fit has a central rms magnetic-field strength, $B_0 = 0.28$\,nT
(2.8\,$\mu$G) and $\mu \approx 0.5$.  The fit is poor: chi-squared for the
5-frequency profile is 35.7 with 6 degrees of freedom.  Fits of similar quality
are obtained for any $\mu$ in the range $0.1 \la \mu \la 0.8$.

The model profiles show an asymmetry in RM fluctuation amplitude, but the
variation (which occurs on a scale $\sim r_c$) is much slower than observed. The
predicted RM fluctuation amplitude is systematically too high (by a factor of
two) in the North. It also falls rapidly in the South lobe at distances $\ga r_c
\approx 150$\,arcsec (following the density profile), and therefore fails to
match the observed value at $\approx$300\,arcsec from the nucleus.

The failure of this model to fit the remarkably low RM fluctuation amplitude in
the North of 3C\,31 is graphically illustrated by a comparison between example
realisations and the observations [Figs~\ref{fig:3Dsim-images}(a) -- (d)].
Corresponding profiles of mean RM at two resolutions and rms RM at 1.5 arcsec
resolution only are shown in Figs~\ref{fig:3Dsim-profiles}(a) -- (f).

\subsection{Cavities}
\label{cavities}

We have estimated the effect of cavities on the RM variation assuming that they
are evacuated volumes that do not perturb the surrounding gas or field.  We
first examined the effect of ``minimal'' cavities that match the inner jets as
modelled by \citet{LB02a}, continued outwards. These are cones with half-opening
angle 13.2\degr\ centred on and extending to a distance of 50\,arcsec from the
nucleus, after which they become cylindrical. The effect of these ``minimal''
cavities on the RM distribution in a spherically symmetric model was too small
to be detectable.  In order for a cavity model to generate the observed
rapid RM gradient across the nucleus, we found that: (a) the initial
half-opening angle of the cavities must exceed that of the inner radio jet, and
(b) the cavities must enclose a volume somewhat larger than that of the
high-brightness structure from which our RM images are derived.

\begin{figure}
\begin{center}
\epsfxsize=7cm 
\epsffile{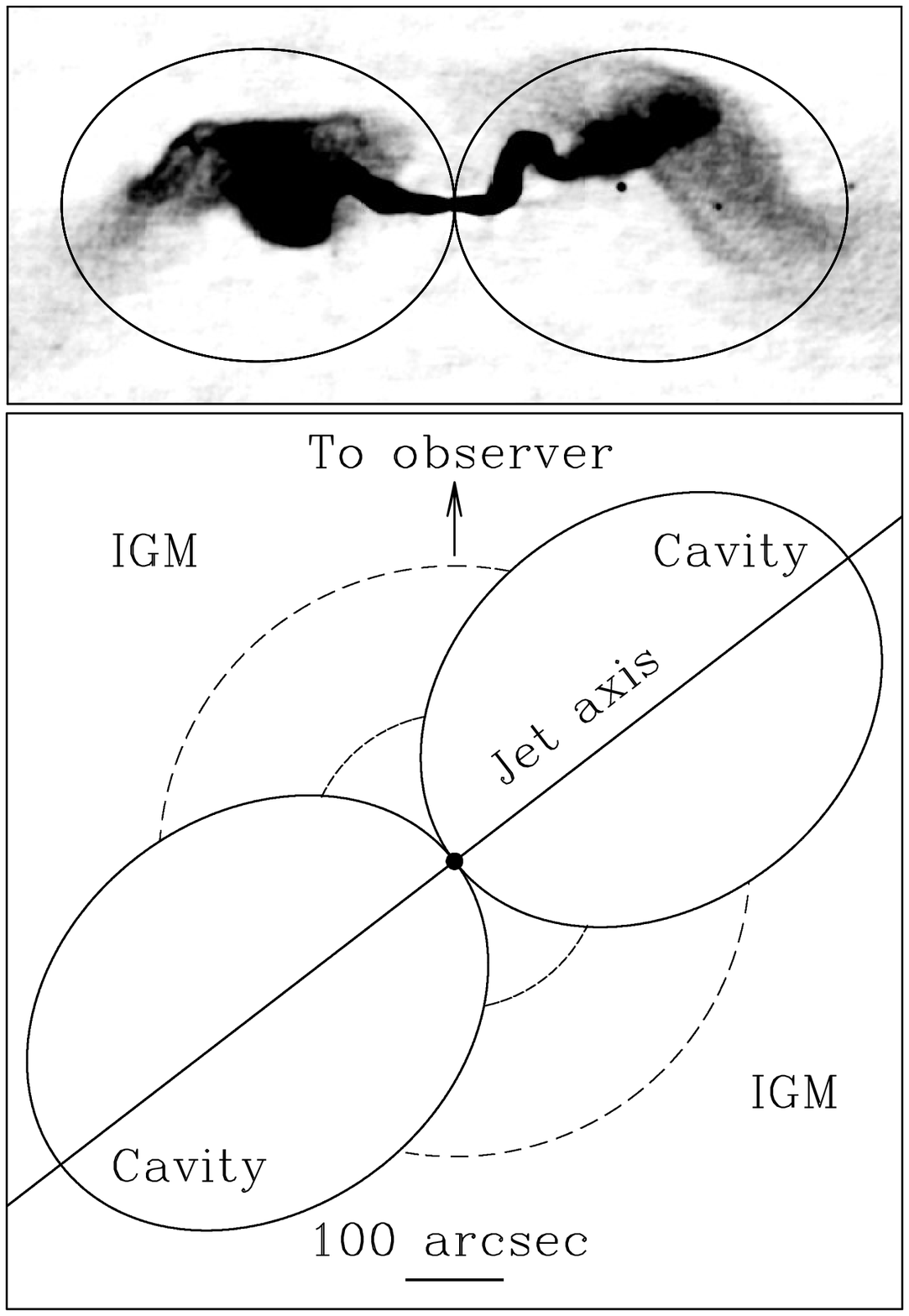}
\caption{The assumed cavity geometry for 3C\,31.  Top: the cavity shape that
produced the best fit to the RM data projected on a total-intensity image and
rotated so the inner jet axis is horizontal.  Bottom: a cross-section through
the cavity in the plane defined by the jet axis and the line of sight. The
dashed arcs of circles represent isodensity contours in the IGM.
\label{fig:cavity}
}
\end{center}
\end{figure}

A simple model which significantly improves the fit has ellipsoidal cavities as
sketched in Fig.~\ref{fig:cavity}. These have semi-major and semi-minor axes of
252\,arcsec (85\,kpc) and 200\,arcsec (68\,kpc), respectively. The semi-major
axis projects to 200\,arcsec on the plane of the sky.  The best fit is for
$\mu \approx 1.0$ [Figs~\ref{fig:3Drms-profiles}(c) and (d)]. Example realisations and
profiles are shown in Figs~\ref{fig:3Dsim-images}(e) and (f) and
Figs~\ref{fig:3Dsim-profiles}(g) -- (i), respectively.  The low level of RM
fluctuations in the North is reproduced quite well, but there is still a
significant difference between the observed 5-frequency and simulated profiles
(chi-squared = 18.6 with 6 degrees of freedom). This is due mostly to one bin,
at 300\,arcsec from the nucleus in the South and coincident with the most
negative RM values in the source [Figs~\ref{fig:RMDP5.5}(b) and (c)].  The
central bins give relatively small contributions to the total chi-squared,
despite the large differences between observed and predicted values, because the
sampling errors are large where the source is narrow.  The 4-frequency profiles
of $\sigma_{\rm RM}$ are formally consistent with the cavity model for any $\mu$
in the range $0.1 \la \mu \la 1.2$ and the chi-squared for $\mu = 1.0$ is 7.0
with 8 degrees of freedom.

For this model, the rms central magnetic-field strength is $B_0 =0.74$\,nT
(7.4\,$\mu$G). This is larger than for the spherically-symmetric model because
there is a lower Faraday depth at large radii for given $B_0$, partly as a
consequence of the steeper radial field dependence ($\mu = 1.0$), but mainly
because there is less gas along the line of sight. Both effects require an
increase in normalization to match the observed Faraday rotation in the South at
large radii.  The X-ray gas parameters were determined by
fitting to azimuthally averaged profiles, however \citep{KB99}. In the presence of a
cavity with the shape we assume, the central density would be underestimated by
a factor $\approx$1.5 and the inferred field strength would be overestimated by
the same factor.

The predicted $\sigma_{\rm RM}$ profiles are generally insensitive to the
precise shape of the cavity provided that it extends to the full extent of the
sampled region and has a half-opening angle comparable with the angle to the
line of sight, $\theta \approx 52^{\circ}$; for example we have generated similar fits
for a cavity which initially expands in a cone, becoming cylindrical far from
the nucleus.  The best-fitting value of $\mu$ depends on the precise shape of
the cavity, since both affect the Faraday depth at large radii.

Although our simple axisymmetric cavity model gives a reasonable fit to the
observed RM distribution, reality is likely to be much more complicated:
\begin{enumerate}
\item The cavity required to fit the RM data appears much larger than the
observed extent of the brightest synchrotron emission at 1.4\,GHz near the
nucleus. We note, however, that there is low-brightness emission from the North
and South spurs close to the nucleus and within the projected envelope of the
cavity. This is visible within 200\,arcsec of the nucleus on the East side of
both jets (Figs~\ref{layout} and \ref{fig:cavity}).
\item It is also possible that earlier epochs of source activity could have
cleared a bigger volume of thermal matter. Such a cavity would not have been
detected in existing {\em Chandra}, {\sl ROSAT} or {\sl XMM-Newton} X-ray images
\citep{Hard02,KB99,Croston08}, but should be clearly visible in a longer {\sl
XMM-Newton} integration or, in the radio, at lower observing frequencies
(cf.\ \citealt{Birzan08}).
\item Alternatively, the surprisingly large size required for the cavities may
just be an artefact of the simplicity of our model.  Clearly, we cannot
constrain the shape of the cavities along lines of sight where there is no
polarized emission.  Our inference of a large transverse size for the cavities
projected on the plane of the sky is purely a consequence of the shape of the
$\sigma_{\rm RM}$ profile in those areas we can sample and the assumptions of
axisymmetry and identical cavities. It is possible that the cavities are
significantly larger along the line of sight than transverse to it or that they
differ radically from each other in shape. In particular, the structure of
3C\,31 as projected on the plane of the sky is not straight, so it is likely
that bends also occur in the plane defined by the source axis and our line of
sight. Such bending might well modify the Faraday depth distribution on large
scales.
\item If significant magnetoionic material is associated with other galaxies in
the NGC\,383 group located in front of the radio emission from 3C\,31, we might
expect to see additional, localized, Faraday rotation. An additional
contribution to the rms RM at a distance of $\approx$300\,arcsec from the
nucleus in the South would greatly improve the $\sigma_{\rm RM}$ fit, so we
searched for foreground galaxies close to the RM anomaly in the South on the
Digital Sky Survey.  No candidate objects are found within $\approx$2\,arcmin,
however.
\item There could be an entirely ``accidental'' large-scale intrinsic asymmetry
in the foreground Faraday screen that produces an excess RM over the outer South
jet and tail, coincidentally reinforcing the RM asymmetry for the inner regions.
Some intrinsic asymmetries must be present in the gas distribution to provide
the observed bending of the radio structure, the different shapes of the radio
spurs and tails in the North and South, and the flat edge of the East side of
the South spur \citep[and this paper, Fig.\ref{layout}]{lb08a}.  A large-scale
asymmetry is also evident in the {\sl ROSAT} image, in which the nucleus of
3C\,31 is displaced by $\approx$1\,arcmin to the North-East of the centre of the
group gas distribution \citep{KB99}.
\item Finally, our interpretation of the RM profiles close to the nucleus might
  be complicated by Faraday rotation associated with the galaxy-scale hot gas
  component imaged by \citet{Hard02}, which has a core radius of only
  3.6\,arcsec. If this had the same field strength and power spectrum as we
  assume for the group gas, then there would be a large excess RM close to the
  nucleus. We have estimated the resulting $\sigma_{\rm RM}$ profile and find
  that there would be a detectable excess between $-20$ and $+15$\,arcsec, a
  factor of two increase between $-12$ and $+10$\,arcsec and a peak of
  $\sigma_{\rm RM} \approx 400$\,rad\,m$^{-2}$ within $\approx$1\,arcsec of the
  nucleus (positive and negative distances refer to the main (North) and counter
  (South) jets, respectively). We have no information about lines of sight
  within a beamwidth of the nucleus, which is unpolarized for our observing
  frequencies and resolutions (perhaps because of such large Faraday rotation
  gradients on small scales). We see no evidence of anomalously high RM's close
  to it, however. In any case, we have no reason to suppose that the field
  strength and power spectrum we have adopted are appropriate for the
  galaxy-scale component, which is smaller than the magnetic autocorrelation
  length we derive for the group gas.  We can therefore justify our neglect of
  the galaxy-scale component, noting that it may affect the innermost bins of
  Fig.~\ref{fig:3Dsim-profiles}.
\end{enumerate}
These considerations suggest that it may not be possible -- or necessary -- to
explain every feature of the large-scale Faraday RM distribution over 3C\,31
with a symmetric model of the magnetoionic medium.  Nevertheless, we have
demonstrated that such a model can provide an approximate fit to our RM
observations and we certainly cannot ascribe the entire asymmetry to
``accidental'' effects and still preserve its correlation with jet sidedness in
the FR\,I population as a whole.

\subsection{Alternative explanations for the RM asymmetry}

Our inference of large cavities surrounding the radio structure is
testable by
deeper radio and X-ray observations. If these cavities are not detected,
alternative ways to produce a Faraday depth distribution
consistent with our data
must be considered.

One possibility is that the magnetic field is anisotropic. A
locally
two-dimensional
structure is expected either if the field is ``draped'' over the surface
of a
subsonically-expanding lobe \citep{DP08} or if it is compressed by a
bow-shock
surrounding the source.  This could introduce a systematic change in $B_z/B$
with position and hence a gradient of RM fluctuation amplitude. For this
mechanism to produce a gradient in the observed sense, a special geometry is
needed: the two-dimensional field sheets in front of the approaching
lobe must
be preferentially orthogonal to the line of sight while those in front
of the
receding lobe are aligned with it. This might occur if the field is enhanced
along surfaces that are oblique to the radio jet axis, as in a bow shock
-- such
a structure might be imposed at some distance from the source by sound/shock
waves created by earlier outbursts from the AGN \citep[as seen in the
Perseus
cluster by][for example.]{Fabian06}.

Alternatively, the Faraday-rotating material might be in a disk whose radius
(after projection) is similar to, or larger than, the sampled area of the
tail. A thin disk with a fairly uniform distribution of Faraday depth would
naturally produce a steep gradient of RM fluctuation amplitude across the
nucleus, coupled with the lack of variation across the outer South
tail.  One
possibility is that the gas distribution
itself
is highly flattened. As an example, we
have simulated an oblate beta-model gas disk with a large core radius
containing
a magnetic field and orientated roughly perpendicular to the radio axis.
Such a
disk could produce the flat profile of $\sigma_{\rm RM}$ in the South,
but would
also generate higher RM than observed in the inner North jet unless it is
extremely thin. Flattened X-ray emitting gas distributions have occasionally
been observed around radio galaxies (e.g.\ \citealt{3C403}), but the
group X-ray
source surrounding 3C\,31 is quite round \citep{KB99,Croston08},
inconsistent
with the appearance of a highly inclined, thin disk.  If the hot gas is
responsible for the Faraday rotation but has a spherical distribution,
then the
only remaining possibility is that the magnetic field has a disk-like
configuration,
a supposition that seems entirely ad hoc.

Finally, the
Faraday rotation might come from a different phase of the ionized IGM
than that visualised by the X-ray data.
Dusty
`superdisks' with diameters $\ga$75\,kpc and thicknesses
$\approx$25\,kpc have
previously been postulated to explain depolarization asymmetry and other
properties of powerful radio galaxies \citep{GKN,GKW}.  Such a disk in
3C\,31
would be too thick to show the rapid change in RM fluctuation amplitude
across
the nucleus that we observe and in any case the observed disk of
molecular and
ionized gas and dust is tiny by comparison with the scale of the observed
Faraday rotation -- $\approx$2\,kpc in diameter
\citep{Mar99,Oku05,Ionized}. This explanation for the Faraday-rotation
asymmetry
also seems extremely unlikely.

\subsection{A note on single-scale models}
\label{1-scale}

A rough, but rapidly calculable, approximation to the profile of rms RM
along the source axis can be derived from a single-scale model of the
magnetic-field structure \citep{Burn66,Felten96}. The basic assumption 
is that the magnetic field is orientated randomly within cells of uniform
size $D$ and both the strength of the field and the thermal electron density are smooth
functions of position.  The rms foreground Faraday rotation $\sigma_{\rm RM}$,
can then be calculated by integrating the expression:
\begin{equation}
\sigma_{\rm RM}^2 = \frac{K^2 D}{3} \int_{z_{\rm em}} ^{z_{\rm obs}} B^2(z) n^2(z) dz
\label{eq-single-scale}
\end{equation}
\citep{Felten96} where $z_{\rm em}$ is the position of the front surface of the
emitting material along the line of sight $z$ and $z_{\rm obs}$ is the position
of the observer (usually $z_{\rm obs} \rightarrow \infty$).  The integral does
not generally have a solution in closed form, but we have verified that our
numerical algorithm reproduces the analytical results derived by
\citet{Felten96} in special cases. This method allows rapid computation of
$\sigma_{\rm RM}$ profiles and we have used it extensively to explore different
field and density combinations.  

For quantitative modelling, the single-scale model has serious
disadvantages, however. Although it can be used to derive a rough value of the field
strength, the appropriate cell size $D$ to use for a realistic power spectrum is
not obvious.  The apparent RM fluctuation scale, or more precisely the RM
autocorrelation length $\lambda_{\rm RM}$ [equations~(\ref{eq-lambdaRM}) and
(\ref{eq-lambdaRM-fourier})], is not the correct value. \citet{Murgia} showed
that the approximation $D \approx \lambda_B$, where $\lambda_B$ is the magnetic
autocorrelation length [equations~(\ref{eq-lambdaB}) and
(\ref{eq-lambdaB-fourier})] holds for power-law magnetic power spectra provided
that $\sigma_{\rm RM}$ is measured over a sufficiently large area -- a
restriction which is not satisfied in our analysis of $\sigma_{\rm RM}$ profiles
at 5.5 arcsec resolution close to the centre of the source.  We have to limit
the size of our averaging bins along the jets in order to resolve the global
variations of path length and RM. Close to the nucleus, the source structure is
dominated by narrow jets, so there are only a few autocorrelation areas per bin
(Table~\ref{tab:autocorr}) and the single-scale model systematically
overestimates the rms given by a full 3-d simulation
(Fig.~\ref{fig:3Drms-profiles}). At distances $\ga$100\,arcsec, where the source
broadens, the single-scale approximation with $D = \lambda_B$ gives a much
better representation of the simulated profiles.  Finally, we note that a full
simulation is required in order to estimate errors for a realistic power
spectrum and the sampling grid defined by the observations.

The single-scale approximation does, however, allow us to estimate the effects
of the finite depth of the simulation volume on the simulated $\sigma_{\rm RM}$
profile by comparing the predictions of equation~(\ref{eq-single-scale}) with
$z_{\rm obs}$ set to the half-width of the simulation box and $z_{\rm obs} =
\infty$. For the cases we have modelled in detail, the two profiles are very
similar (Fig.~\ref{fig:3Drms-profiles}), so almost all of the RM fluctuations
will be accounted for by the 3-d simulation.

\section{Hydra A}
\label{Hydra}

Aside from 3C\,31, Hydra\,A is the only FR\,I radio galaxy with a well-imaged RM
distribution in which there is a pronounced asymmetry across the nucleus
\citep{HydraA}.  In 3C\,31, the orientation is well determined (at least near
the nucleus) and the group gas is well characterised by X-ray observations, but
our preferred explanation for the observed global RM distribution -- that the
radio lobes have inflated cavities which are devoid of thermal plasma -- has not
yet been adequately tested using X-ray observations.  For Hydra\,A, the radio
source orientation is less well determined, but the cavities associated with the
inner radio lobes have been directly imaged using {\sl Chandra}
\citep{McN2000,Wise2007}.  It is therefore of interest to to examine how well a
cavity model can account for the global RM variations over this source.  As for
3C\,31, we start by deriving the RM structure function (Section~\ref{hydrasf})
and use the results to simulate the RM variations (Section~\ref{hydrasim}).

\begin{figure}
\begin{center}
\epsfxsize=8.5cm
\epsffile{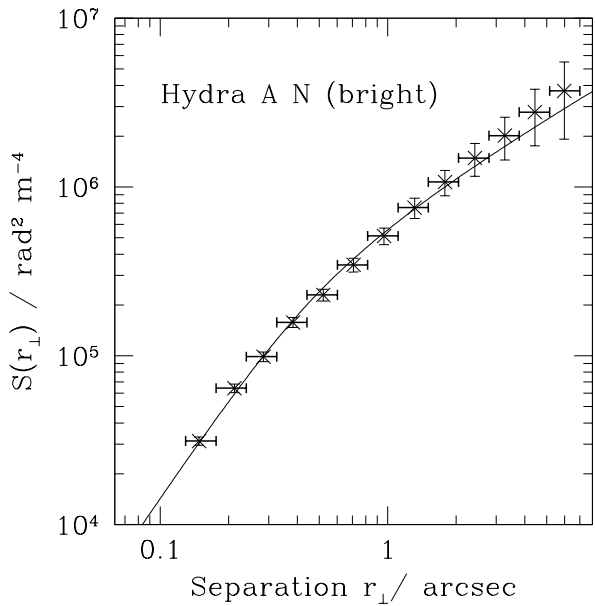}
\caption{The RM structure function for the brightest region of the North
lobe in
  Hydra\,A. The curve shows the predicted structure function, $S(r_\perp)$, for a
  power-law RM power spectrum with slope $q = 2.77$ and a convolving beam of
  0.3\,arcsec FWHM.  Error bars show the rms spread for multiple
realisations of
  the model power spectrum.
\label{fig:hydra_sf}
}
\end{center}
\end{figure}

\begin{figure}
\begin{center}
\epsfxsize=8.5cm
\epsffile{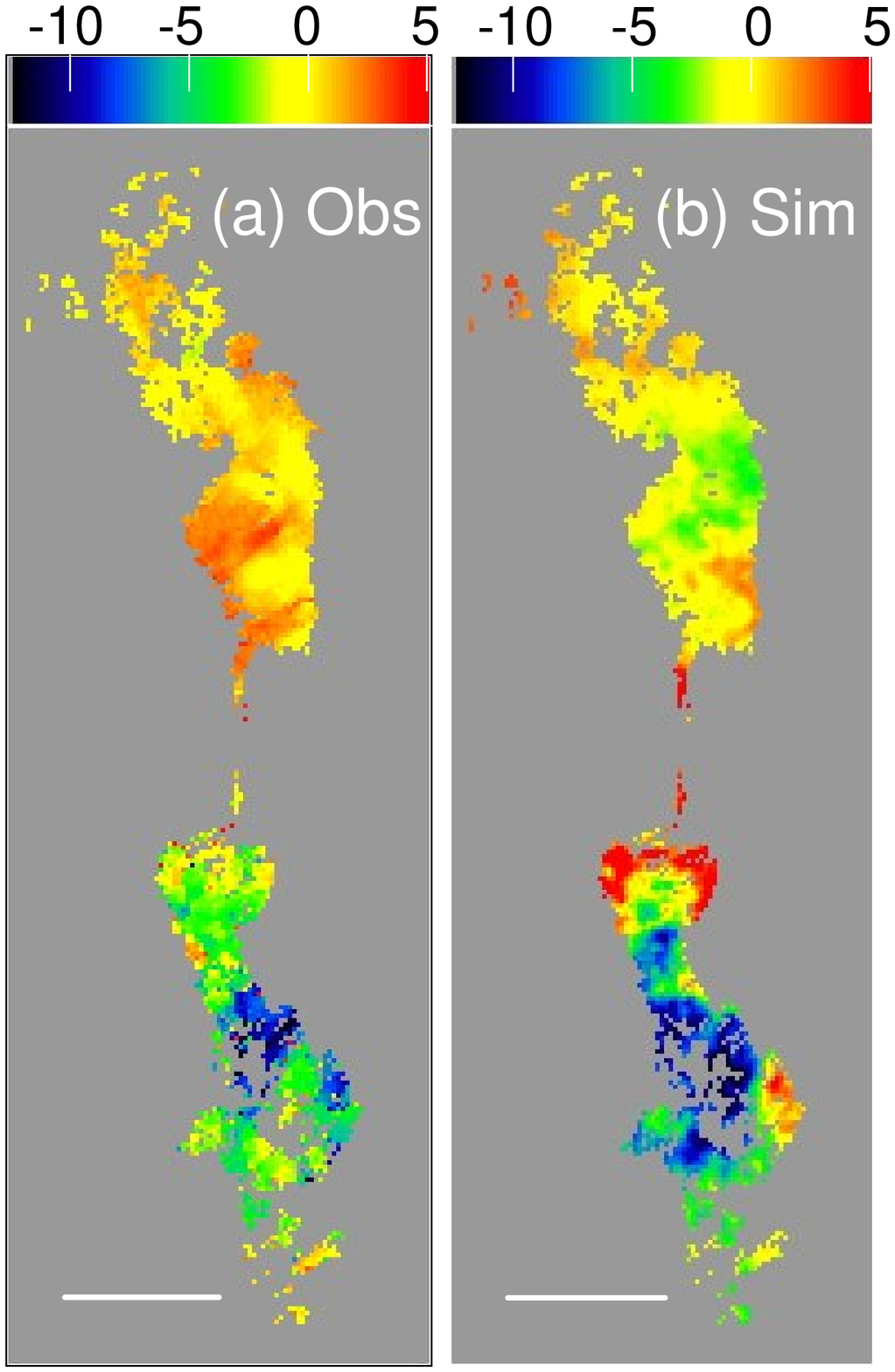}
\caption{Observed and simulated RM images for Hydra\,A. (a) Observed RM, using
data from \citet{HydraA}. The RM's for the North and South lobes are from
\citet{Pacerman2} and \citet{HydraA}, respectively. (b) A simulation using a
power-law RM power spectrum with index $q = 2.77$. The assumed angle to the line
of sight is 45$^\circ$, $B \propto n^{0.25}$ ($\mu = 0.25$) and the density
model includes cavities A and B from \citet{Wise2007} as described in the
text. The RM scale is from -12500 to 5000\,krad\,m$^{-2}$.  The images have been
rotated clockwise by 26$^\circ$ compared with those in \citet{HydraA} and
\citet{Pacerman2} and the resolution is 0.3\,arcsec FWHM.  The horizontal bars
indicate a scale of 20\,kpc (18.7\,arcsec).
\label{fig:hydra_images}
}
\end{center}
\end{figure}

\subsection{Structure function}
\label{hydrasf}

In order to derive an RM power spectrum for our simulations and to compare our
results with the Bayesian maximum-likelihood method of \citet{VE05}, we measured
the RM structure function as in Section~\ref{2d}.  We used RM images made using
the {\sc aips rm} task \citep{HydraA} and the {\sc Pacerman} algorithm
\citep*{Pacerman2}, as in Section~\ref{RM-variations}. We evaluated the structure
function for the North lobe of Hydra\,A between 5 and 15\,arcsec from the
nucleus, where both algorithms are in good agreement.\footnote{In the South
lobe, the RM's are very large, there is significant depolarization and the RM
distributions calculated using the two algorithms differ significantly.  We have
not, therefore, calculated the RM structure function in this part of Hydra\,A.}
The results derived from the {\sc pacerman} RM image are shown in
Fig.~\ref{fig:hydra_sf}. As noted by \citet{Pacerman2}, there is significant
uncorrelated noise in the RM image, and we have corrected the structure function
for this using the noise estimates from that reference, as in
Section~\ref{Sfunc_observe}.

We fit BPL and CPL models, again as in Section~\ref{Sfunc_observe}. Both fits
gave the same solution: a pure power law with $q_{\rm low}$ (or $q$) = $2.77 \pm
0.10$ with no requirement for a break in slope or high-frequency cut-off within
our (relatively limited) sampling range.  The fitted structure function is
shown in Fig.~\ref{fig:hydra_sf}. BPL models with $f_b > 0.9$\,arcsec$^{-1}$
(0.8\,kpc$^{-1}$) are allowed by the data for $q_{\rm low} \approx 2.7$,
however.  In contrast, \citet{VE05} found a Kolmogorov spectrum for $f \ga
0.32$\,kpc$^{-1}$, flattening at lower frequencies (but they did not include the
effect of the convolving beam, FWHM = 0.32\,kpc, in their analysis).

Thus we find that the index of the power spectrum for Hydra\,A ($q \approx 2.8$)
is steeper than the low-frequency index for 3C\,31 ($q_{\rm low} \approx
2.3$). Assuming a BPL power spectrum, the break frequency is at $f_b \approx
0.2$\,kpc$^{-1}$ for 3C\,31 compared with $f_b \ga 0.8$\,kpc$^{-1}$ for
Hydra\,A.

\subsection{Distribution of magnetoionic material}
\label{hydrasim}

We attempted to reproduce the large-scale variations over a composite RM image
of Hydra\,A made using the {\sc pacerman} algorithm for the North lobe, as in
Section~\ref{hydrasf}, and the {\sc aips rm} task for the South lobe, where the
{\sc pacerman} solution is questionable \citep{Pacerman2}.  There is substantial
depolarization in the South, and the {\sc aips rm} image also contains a number
of artefacts such as large discontinuities on scales smaller than a beam. In
order to reduce these problems, we blanked all points with RM outside the range
$-15000$ to $+10000$\,rad\,m$^{-2}$ and those below a total-intensity
threshold. The composite RM image is shown in Fig.~\ref{fig:hydra_images}(a).

We characterised the background density distribution using the double beta model
of \citet{Wise2007}, normalized to the deprojected profile of
\citet{David01}. The parameters of the model are given in
Table~\ref{tab:hydradens}.  We approximated the shapes of the inner two cavities
A and B \citep{Wise2007} as identical ellipsoids with projected semi-major axes
of 19.8\,arcsec exactly along the jet and semi-minor axes of 11.8\,arcsec
centred at projected distances of 24.1\,arcsec on either side of the
nucleus. These are the mean values for the two cavities given by
\citet{Wise2007}; A and B are on opposite sides of the nucleus, but otherwise
very similar in geometry and location. Our simulations were as described for
3C\,31 in Section~\ref{3Dsim}, but taking a power-law magnetic power spectrum
with $q = 2.77$, as derived in Section~\ref{hydrasf} and an upper frequency
limit of $f_{\rm max}$ = 1.67\,arcsec$^{-1}$, set by the size of the simulation
grid (512$^3$ with 0.3-arcsec pixels).  We also fixed the lower frequency limit
to be $f_{\rm min}$ = 0.012\,arcsec$^{-1}$. $f_{\rm min}$ is poorly constrained
by our structure-function data, but the presence of large-scale RM variations
across the 80-arcsec inner structure of Hydra\,A \citep{HydraA} requires the
power spectrum to have significant amplitude at this frequency.

As the inclination of Hydra\,A is not well determined, we simulated angles to the
line of sight in the range $37.5^\circ \leq \theta \leq 60^\circ$ -- larger
values failed to generate the observed RM asymmetry and smaller ones are
improbable given the size of the source and its relatively low jet/counter-jet
ratio \citep{Taylor90}.  We also varied the dependence of magnetic field
strength on density, taking $B \propto n^\mu$ with $0.1 \leq \mu \leq 1.0$ For
each pair of values $\theta, \mu$, we generated 25 realisations.  We then
compared the observed and simulated profiles of $\sigma_{\rm RM}$, binning in
boxes of length 4.2\,arcsec along the jet axis (PA 26$^\circ$ on the sky) and
extending to cover the entire width of the source.  Finally, we adjusted the
central rms field strength to optimise the agreement between observed and
simulated profiles, using a combination of fitting and sampling errors to define
chi-squared, just as for 3C\,31.

\begin{figure}
\begin{center}
\epsfxsize=6cm
\epsffile{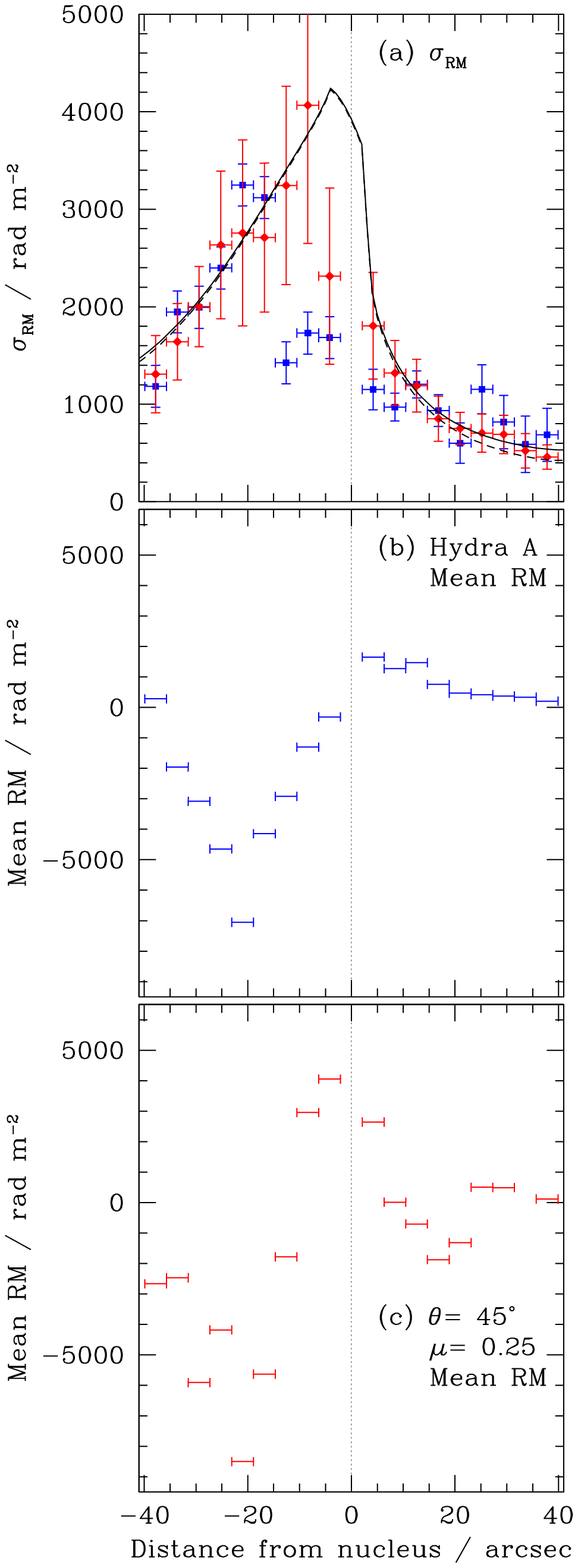}
\caption{Profiles of RM along the inner jet axis for observed and simulated
  images of Hydra\,A (Fig.~\ref{fig:hydra_images}).  The simulations assumed
  $\theta = 45^\circ$, $B_0 = 1.9$\,nT and $\mu = 0.25$, with the geometry
  described in the text. Data have been averaged in boxes of length 4.2\,arcsec
  along the jet axis, extending to cover the entire width of the source. The jet
  axis is taken to be vertical in Fig.~\ref{fig:hydra_images} or PA 26$^\circ$
  on the sky. Positive and negative distances refer to the North and South
  lobes, respectively (the former is approaching in the simulations). (a) rms
  RM. The observed values are indicated by the blue squares, with errors derived
  from the RM fits. The simulation results (red diamonds) represent the mean and
  rms scatter from 25 realisations. The curves show the predicted values of
  $\sigma_{\rm RM}$ for a single-scale model with $D = \lambda_B$ = 3.1\,kpc, as
  calculated for the model power spectrum. Full line: infinite integration
  range; dashed line: integration truncated at the near surface of the
  simulation box.  (b) Observed mean RM. (c) Mean RM for the simulation shown in
  Fig.~\ref{fig:hydra_images}(b). This particular realisation (one out of 25)
  was chosen to illustrate that a large-scale change in sign of RM can be
  generated naturally by an isotropic magnetic field with the power spectrum we
  assume.
\label{fig:hydra_profiles}
} 
\end{center}
\end{figure}

There is a well-defined minimum in chi-squared, with best-fitting values of
$\theta = 45^\circ$ and $\mu = 0.25$. The 68\% confidence region is an ellipse
in the $\theta, \mu$ plane with semi-major axes $\sigma_\theta \approx 7^\circ$
and $\sigma_\mu \approx 0.2$.  The agreement between observed and predicted
profiles of $\sigma_{\rm RM}$ for the best fit is good
[Fig.~\ref{fig:hydra_profiles}(a); chi-squared = 12.7 with 14 degrees of
freedom].  In particular, the large difference between the fluctuation
amplitudes in the two lobes is reproduced well, whereas a spherically-symmetric
model without cavities gives an unacceptable fit for any inclination.  As in
3C\,31, the discrepancies close to the nucleus are not significant: errors on
the predicted values are large because of poor sampling in the region of the
narrow inner jets.  We show an example realisation in
Fig.~\ref{fig:hydra_images}(b). This has been chosen to illustrate that the
change in sign of RM observed on large scales can be generated naturally by our
assumed power spectrum without the need to postulate an additional ordered
component of magnetic field (cf.\
\citealt{HydraA}). Figs~\ref{fig:hydra_profiles}(b) and (c) show profiles of
observed and predicted mean RM.

The changes in RM amplitude across Hydra\,A in this picture are due almost
entirely to the inner beta model, whose core radius is 27.7\,kpc (26.5\,arcsec)
and to the cavity structure.  Any RM fluctuations due to the larger and more
tenuous gas component would be essentially constant in amplitude across the
central 80\,arcsec region which we consider.  The observed fall-off in RM at
distances $\ga$20\,arcsec in the South lobe is consistent with the drop in
density, as predicted but not seen in the corresponding (South) lobe of
3C\,31.

The central magnetic field strength required to match the RM fluctuation
amplitude for the $45^\circ$ simulation with cavities is 1.9\,nT, significantly
larger than the 0.7\,nT deduced from modelling of the North lobe alone by
\citet{VE05}.  We have used the single-scale approximation with scale-length
$\lambda_B$ to compare our predicted profiles of $\sigma_{\rm RM}$ with those of
\citet{VE05} and find very good agreement for the North lobe. The discrepancy in
field strength therefore results from differences in the field and density
models, as follows: 
\begin{enumerate}
\item The spherically-symmetric model used by \citet{VE05} predicts a
  significantly higher density than that of \citet{Wise2007} over the range of
  radii responsible for the bulk of the Faraday rotation.
\item Our inclusion of cavities further reduces the amount of gas along the line of
  sight compared with the expectation from a spherically symmetric model.
\item We find that $\mu \approx 0.25$ gives a better fit to the RM fluctuation
  profile for a density model with cavities, whereas \citet{VE05} estimate $\mu
  = 0.5$ for a spherically symmetric model.
\item The power spectrum derived by \citet{VE05} is significantly different from
  ours. In order to estimate the effect on the derived central field strength,
  we can compare the magnetic autocorrelation lengths.  \citet{VE05} derive
  $\lambda_B$ = 1.5\,kpc [dividing by 2 to match our definition from
  equation~(\ref{eq-lambdaB})], compared with 3.1\,kpc for our model power
  spectrum.
\end{enumerate}

Our results for Hydra\,A provide quantitative support for our picture of
magnetic field distributed through a spherical background gas distribution with
embedded cavities. In this source (the only case where the gas density and
cavity structure are both well determined), we can match the global variation of
RM fluctuation amplitude for plausible angles to the line of sight.

\begin{table}
\caption{Two-component density model assumed for simulations of RM in
  Hydra\,A. $n_0$ is the central density of a component, $r_c$ is its core radius
  and $\beta_{\rm atm}$ is the form factor. $n(r) =
  n_0(1+r^2/r_c^2)^{-3\beta_{\rm atm}/2}$. \label{tab:hydradens}} 
\begin{tabular}{rrr}
\hline
&&\\
$n_0$ / m$^{-3}$ & $r_c$ / kpc & $\beta_{\rm atm}$  \\
&&\\
\hline
&&\\
$6.2 \times 10^4$ & 27.7 & 0.686 \\
$2.6 \times 10^2$ & 235.6 & 0.907 \\
&&\\
\hline
\end{tabular}
\end{table}

\section{Summary and Further Work}
\label{Conclusions}

\subsection{Summary}

We have analysed images of linearly polarized emission in the FR\,I radio galaxy
3C\,31 with resolutions of 5.5 and 1.5\,arcsec FWHM at six frequencies between
1365 and 8440\,MHz.  Our use of all six frequencies in an analysis of the
variation of the degree of polarization with wavelength has enabled us to
measure very small depolarizations, which we characterize using the Burn law
parameter $k$.  At 1.5-arcsec resolution, we find $\langle k \rangle =
3$\,rad$^2$\,m$^{-4}$ (DP$^{\rm 22cm}_{\rm 6cm}$ = 0.99) within 60\,arcsec of
the nucleus in the North jet and $\langle k \rangle = 55 $\,rad$^2$\,m$^{-4}$
(DP$^{\rm 22cm}_{\rm 6cm}$ = 0.88) in the South.  The low depolarization
and absence of deviation from $\lambda^2$ rotation confirm that almost all of
the observed Faraday rotation is due to foreground material.  The RM fluctuation
amplitude is significantly larger in the South.

The amplitude and scale of the RM fluctuations in 3C\,31 appear qualitatively
similar to those derived for other well-observed FR\,I sources in comparable
environments, e.g.\ NGC\,6251 and 3C\,449 \citep{PBW84,Fer99}, but are much
smaller in amplitude than those in FR\,I sources located in cD galaxies with
cooling cores such as M\,87 and Hydra\,A \citep{OEK,HydraA} and much larger than
for NGC\,315, which is in a sparse group \citep{LCCB06}.

We have determined the RM structure function for five regions in 3C\,31 over
which the fluctuation amplitude is roughly constant.  We model the fluctuations
on the assumption that the magnetic field responsible for the foreground
rotation is an isotropic Gaussian random variable (characterised entirely by its
power spectrum). For the range of foreground Faraday rotation seen in 3C\,31,
the observed RM image is very close to the true image convolved with the
observing beam. We have used this approximation to incorporate the effects of
the beam in the calculation of theoretical structure functions for model power
spectra by numerical integration of a Hankel transform relation. We evaluate
errors due to imperfect sampling by making multiple realisations of the model
power spectrum on the observed grid.  For the four regions with the most
reliable RM measurements, our data are consistent with an RM power spectrum
which has the same form everywhere but whose amplitude varies with position. The
power spectrum cannot be a single power law over the entire range of sampled
spatial frequencies. A simple functional form that fits our data is a broken
power law $\hat{C}(f_\perp) \propto f_\perp^{-q}$ with $q = 2.32$ for $f <
0.062$\,arcsec$^{-1}$ and $\hat{C}(f_\perp) \propto f_\perp^{-11/3}$, as
expected for Kolmogorov turbulence, at higher frequencies. A power spectrum with
a similar low-frequency slope and an abrupt high-frequency cut-off gives almost
as good a fit to the observed structure functions, however.  We can only
determine an approximate lower limit to the outer scale: in terms of spatial
frequency, $f_{\rm min} \la 0.005$\,arcsec$^{-1}$ (0.015\,kpc$^{-1}$).  We
conclude that these (and other, similar) models of the power spectrum cannot be
distinguished using the available RM data alone. Our depolarization measurements
at 1.5-arcsec resolution are also consistent with extrapolations of either power
spectrum to higher spatial frequencies. In the South lobe of 3C\,31, the model
power spectra (normalized using the RM structure function) predict more
depolarization than is observed at low resolution, however.

In Hydra\,A, we again fit the RM power spectrum with a power law, but measure a
steeper slope, $q = 2.8$, than in 3C\,31. We find no evidence for a
high-frequency cut-off or for a portion of the power spectrum with the
Kolmogorov slope at spatial frequencies $\la$0.8\,kpc$^{-1}$ (cf.\
\citealt{VE05}).

RM analyses for sources in Abell clusters give qualitatively similar
results. \citet{Murgia} found a spectral slope $q \approx 2$ extending to large
spatial scales in Abell 119 and for Abell 2255, \citet{Govoni06} argued for a
spectral slope which steepens from $q \approx 2$ to $q \approx 4$ with
increasing distance from the cluster centre. In contrast, \citet{Guidetti08}
found that a power-law power spectrum with a Kolmogorov slope and an abrupt
long-wavelength cut-off at $f_{\rm min}$ = 0.014\,kpc$^{-1}$ gave a very good
fit to their RM and depolarization data for A2382, although a shallower slope $q
\approx 2$ extending to longer wavelengths was not ruled out.  All three of
these studies assumed high-frequency cut-offs $f_{\rm max} \approx
0.1$\,kpc$^{-1}$. 

In all cases except Hydra\,A (where the range of spatial scales we have
investigated is very small), there is evidence for a change in the power-law
slope of the power spectrum, but ambiguity about its precise form. Two simple
forms of the power spectrum which fit all of the datasets are:
\begin{enumerate}
\item a broken power-law with a high-frequency slope $q_{\rm high} = 11/3$ (the
  Kolmogorov value) and no high-frequency cut-off, but flattening to a slope
  $q_{\rm low} \approx$ 2 -- 3 or even cutting off below a frequency $f_{\rm
  b} \sim$ 0.01 -- 1\,kpc$^{-1}$ and 
\item a power-law with index $q \approx$ 2 -- 3 extending to very low
  frequencies and a high-frequency cut-off at $f_{\rm max} \sim$ 0.1 --
  2\,kpc$^{-1}$.
\end{enumerate}
Other qualitatively similar models would also fit the data.  Numerical
simulations predict a wide range of spectral slopes \citep{Dolag06}, and a
detailed comparison between theory and observation seems premature.

Our observation of a change in RM fluctuation amplitude across the nucleus of
3C\,31 qualitatively supports the relativistic-jet interpretation of the
brightness asymmetry between the jets, wherein the fainter (receding) jet is
observed through a greater path length of magnetoionic medium than the nearer
(approaching) jet. To test this idea quantitatively, we have modelled the global
RM variations using three-dimensional simulations, again treating the magnetic
field as an isotropic Gaussian variable with a broken power law power
spectrum. We assume that the entire structure of 3C\,31 has an inclination of
52$^\circ$ as in our kinematic models of the inner jets.  The profile of the RM
variance as a function of distance from the nucleus of the galaxy is
incompatible with a spherically symmetric distribution for the magnetoionic
medium as derived from X-ray observations: this would cause too gradual a
variation.  We expect, however, that the radio source will have evacuated
cavities in the surrounding hot gas. A model in which the edges of the cavities
coincide with the boundaries of the radio emission on large scales is more
compatible with the observed RM profile, but current X-ray images are not deep
enough to reveal any cavities around 3C\,31. The derived central rms
magnetic-field strength $B_0 \approx 0.7$\,nT (7\,$\mu$G) and the field varies
roughly linearly with density.  The magnetic field is not dynamically dominant
over the volumes we model. The ratio of thermal to magnetic pressure,
$\beta_{\rm P} = p_{\rm thermal}/(B^2/2\mu_0) \approx 10$ at the core radius of
the group gas, $r_c = 154$\,arcsec (52\,kpc).

We have also modelled the RM distribution in Hydra\,A, where cavities {\em are}
seen in X-ray images.  A three-dimensional simulation including the cavities
gives a good representation of the RM fluctuation profile for an inclination of
$45 \pm 7$\,deg with $B \propto n^{0.25 \pm 0.2}$ and a central magnetic field
strength of 1.9\,nT (19\,$\mu$G), so $\beta_{\rm P} \approx 30$ at $r_c =
26$\,arcsec (27.7\,kpc).

Close to the nuclei of both sources, our models predict higher RM fluctuation
amplitudes than are observed. Although the sampling variances are large, this
may be an indication that our assumption of a power-law dependence of magnetic
field strength on density is oversimplified and that we have therefore
overestimated the field strengths over these volumes.

Few other RM profiles have been published for FR\,I sources with jets, and it is
not yet clear whether the abrupt changes in RM fluctuation amplitude across the
nucleus observed in 3C\,31 and Hydra\,A are common. In 3C\,296 \citep{LCBH06},
there is a smooth variation of rms RM across the nucleus, consistent with the
inferred inclination of $58^\circ$ and Faraday rotation produced by a
spherically symmetric hot gas component with a large core radius, as observed
using {\sl XMM-Newton} \citep{Croston08}. The RM fluctuations observed in
NGC\,315 \citep{LCCB06} are too small to determine a reliable profile, although
clearly larger on the counter-jet side.

\subsection{Further work}

Our modelling of the RM variations in 3C\,31 and Hydra\,A suggests a number of
observational and theoretical investigations. 
\begin{enumerate}
\item It is clearly necessary to sample a larger range of spatial scales to
  improve constraints on the RM power spectrum and in particular to decide
  whether it has the form expected for Kolmogorov turbulence on small
  scales. Higher {\em resolution} would probe small scales directly and remove
  the need to rely on crude estimates from depolarization. Higher {\em
  sensitivity} would provide better sampling of large spatial scales by allowing
  study of low-surface-brightness regions.  Use of analysis techniques such as
  Bayesian maximum likelihood would give more rigorous error estimates.  We have
  shown (Fig.~\ref{fig:RMDP5.5}b and c) that it is possible to derive accurate
  RM images from a small number of discrete frequencies in a suitably chosen
  observing band: the new generation of wideband correlators used by EVLA and
  eMERLIN will enable simultaneous observations over much larger frequency
  ranges, greatly improving the speed and accuracy of RM determination. 
\item Large changes in RM fluctuation amplitude are seen across the nuclei in
  3C\,31 and Hydra\,A, but not in 3C\,296. The prevalence of this phenomenon and
  its dependence on orientation, large-scale radio structure and environment are
  unknown. It is clear that the RM fluctuation amplitude scales with density,
  ranging from a few rad\,m$^{-2}$ in sparse environments (e.g.\ NGC\,315) to
  $\sim 10^4$\,rad\,m$^{-2}$ in cooling cores such as Hydra\,A.
  Observations of a larger sample, employing resolutions and frequency ranges
  matched carefully to the environments, will be needed.
\item We have argued that the presence of cavities in the external medium must
  be taken into account when modelling RM distributions: failure to do so will
  lead to biased estimates of the magnetic field strength and its variation with
  density. Accurate modelling requires knowledge of the source geometry (which
  we can constrain for 3C\,31 but not Hydra\,A) and observations of the cavity
  structure (which are available for Hydra\,A but not yet for 3C\,31).  Deep
  {\sl XMM-Newton} observations of the 3C\,31 group are needed, as are
  comparable data for further sources. Kinematic models of the radio jets in
  Hydra\,A in other sources could be used to constrain their orientations.
\item Deviations in source structure and density from axisymmetry, Faraday
  rotation from foreground galaxies, X-ray emission from other group members in
  the field and from relativistic electrons deposited in the cavity by the radio
  source (via the inverse Compton process) and variations in the Galactic
  magnetic field across the source structure are potential complications which
  need to be better understood.
\item The assumption that the magnetic field is an isotropic, Gaussian random
  variable is questionable. RM images of some radio sources show evidence for
  preferred directions and there are theoretical reasons to suppose that fields
  at the edges of cavities and behind radio-source bow shocks should be
  two-dimensional.  More detailed simulations of radio-source evolution in a
  magnetised intergalactic medium are needed.
\end{enumerate} 

\section*{Acknowledgements}

RAL would like to thank the Istituto di Radioastronomia, NRAO and Alan and Mary
Bridle for hospitality during the course of this work.  We also thank Greg
Taylor for allowing us to use his {\sc aips} software for the analysis of
rotation measure and for the images of Hydra\,A.  Klaus Dolag, Torsten En\ss lin
and Corina Vogt kindly provided the {\sc pacerman} RM images of 3C\,31 and
Hydra\,A.  We are particularly grateful to the referee, John Wardle, whose
comments led to significant improvements in the paper, especially in the error
analysis. We acknowledge travel support from NATO Grant CRG931498.  The National
Radio Astronomy Observatory is a facility of the National Science Foundation
operated under cooperative agreement by Associated Universities, Inc.

\newpage

\appendix

\section{Notation and basic theory}
\label{App-theory}

In this Appendix, we summarise the basic theory of RM fluctuations in a
foreground screen as given in the literature \citep{Tribble91a,EV03,Murgia},
primarily to establish notation and conventions for comparison with other work.

\subsection{Rotation measure}
\label{RMbasics}

The rotation $\Delta\chi$ of the {\bf E}-vector position angle of linearly-polarized
radiation by a foreground magnetized thermal plasma is given by:
\begin{equation}
\Delta\chi =  K \lambda^2\int
n B_z dz = {\rm RM} \lambda^2 \label{eq-RMdef}
\end{equation}
where $n$ is the thermal electron density, $B_z$ is the magnetic
field component along the line of sight, RM is the rotation measure and 
\begin{equation}
K = \frac{e^3}{8\pi^2m_e^2c^3\epsilon_0} \label{eq-RMconst}
\end{equation}
in SI units. Alternatively, 
\begin{eqnarray}
\Delta\chi/{\rm rad} & = & 8.1193 \times 10^{-3} \nonumber \\
                     & \times & (\lambda/{\rm m})^2 \int (n/{\rm
  m}^{-3}) (B_z/ {\rm nT}) d(z/{\rm kpc})  \label{eq-RMnumbers}\\ \nonumber 
\end{eqnarray}

\subsection{Theory: real space}
\label{realspace}

In what follows, we make the simplifying assumptions of Section~\ref{RMstats}:
Faraday rotation is due entirely to a foreground medium in which the magnetic
field is a Gaussian isotropic random variable and whose density and path length
are uniform over the area of interest. 

Our notation is similar to that of \citet{EV03}. ${\bf r} = (x, y, z)$ is a
vector of magnitude r with $z$ along the line of sight and ${\bf r_\perp} = (x,
y)$ (magnitude $r_\perp$) is a vector in the plane of the sky.
 
The magnetic autocorrelation function $w(r)$ and the RM autocorrelation function
$C(r_\perp)$ are defined by: 
\begin{eqnarray*}
w(r)&=& \langle{\bf B}(\bar{{\bf r}}){\bf \cdot}{\bf B}(\bar{{\bf r}} + {\bf
           r})\rangle \\ 
C(r_\perp) &=& \langle{\rm RM}({\bf r_\perp} +{\bf r_\perp^\prime}){\rm RM}({\bf
r_\perp^\prime})\rangle \\
\end{eqnarray*}
and are related by:
\begin{eqnarray*}
C(r_\perp) &=&  K^2 n^2 L \int^\infty_{r_\perp} dr  w(r) r (r^2-r_\perp^2)^{-1/2}  \\
\end{eqnarray*}
The autocorrelation lengths for the magnetic field ($\lambda_B$) and RM
($\lambda_{\rm RM}$) are of interest for comparison with other work.  We adopt
the definitions:
\begin{eqnarray}
\lambda_B &=& \int^\infty_0 dr \frac{w(r)}{w(0)} \label{eq-lambdaB}\\
\lambda_{\rm RM} &=& \int^\infty_0 dr_\perp \frac{C(r_\perp)}{C(0)}
\label{eq-lambdaRM}\\ \nonumber
\end{eqnarray}
These are equivalent to the expression given by \citet{Murgia}. The values given
by \citet{EV03} are a factor of 2 larger.\footnote{\citet{EV03}'s definitions
of autocorrelation lengths are problematic, as they integrate from $-\infty$
to $+\infty$ in $r$ and $r_\perp$, both of which are intrinsically positive
quantities.}  We also note that the mean square field $\langle B^2 \rangle =
w(0)$ and that the average magnetic energy density $\langle \epsilon_B \rangle
= w(0)/2\mu_0$ in SI units.

\subsection{Theory: Fourier space}
\label{fourier}

Here, ${\bf f} = (f_x^2+f_y^2+f_z^2)^{1/2}$ (magnitude $f$) is a vector in frequency
space with the $f_z$ coordinate along the line of sight. ${\bf f_\perp} = (f_x,f_y)$
(magnitude $f_\perp$) is a frequency vector in the sky plane. We use the Fourier
transform convention of \cite{Bracewell}, so for
one dimension, spatial coordinate $x$ and frequency $f_x$ the transform pair
$g(x)$ and $\hat{g}(f_x)$ are related by:
\begin{eqnarray*}
\hat{g}(f_x) & = & \int^\infty _{-\infty} g(x) \exp(-2\pi if_x x) dx \\
g(x) & = & \int^\infty _{-\infty} \hat{g}(f_x) \exp(2\pi i f_x x) df_x \\ 
\end{eqnarray*}

In the isotropic case, we can define the two-dimensional RM power spectrum
$\hat{C}(f_\perp)$ such that $\hat{C}(f_\perp)df_x df_y$ is the power for an
element $df_x df_y$ in frequency space.  The RM power spectrum and
autocorrelation function then form a Hankel transform pair
[equation~(\ref{eq-Hankel})].  We define the three-dimensional magnetic power
spectrum $\hat{w}(f)$ such that $\hat{w}(f)df_x df_y df_z$ is the power in a
volume $df_x df_y df_z$ of frequency space. This is the Fourier transform of the
magnetic autocorrelation function and is very simply related to the RM power
spectrum:
\begin{eqnarray}
\hat{w}(f_\perp) & = & \frac{2\hat{C}(f_\perp)}{K^2n^2L} \label{eq-B-RM-spec}
\nonumber \\
\label{eq-what-chat}
\end{eqnarray}

For our model power spectra, it is most straightforward to evaluate
autocorrelation lengths from integrals over $\hat{w}(f)$:
\begin{eqnarray}
\lambda_B &=& \frac{1}{4}\frac{\int_0^\infty \hat{w}(f) f df}{\int_0^\infty
  \hat{w}(f) f^2 df} \label{eq-lambdaB-fourier}\\
\lambda_{\rm RM} &=& \frac{1}{2\pi} \frac{\int_0^\infty \hat{w}(f) df}{\int_0^\infty
  \hat{w}(f) f df} \label{eq-lambdaRM-fourier}\\ \nonumber
\end{eqnarray}
The constants in equations~(\ref{eq-lambdaB-fourier}) and (\ref{eq-lambdaRM-fourier}) differ by a
factor of $4\pi$ from those in the equivalent expressions in \citet[their
equation 39]{EV03}. $2\pi$ of this is because we use spatial frequency $f$
rather than wavenumber $k$ ($k = 2\pi f$) and the remaining factor of 2 is due
to the definition of autocorrelation length, as explained above.
 
\section{Analytical formulae for structure functions}
\label{psformulae}

\subsection{Power-law power spectrum}
\label{PL-analytic}

In order to validate our numerical approach to calculation of autocorrelation
functions, we first consider a power spectrum
which has a power-law form $\hat{C}(f_\perp) = f_\perp^{-q}$ for all frequencies
$f_\perp$.  The structure function then has a simple analytical form
\citep{MS96,EV03}. In our notation and using equation~(\ref{eq-Hankel}):
\begin{eqnarray}
C(r_\perp) & = & 2\pi \int^\infty _0 f_\perp^{1-q} J_0(2\pi f_\perp r_\perp)
df_\perp \nonumber \\
S(r_\perp) & = & 2[C(r_\perp)-C(0)] \nonumber \\
     & = & 4\pi \int^\infty_0 f_\perp^{1-q} [1 - J_0(2\pi f_\perp r_\perp)]
  df_\perp \nonumber  \\
     & = & \frac{4\pi^{q-1}}{q-2}\frac{\Gamma(2-q/2)}{\Gamma(q/2)}r_\perp^{q-2}
  \label{eq-asymp}\\ \nonumber 
\end{eqnarray}
for $2.5 < q < 4$.

For Gaussian convolution [equation~(\ref{eq-Hankel-convolve})]:
\begin{eqnarray}
C(r_\perp) & = & 2\pi \int^\infty _0 f_\perp^{1-q} J_0(2\pi f_\perp r_\perp) 
\exp(-2\pi \sigma^2 f_\perp^2) df_\perp \nonumber \\
S(r_\perp) & = & 4\pi \int^\infty_0 f_\perp^{1-q}\exp(-2\sigma^2f_\perp^2) 
[1 - J_0(2\pi f_\perp r_\perp)] df_\perp \nonumber \\
     & = &  \frac{2^{1+q/2}\pi^{q/2}\sigma^{q-2}\Gamma(2-q/2)}{q - 2} \nonumber \\
     & \times &\left[_1F_1 \left( 1-\frac{q}{2},1;\frac {-\pi
    r^2_\perp}{2\sigma^2}\right) -1 \right] \label{eq-hyper}\\ \nonumber  
\end{eqnarray}
where we have integrated once by parts. We obtain three terms: two of these can
be integrated using equation 6.631.1 of \citet{GR} and combined with equation
9.212.2 from the same reference to give a single confluent hypergeometric
function $_1F_1$ multiplied by a $\Gamma$ function; the third is the integral
representation of the same $\Gamma$ function.

\subsection{The small-separation limit}
\label{smallsep}

Quite generally, the structure function for a power spectrum which is bounded in
frequency (unlike the asymptotic case of equation~\ref{eq-asymp}) will be
quadratic, $S(r_\perp) \propto r_\perp^2$ for sufficiently small separations $r_\perp$
\citep{Tatarskii,Tribble91a}. For a power spectrum $\hat{C}(f_\perp)$, the
structure function is:
\begin{eqnarray*}
S(r_\perp) & \approx & 4\pi^3 r_\perp^2  \int^\infty_0\hat{C}(f_\perp) f_\perp^3
df_\perp \\
\end{eqnarray*}
provided that $f_\perp r_\perp \ll 1$ over the full range of the integral.
Similarly for a Gaussian convolving beam, 
\begin{eqnarray*}
S(r_\perp) & \approx &4\pi^3 r_\perp^2  \int^\infty_0\hat{C}(f_\perp) f_\perp^3
\exp(-2\pi\sigma^2f_\perp^2) df_\perp \\
\end{eqnarray*}
For the specific example
of the cut-off power law spectrum of equation~\ref{eq-cutoff-pl} without convolution, 
\begin{eqnarray*}
S(r_\perp) & \approx & \frac{4\pi C_0 f_{\rm max}^{4-q} r_\perp^2}{4-q}    
\end{eqnarray*}
The structure function at small separations then has the same functional form as
equation~\ref{eq-asymp} for a power-law index of $q = 4$, leading to possible
confusion between power spectra with high-frequency cut-offs and steep power
laws extending to high frequencies.

\section{Rotation and depolarization at longer wavelengths}
\label{SP3depol}

In this Appendix, we simulate the Faraday rotation and depolarization expected
from our model RM power spectrum under conditions where the short-wavelength
approximation does not hold, primarily to aid interpretation of other
observations.

\begin{figure}
\begin{center}
\epsfxsize=6.5cm
\epsffile{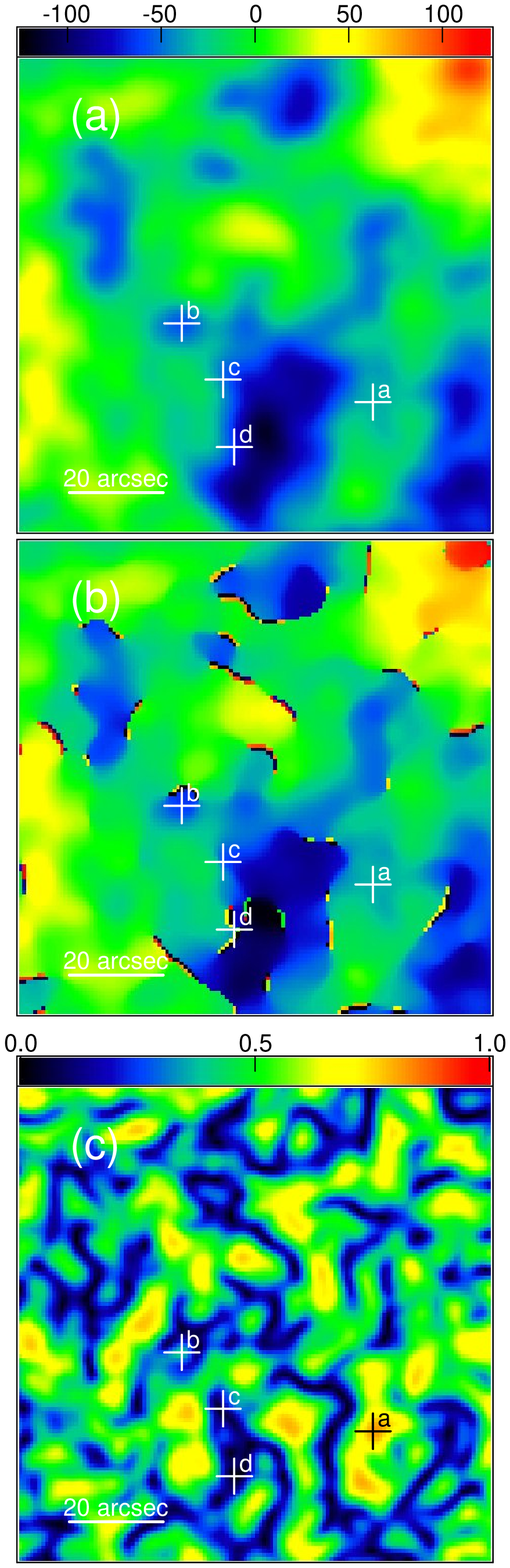}
\caption{Simulated RM and $p$ images generated from a realisation of the BPL 
power spectrum described in the text, computed on an $8192 \times
8192$ grid with 0.05\,arcsec pixels. The amplitude is $D_0 = 0.38$ and the
remaining parameters are as given in Table~\ref{tab:fitparams}.  The crosses
show the locations for the $\chi$ -- $\lambda^2$ plots in
Fig.~\ref{fig:SP3simPA}. (a) Model RM image convolved with a 5.5-arcsec FWHM
Gaussian. (b) Model RM image of the same area derived by generating $Q$ and $U$
images at 5 observing frequencies, convolving them to 5.5-arcsec resolution and
fitting $\Delta\chi = {\rm RM}\lambda^2$ as for the observations. (c) Ratio of
the degree of polarization at 1365\,MHz to the value at infinite frequency,
derived from from the convolved $Q$ and $U$ images.
\label{fig:SP3sim}}
\end{center}
\end{figure}

\begin{figure}
\epsfxsize=8cm 
\epsffile{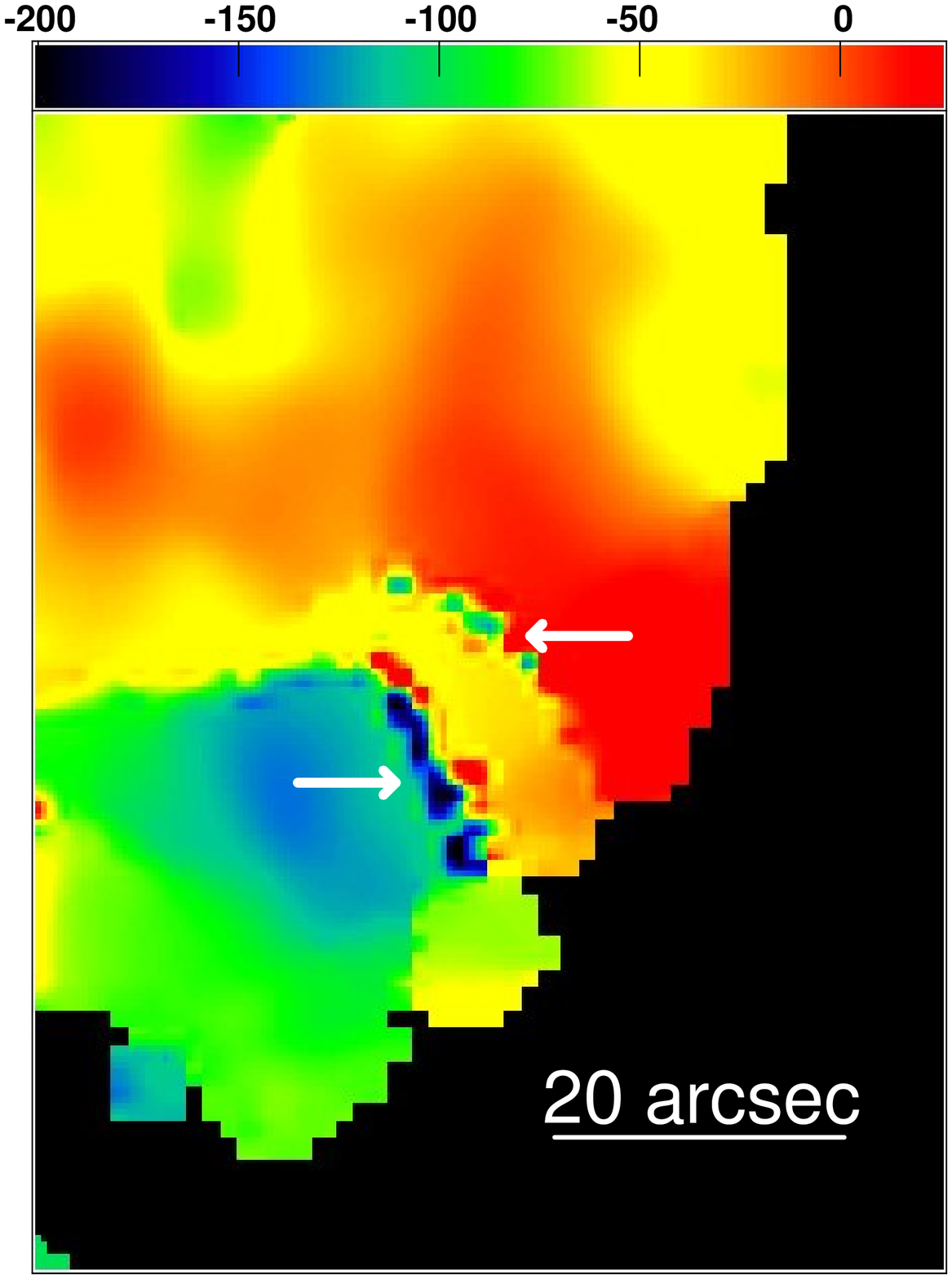}
\caption{Part of the {\em observed} RM image of the South lobe of 3C\,31 from
Fig~\ref{fig:RMDP5.5}(c).  Two examples of artefacts in regions of high RM
gradient are indicated by white arrows.
\label{SP3_RM_artefact}}
\end{figure}

\begin{figure}
\epsfxsize=8.5cm
\epsffile{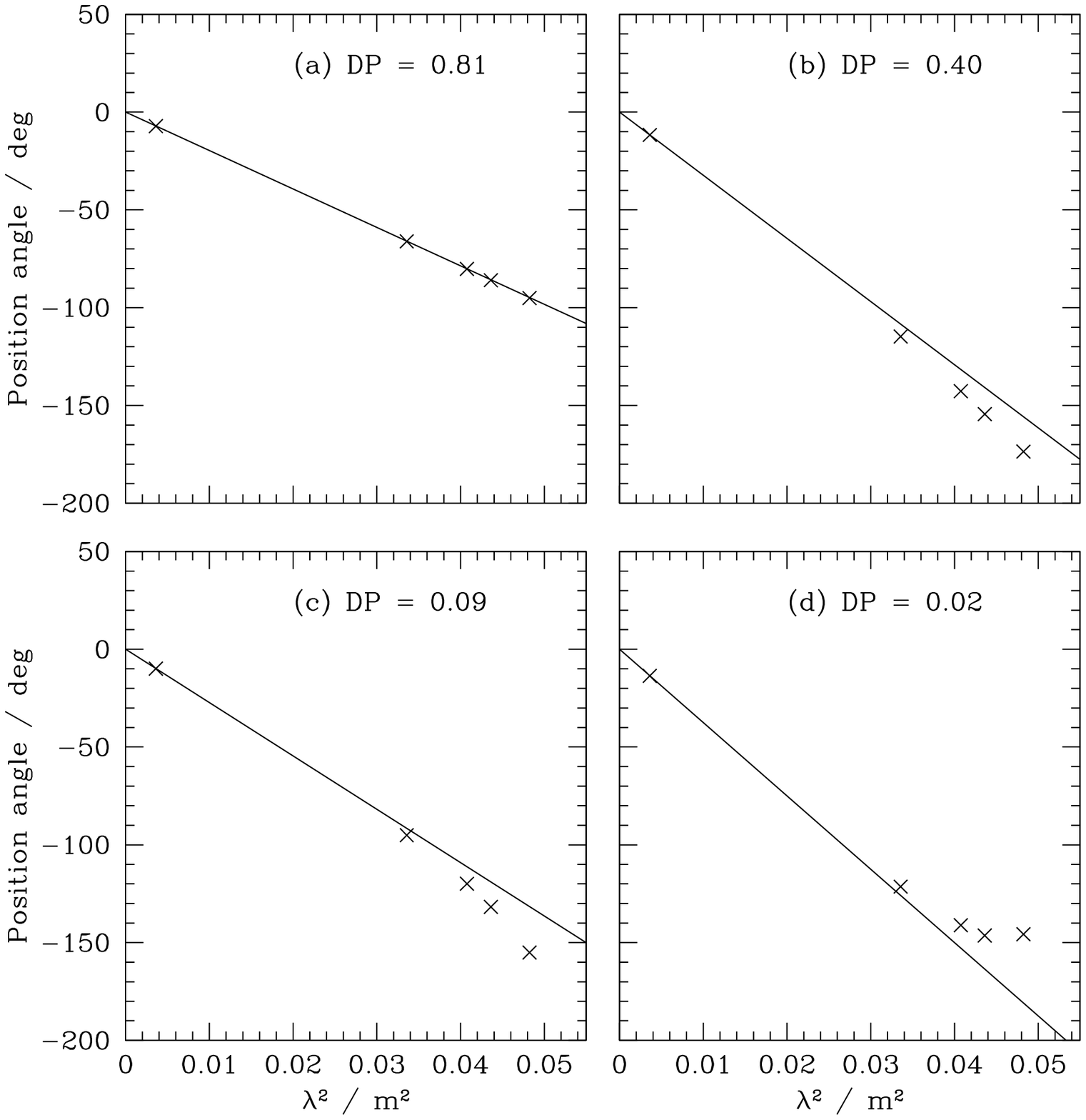}
\caption{Plots of position angle against $\lambda^2$ for simulated data with the
  BPL RM power spectrum and an amplitude $D_0 = 0.38$, as required to match the
  depolarization in region SP3 of 3C\,31.  The corresponding RM and
  depolarization images are shown in Fig.~\ref{fig:SP3sim}, where the locations
  corresponding to individual panels are marked.  The full lines correspond to
  the RM's from Fig.~\ref{fig:SP3sim}(a) and DP is the ratio of the polarization
  at 1365\,MHz to its intrinsic value.  The panels are arranged from least to
  most depolarized. Panel (a) has DP = 0.81 and shows accurate $\lambda^2$
  rotation at the correct RM. Panels (b) -- (d) correspond to positions with
  higher RM gradients, stronger depolarization (DP $\ll$ 1) and significant
  deviations from $\lambda^2$ rotation. At position (d), the RM fitting
  algorithm has become confused, leading to the RM anomaly visible in
  Fig.~\ref{fig:SP3sim}(b).
\label{fig:SP3simPA}}
\end{figure}

In Fig.~\ref{fig:SP3sim}, we show the results of a simulation using the method
described in Section~\ref{innerscale}. The RM power spectrum has the BPL form
with the parameters given in Table~\ref{tab:fitparams} and an amplitude, $D_0 =
0.38$, chosen to match the observed depolarization in region SP3.
Fig.~\ref{fig:SP3sim}(a) shows the simulated RM image convolved to a resolution
of 5.5\,arcsec.  The image calculated by evaluating $Q$ and $U$ on a fine grid,
convolving them to 5.5\,arcsec resolution, deriving the position angle $\chi$
and fitting for the RM is shown in Fig.~\ref{fig:SP3sim}(b). Although this is
close to the convolved RM image in most locations, it also shows discrepant
values associated with large RM gradients across the beam. These artefacts are
extended perpendicular to the direction of maximum RM gradient and are always
accompanied by strong depolarization, as shown in Fig.~\ref{fig:SP3sim}(c). We
see a few artefacts of the predicted type in the observed 5.5\,arcsec RM image,
again associated with high RM gradients: two examples are shown in
Fig.~\ref{SP3_RM_artefact}.  The reason for the artefacts is that there are
large deviations from $\lambda^2$ rotation, and the RM fitting algorithm does
not correct properly for the $n\pi$ ambiguities in ${\bf E}$-vector position
angle.  This is illustrated in Fig.~\ref{fig:SP3simPA} by plots of $\chi$
against $\lambda^2$ at the points labelled a -- d in Fig.~\ref{fig:SP3sim}
(these are ordered from least to most depolarized). At position a, the RM
gradient is small and there is negligible deviation from $\lambda^2$
rotation. Positions b and c are intermediate cases: $\Delta\chi$ is not perfectly
proportional to $\lambda^2$, but the deviations are small enough for the RM
fitting algorithm to find a reasonable approximation to the correct
solution. Finally, position d is in a region of high RM gradient with large
deviations from $\lambda^2$ rotation. The fitting algorithm then introduces
spurious $n\pi$ corrections to the position angles and finds an incorrect RM, as
is clear from Fig.~\ref{fig:SP3sim}(b).  Such deviations from $\lambda^2$
rotation can occur even at points with high signal-to-noise ratio, and may not,
therefore, be corrected even by fitting algorithms like {\sc pacerman}
\citep{Pacerman}, which take account of the spatial coherence of RM. 

For the range of RM, observing frequency and resolution appropriate to our
observations of 3C\,31, such artefacts do not pose a serious problem, however.
Firstly, they should cover a small fraction of the area, even in SP3. Secondly,
they inevitably have low polarized flux density at low frequency and are
therefore usually excluded from RM fits on grounds of low signal-to-noise ratio.

\end{document}